\documentclass[1p]{elsarticle}

\usepackage{subcaption}

\usepackage[utf8]{inputenc}
\usepackage{amsmath,amssymb}

\usepackage{graphicx}
\usepackage{grffile}

\DeclareGraphicsExtensions{.pdf}

\begin{document}
\begin{frontmatter}

\title{On the Modeling of Musical Solos as Complex Networks}%\tnoteref{label1}}
%\tnotetext[label1]{}
\author{Stefano Ferretti}%\corref{cor1}\fnref{label2}}
\ead{s.ferretti@unibo.it}
\ead[url]{http://www.cs.unibo.it/sferrett}
% \fntext[label2]{}
%\cortext[cor1]{Department of Computer Science, University of Bologna}
\address{Department of Computer Science and Engineering, University of Bologna\\ Mura Anteo Zamboni 7, 40127 Bologna, Italy}%\fnref{label3}}
%% \fntext[label3]{}

\begin{abstract}
Notes in a musical piece are building blocks employed in non-random ways to create melodies. It is the ``interaction'' among a limited amount of notes that allows constructing the variety of musical compositions that have been written in centuries and within different cultures. Networks are a modeling tool that is commonly employed to represent a set of entities interacting in some way. Thus, notes composing a melody can be seen as nodes of a network that are connected whenever these are played in sequence. The outcome of such a process results in a directed graph. By using complex network theory, some main metrics of musical graphs can be measured, which characterize the related musical pieces. In this paper, we define a framework to represent melodies as networks. Then, we provide an analysis on a set of guitar solos performed by main musicians. Results of this study indicate that the presented model can have an impact on audio and multimedia applications such as music classification, identification, e-learning, automatic music generation, multimedia entertainment.
\end{abstract}
%\nodate{}


\end{frontmatter}

%%%%%%%%%%%%%%%%%%%%%%%%%
\section{Introduction}
%%%%%%%%%%%%%%%%%%%%%%%%%

Nowadays, there is a common trend in research to model everything as a network, i.e., systems or data which can be represented by graphs. 
In particular, complex network theory is a mathematical tool that connects the real world with theoretical research, and is employed in many fields. Networks are employed across a multitude of disciplines ranging from natural and physical sciences to social sciences and humanities 
\cite{Boccaletti2006175,newmanHandbook}.

Technological, biological, economic systems, disease pathologies, protein-protein interactions, 
can be modeled in the same way.
Focusing on multi\-me\-dia contents, it has been proved that language, for instance, can be seen as a system that can be represented as a complex network 
\cite{BiemannRW12,Cancho2261,Cong2014598,Grabska,pardo}.
As human language has a non-random structure, since it is used by humans to construct sentences from a limited amount of discrete units (words), also music is created by combining notes played by a set of instruments.
In this paper, we show that musical pieces can be treated as complex networks as well (we will focus on melodic lines played by a single instrument).

When dealing with audio, the main concern has been on the issue of di\-gi\-talizing it, in the most efficient way, or to synthesize, represent, reproduce sounds, by employing a variety of sound generation techniques.
Attention has been paid on transmitting, indexing, classifying, clustering, summarizing music 
\cite{Anglade09genreclassification,doi:10.1080/09298210802479268,Li2011,McKayF06,Patra:2013,SchnitzerEtAl_2009,Su:2012,Yang:2006}.
However, the idea of capturing some general characteristics of a melody (and harmony) is somehow an overlook aspect.
In literature, there are works in the field of computer science that focus on musical scores
\cite{Prather:1996,ASI:ASI20876}. 
There are studies on the development of di\-gi\-tal libraries and on human interaction with musical scores and more general notational/instructional information objects \cite{ASI:ASI20876}.
Other works are on the automatic transcription of the melody and harmony \cite{Ryynanen:2008}.
As concerns music information retrieval, a goal is to devise automatic measurements of the similarity between two musical recordings, based on the analysis of audio contents.  
Techniques worthy of mention are acoustic-based similarity measures \cite{Berenzweig:2004}, compression-based classification methods \cite{Cilibrasi:2004}, statistical analyses and artificial neural networks \cite{Manaris:2005}.
Finally, artificial intelligence techniques have been employed to capture statistical proportions of music attributes, such as pitch, duration, melodic intervals, harmonic intervals, etc.~\cite{Manaris:2005}; outcomes confirm that several essential aspects of music aesthetics can be modeled through power law distributions.

Studies on music can be based on symbolic data (music score sheets) or on audio recordings. Symbolic music data eases the analysis in several music application domains.
For example, finding  the  notes  of a melody in an audio file can  be  a  difficult  task, while with symbolic music, notes are  the  starting  point  for  the analysis.  
Thus, in general traditional musicological concepts such as melodic and harmonic structure are easier to investigate in the symbolic domain, and usually more successful \cite{Knopke2011}.

In this paper, we develop a model that allows capturing some essential features of a musical performance (a music track).
We will focus on melodies, and specifically on musical ``solos'', which are a part of a song where a performer plays (often improvises) a melody with accompaniment from the other instruments \cite{icme16}.
It is quite common in music theory asserting that solos performed by musicians are bound to their technical and artistic skills. Indeed, musicians are recognized for their own ``style'' in playing a solo over a music piece, that identifies a sort of musical ``language'', typical of that musician. 
It is not by chance that an artist can be recognized from others, and that we can classify artists in categories and hierarchies.
Moreover, since during a solo a player (quite often) improvises and creates a melody ``in real time'', he employs typical patterns (licks) he is used to utilize.
The goal of this work is to make a step further toward the identification of the rules and characte\-ri\-stics of the music style of a certain performer. 
If a music line is conceived of as a complex network of musical units (notes, rests) and their relations, it is expected to exhibit emergent properties due to the interactions between such system elements. 
Complex networks provide appropriate modeling for music as a complex system and powerful quantitative measures for capturing the essence of its complexity. 

As a proof of concept, we retrieved and analyzed different solos of some main guitar players.
Namely, the artists are Eric Clapton, David Gilmour, Jimi Hendrix, Allan Holdsworth, B.B.~King, Pat Metheny, Steve Vai, Eddie Van Halen.
The selection of guitar as instrument and these particular artists is motivated by the fact that there is a quite active community of guitar enthusiasts that share musical scores on the Web. Scores are published and formatted, usually, by employing description schemes that are alternative, easier and more intuitive to read with respect to the classic musical sheets.
These schemes are based on guitar tablatures, and there is a wide list of software applications and libraries to handle digital representations of such scores.
The considered artists are prominent musicians; thus, several scores of their music are available online. 
This simplified the creation of the database.

It is worth mentioning that a previous work was published in \cite{Liu2010126}, presenting a methodological approach that is similar to that presented in this work. Nonetheless, the use of the model and the application were different to those considered in this work. In \cite{Liu2010126}, complete scores of classic and Chinese music are considered. Networks are concatenations of a number of different music pieces by the same author, having the same style, to reach 18K notes per composer. Using these aggregate networks, they find scale-free properties, small-world phenomenon, mean shortest distances around $3$ and clustering coefficients around $0.3$.
The application approach followed in this work is different. In fact, the analysis is on solos rather than aggregate music pieces. The focus on solos has the specific goal to eliminate repetitions typical of the main melodies of contemporary pop/rock/blues songs, the main theme of jazz compositions, and above all the rhythmic parts. 
In any case, following the approach employed in \cite{Liu2010126}, in the last part of Section \ref{sec:res} an analysis is performed of concatenated solos of each artist. Obtained results are comparable with those obtained in \cite{Liu2010126}, i.e.~networks are small worlds, with a mean shortest distances around $3$ and clustering coefficients around $0.3$. However, we show that results obtained for concatenated networks are different to the average results for separate solos.

The presented analysis provides novel results and promotes novel applications in the artificial intelligence, didactics and multimedia domains. In particular, the contributions of this work are the following:
\begin{itemize}
 \item It is proposed to represent a melodic track as a network (but this can be extended to a whole instrumental music track); this provides a representation of the entire track and allows calculating some general measures that characterize it. Such a representation provides novel insights into understanding the music composition process and fosters novel applications in this domain.
 \item We show how complex network theory can be profitably employed in a novel application scenario.
 \item The proposed representation network model has been applied to a set of guitar solos. We show that networks associated to different solos do have different topologies; this allows discriminating among different music solos and music styles of different artists, in general. An accurate analysis of such tracks can lead to the extrapolation of general characteristics of a given performer.
 \item We calculate different measures, typical of complex network theory, on the considered networks, and present some aggregate results to characterize these performers. We measure the length of solos, the dimensions of the networks, the degree distribution, distance metrics, clustering coefficient, centrality measures (betweenness and eigenvector centralities) and, finally, we identify that the network representation of certain solos are small worlds. The paper discusses how these metrics are related to the ``style'' of the performer.
 \item Through statistical tests, we show that certain musicians have statistically significant differences, by looking at the considered metrics. 
\end{itemize}

As mentioned, the outcomes of such study can have an impact on multimedia applications, intelligent systems and on studies of music classification and identification, in general.
While probably a music track cannot be fully described via mathematical measurements, nonetheless, these measures can help in discriminating among the main features of a performer and a music track.
Such results can be employed as building blocks inside media applications for the automatic generation of digital music with certain specific characteristics (e.g., the generation of a solo ``à la'' Miles Davis). Such applications could be extensively exploited in didactic scenarios, automatic music generation applications, and multimedia entertainment.

The reminder of this paper is organized as follows.
Section \ref{sec:terminology} presents some basic terminology employed in the rest of the paper. 
Section \ref{sec:model} describes the network model for music solos. 
Section \ref{sec:metrics} presents the main metrics, typically employed in complex network theory, that are used to characterize musical solos.
Section \ref{sec:assessment} discusses on an assessment on a list of solos performed by a set of prominent guitar players.
Section \ref{sec:res} presents the obtained results.
Finally, Section \ref{sec:conc} provides some concluding remarks.

%%%%%%%%%%%%%%%%%%%%%%%%%
\section{Terminology}
\label{sec:terminology}
%%%%%%%%%%%%%%%%%%%%%%%%%

Here, a brief terminology is introduced to avoid potential ambiguities.
A song is a musical piece, which is composed of multiple, simultaneous sounds played by different instruments.
The part of the song played by a single instrument is referred as a track. Notice that an instrument can play different tracks in the song (e.g.~there are multiple instruments of the same type, or the tracks have been overdubbed).

Instruments (e.g.~guitar) can have a rhythmic (accompaniment) and/or a melodic role. These roles are not exclusive and in the same track an instrument can play both roles during different moments of the song.

The solo is a part of a track where a performer is playing with unobtrusive accompaniment from the other instruments. 
(Note that even if the solo is a subset of the track, in the following these two terms are used as synonyms, since they both represent a sequence of notes, chords and rests.)
It is employed quite often in jazz, blues, rock songs, where the solo has, usually, the twofold role of: i) creating a melody (in certain cases an improvised melody), which is alternative to the main melody of the song; and ii) showing off the skills of the performer, due to the technical difficulties to play that solo, or due to the ability of the performer to create an intense, touching melody.
Needless to say, solos are melodies composed of notes (or groups of simultaneous notes). A note can be a pitched sound or a rest, lasting for a certain duration.
Thus, notes co-occur in melodies.
We define links as co-occurrences of notes in the melody. 
More specifically, the approach considers that there is a strict relation (i.e.~link) between two notes if these are neighbors (i.e.~subsequent).
This allows capturing a quantitative measure of correlation among notes in the melody.

%%%%%%%%%%%%%%%%%%%%%%%%%
\section{A network model for musical solos}
\label{sec:model}
%%%%%%%%%%%%%%%%%%%%%%%%%

We mentioned that in this model, a note can be a pitched sound or a rest, lasting for a certain duration.
Thus, two notes of the same pitch with different durations are considered as two different network nodes.
Rests and chords are other possible nodes of a network. 
In fact, it is quite usual to hear performers playing multiple simultaneous notes to create multiple voices in the melody they are composing in the solo (when the employed instrument allows this, such as with a guitar or a piano).

We can represent a track as a directed network, whose nodes are the notes played by the performer. When a performer plays a note $x$, followed by a subsequent note $y$, we add the two nodes $x, y$ in the network and a directed link $(x, y)$ from $x$ to $y$. 
If, for instance, the player subsequently plays another note $z$, we add another node $z$ and a link $(y, z)$ leaving from the already existing node $y$ to $z$.
Networks can have cycles, i.e~a performer can play two subsequent notes of the same type.
Weights can be associated to links $(x, y)$, depending on how many times that link $(x, y)$ is present in the solo, i.e.~how many times the performer plays a sequence of the two $x, y$ notes.

From a mathematical point of view, the simplest form to represent the network is through the use of the adjacency matrix $A$, i.e., a $n \times n$ matrix (where $n$ denotes the amount of nodes of the network) with elements 
  \[
    A_{xy}=\left\{
                \begin{array}{ll}
                  1 & \text{when there is a link from vertex $x$ to vertex $y$}\\
		  0 & \text{otherwise}
                  \end{array}
              \right.
  \]
For weighted networks, we can replace the binary possible values of $A_{xy}$ with a (non-negative) associated weight.
  
\begin{figure}
\centering
  \includegraphics[width=.7\linewidth]{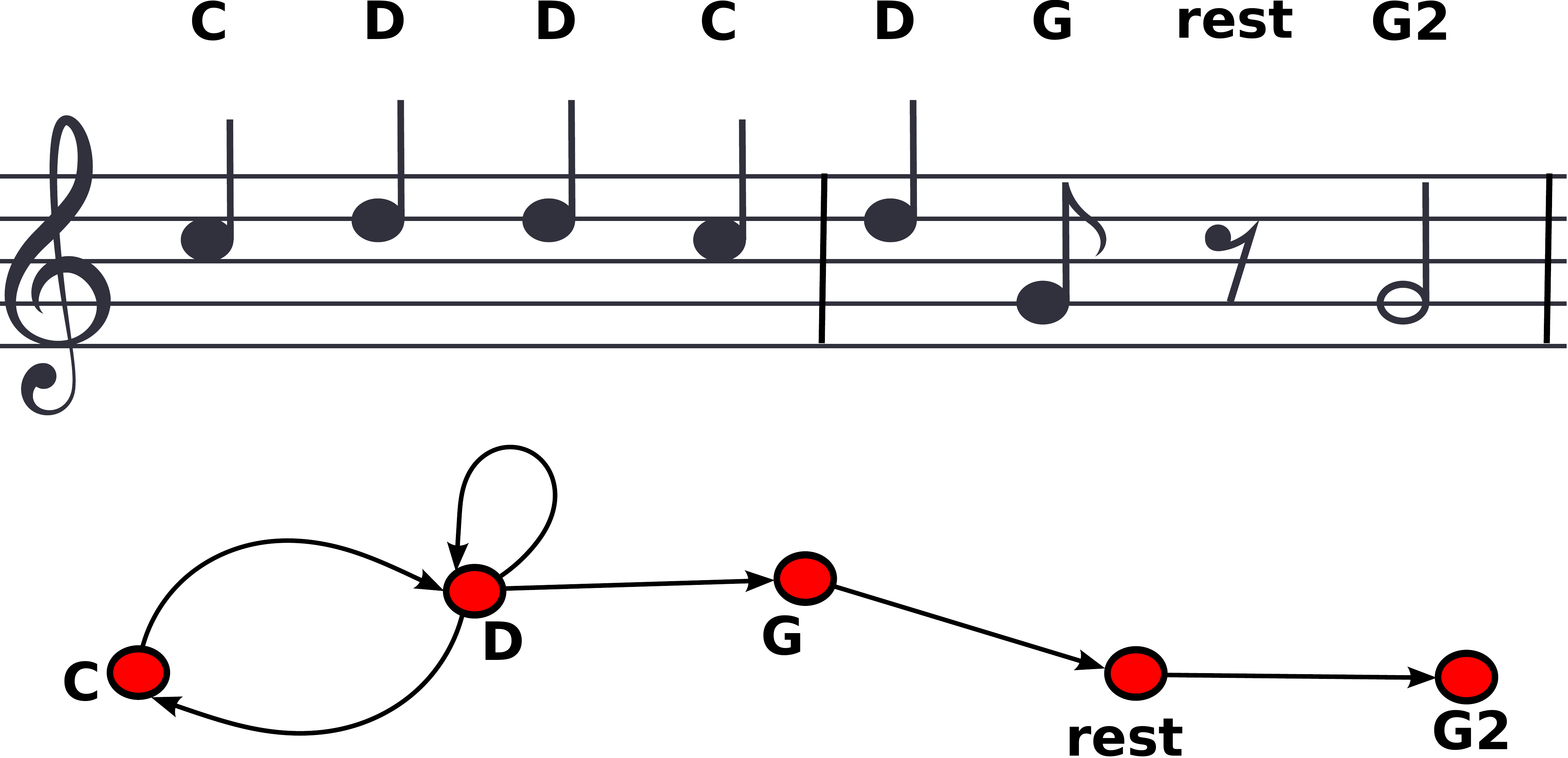}
\caption{Example of melodic line mapped to a network}
\label{fig:netExample}
\end{figure}

Figure \ref{fig:netExample} shows an example of a simple music score sheet of a melodic line, and the associated network. On top of the score sheet, the text label over each note represents the name of the note. The same label is shown next to the network node associated to the note.
In the network, a link is created from the $C$ node to $D$, $(C, D)$, since the first note on the sheet is a $C$, followed by a $D$. Then, a self loop $(D, D)$ is added to the network, since the third note on the sheet is a $D$, again. The fourth note is a $C$, that corresponds to the $(D, C)$ link. 
Then, there is a sequence of links $(D, G), (G, \text{rest}), (\text{rest}, G2)$. Note that there are two different nodes for the two $G$ notes, since their pitch is the same (i.e.~$G$), but they have different duration (the first $G$ is a eighth note, while the second one is a half note).
It is possible to observe that there is a single link $(C, D)$, while the sequence of a $C$ note to a $D$ note is played twice in the score sheet.
To mark this, a weight might be added to links, so that the $(C, D)$ link has a double weight with respect to other links, which correspond to note pairs that are played just once.

To sum up, according to the employed model nodes correspond to specific notes. Labels associated to nodes vary depending on the type of note.
In case of a single note, the related node has a label composed of the note pitch, octave and duration. A ``rest node'' is labeled with the duration of the rest. Finally, nodes corresponding to chords are labeled with the pitch, octave and duration of each note composing the chord.

% fai vedere qualche esempio di rete e di assolo
\begin{figure}
\centering
\begin{subfigure}{0.45\textwidth}
   \centering
  \includegraphics[width=\linewidth]{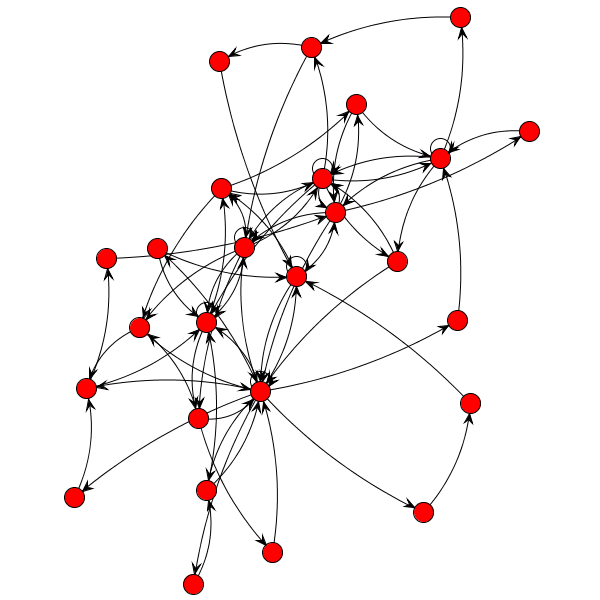}
   \caption{B.B.~King -- Rock me baby}
  \label{fig:bbking}
\end{subfigure}%
\hspace*{\fill}%
\begin{subfigure}{.45\textwidth}
  \includegraphics[width=\linewidth]{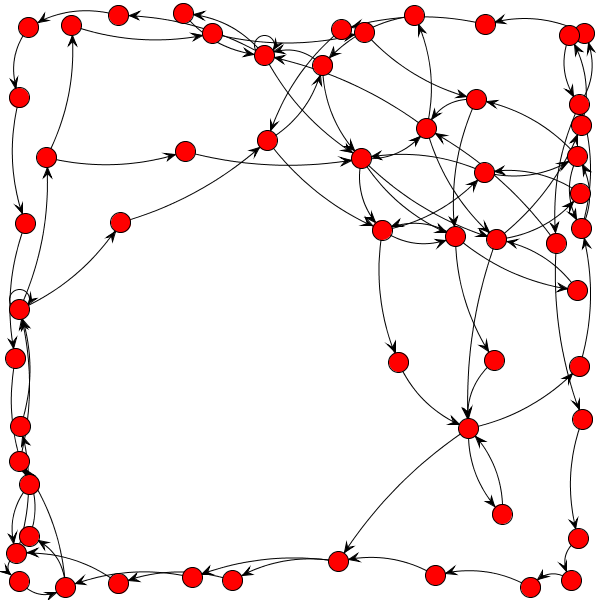}
  \caption{Pink Floyd -- Comfortably numb (first solo)}
  \label{fig:confNumb}
\end{subfigure}
\begin{subfigure}{.45\textwidth}
  \centering
  \includegraphics[width=\linewidth]{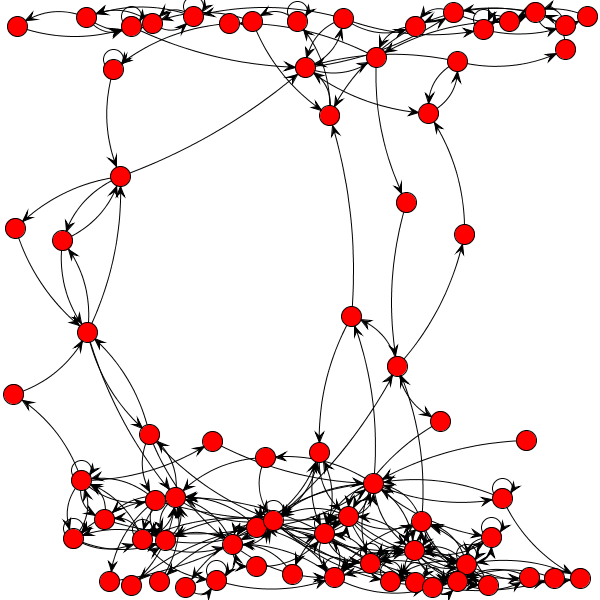}
  \caption{Eric Clapton (Cream) -- Crossroads (second solo)}
  \label{fig:crossroads}
\end{subfigure}%
 \hspace*{\fill}%
 \begin{subfigure}{.45\textwidth}
  \centering
   \includegraphics[width=\linewidth]{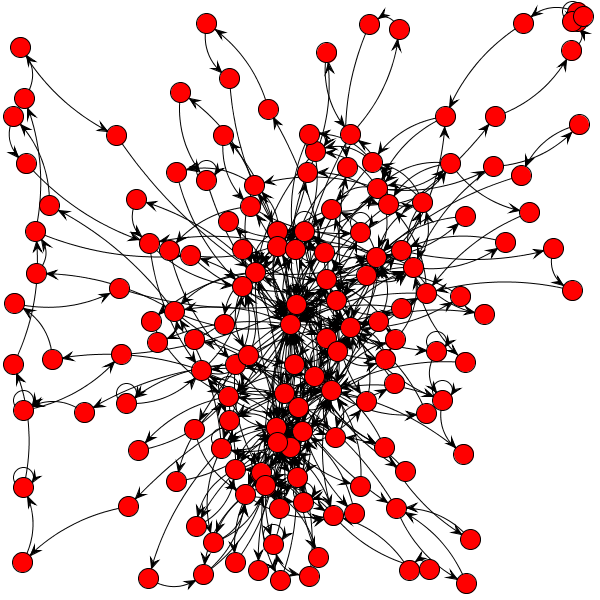}
  \caption{Jimi Hendrix -- Red House}
  \label{fig:redHouse}
\end{subfigure}
\caption{Four examples of networks from different scores}
\label{fig:someNets}
\end{figure}

As concerns the amount of possible nodes in a network, in the occidental music an octave (the interval between one musical pitch and another with half or double its frequency, which is the same note lowered or raised of an octave), is composed of twelve sounds.
Focusing on electric guitar, as an example, it is usually possible to create sounds belonging to four octaves. 
(Actually, this is a simplified measure, just to give an idea on the amount of possible nodes; in fact, 
a higher amount of sounds can be produced if one considers the possibility of playing the so called ''harmonics``, i.e.~pitch notes quite higher than ''usual ones``. Moreover, in certain cases strings can be tuned differently to the standard tuning, so as to produce lower pitches).
Then, each note has an associated duration. 
Furthermore, it is possible to create and employ chords (e.g.~bi-chords, two notes played simultaneously, are quite often exploited in guitar melodies). 
Therefore, according to this model a guitar solo (and, in general, a track) can be associated to a network composed of hundreds of nodes.

Figure \ref{fig:someNets} depicts four examples of networks derived from four famous blues/rock guitar solos. 
Network image representations have been generated using the APIs provided by the open-source software JUNG (Java Universal Network/Graph) Framework \cite{jung}. In particular, given a network a graph image is created, using a layout (i.e.~position of nodes) that tries to avoid (or minimize) nodes and links overlapping. Thus, the spatial position of nodes has no meaning with respect to musical aspects, and spatially near nodes do not necessarily refer to similar notes.

It is interesting to observe that these networks are quite different one from the other.
Figure \ref{fig:bbking} shows a simple network, with some nodes that have higher in/out degrees (i.e., number of links entering or leaving a node). Figure \ref{fig:confNumb} has a more linear structure, suggesting that the melodic line was ``simple'', with poor repetitions of single notes.
Figure \ref{fig:crossroads} appears to be a clustered network, with some few nodes connecting the two clusters.
Finally, Figure \ref{fig:redHouse} has a quite complex structure, with many nodes and with the presence of several hubs.
% It is worth noting that every artist ''produces`` networks with different structures and shapes; anyhow, 
This suggests that it might be interesting to assess if different artists do have different characteristics that, statistically, produce different types of networks.

An interesting feature of this model is that it describes the track by focusing on its structure and its topology. The network is composed by different notes, but the main aspect is that these notes create a novel node when they are different from others played in the solo. 
The structure of the network is not influenced, for instance, by the tonality of the song, i.e.~the obtained network would be the same if a performer plays the same solo but in different keys. Similarly, we would obtain the same network if the same solo is played at different tempos.

It is worth noticing that usually a melodic line is strongly influenced by the underlying harmonic, chord progression. Modulations and chords' alterations are quite common in jazz compositions, where improvisations use and outline the harmony as a foundation for melodic ideas.
On one hand, it is recognized that complex chord progressions do offer performers the ability to vary their melodic lines. On the other hand, 
some other artists assert that a complex chords structure represents a constraint for the improviser, which is forced to follow that structure, while a simpler harmony offers much more freedom for improvising on each individual chord change \cite{barrett,Boothroyd}.
Without going into this debate, we simplify this matter by simply ignoring the chord progression of the considered compositions.
Hence, while important, in this study we do not focus on the harmonic aspects, just taking the melodic line to create the networks.
This can be regarded as a further work.
Anyhow, the harmonic structure implicitly influences the melody that has been created by the artist.

%%%%%%%%%%%%%%%%%%%%%%%%%
\section{Metrics of Interest}
\label{sec:metrics}
%%%%%%%%%%%%%%%%%%%%%%%%%

This section presents the main metrics, typically employed in complex network theory, that are used to characterize musical solos.

\subsection{Length of solos}

The length of a musical track is the amount of notes, chords or rests composing the track. 
This metrics is different to the temporal duration time of the solo, even if these two metrics are related. In fact, during a bar, one might play a fast sequence of notes with short duration (e.g.~a set of $16$ sixteenth notes) that would occupy the same 
temporal time of a single whole note.

This metrics expresses how much a performer is inclined to elaborate the melodic line he is creating during the solo. But it is also strongly related to the music genre. For instance, solos in jazz compositions are usually quite longer than modern pop-rock ones.

\subsection{Number of nodes}

This measure is the total amount of nodes in a given network, that is the total amount of different notes that have been played during the solo. 
The number of nodes differ from the length of solos, since the length of solo counts multiple times the same note $x$ during a solo, i.e.~$x$ is counted each time the note is played. Conversely, the amount of nodes counts a note $x$ once, if that note has been played during the solo.

\subsection{Degree distributions}

The degree of a node $x$ is the amount of links that connect $x$ with other nodes in the network (included $x$ itself, when a loop is performed). The degree counts how many times the performer decides playing a note, after (and before) playing another one.
Since the graph is directed, it is worth considering also the in-degrees (number of links arriving at a certain node) and out-degrees (number of links leaving a certain node). The in-degree $d^{in}_x$ of a node $x$ can be written in terms of the adjacency matrix as 
$$d^{in}_x = \sum_y A_{xy},$$
while the out-degree is 
$$d^{out}_x = \sum_y A_{yx}.$$ 
Thus, the degree of a node is the summation between its in-degree and out-degree, i.e.
$$d_x = d^{in}_x + d^{out}_x.$$ 

A normalized degree is the degree of a node divided by the maximum degree value, so as to obtain a measure between $0$ and $1$. 

Weights can be associated to links and exploited to measure the so called weighted degrees. In this case, a weight is assigned to each link, measuring the amount of times the solo ``traverses'' that link (i.e., how many times the performer played those two notes in sequence). Thus, the weighted degree is the summation of the weights of links of a given node.

\subsection{Distances}

The average distance is the average shortest path length in a network, i.e.~the average number of steps along the shortest paths for all possible pairs of nodes.
The distance between two nodes $x, y$ is the shortest sequence of notes played in a solo starting with $x$ and ending with $y$.

It is worth noting that when considering distances in most directed networks obtained from the considered solos, the average path length is infinite, since such directed networks are not strongly connected, i.e.~in these networks it is not possible to create a path from a node to another, for all possible pair of nodes, by employing their directed links. 
However, these nets are typically weakly connected; that is, the corresponding undirected network, obtained by removing the direction information and considering links as bidirectional ones, is connected. Hence, it is possible to find an undirected path from each network node to any other node.
For instance, in the sample network of Figure \ref{fig:netExample} it is not possible to go from $G$ to $C$ when we consider the directed net, but a path exists if we consider the undirected network.
Distances among nodes are calculated using the standard breadth-first search algorithm, which finds the shortest distance from a single source node to every other node in the network.

In this application domain, small networks have a low distance (as we will see for for the solo depicted in Figure \ref{fig:bbking}). Thus, a higher distance reveals a higher complexity of the solo, in terms of amount of notes.
When we compare networks of the same size, a higher average path length means that going from one note to another a high amount of notes should be traversed, on average. In some sense, similarly to geographical networks where higher distances suggest that only local connections exists, without high jumps from one geographical area to another \cite{watts1998cds}, in the case of musical networks one might assume that the player is used to play ``near'' notes, meaning that he prefers combining certain groups of notes in his solo.

This metrics should be considered together with the clustering coefficient. In fact, these two metrics allow determining if the network is a small world or not (see next subsections).

\subsection{Clustering coefficient}

The clustering coefficient is a measure assessing how much nodes in a graph tend to cluster together.
It measures to what extent friends of a node are friends of one another too.
When two connected nodes have a common neighbor, this triplet of nodes forms a triangle. The clustering coefficient is defined as 
$$C = \frac{3 \times \text{number of triangles in the network}}{\text{number of connected triplets of nodes}}$$
where a ``connected triple'' consists of a single node with links reaching a pair of other nodes; or, in other words, a connected triple is a set of three nodes connected by (at least) two links \cite{newmanHandbook}. A triangle of nodes forms three connected triplets, thus explaining the factor of three in the formula.

In this context, a triangle of nodes means that the performer played in a solo multiple times the corresponding three notes in sequence, but in different orders. In other words, the presence of many triangles (i.e.~high clustering coefficient) indicates that the performer is used to play different clusters of notes in whatever order. Conversely, a low clustering coefficient might indicate that the performer prefers following a specific melodic line with a specific order on triplets.

\subsection{Small worlds}
\label{sec:sw}

Small world networks are networks that are ``highly clustered, like regular lattices; yet, they have small characteristic path lengths, like random graphs'' 
\cite{watts1998cds}.
In a small world, most nodes are not linked with each other, but most nodes can be reached from every other by a small number of hops. Indeed, in a small-world network the typical distance between two randomly chosen nodes grows proportionally to the logarithm of the number of nodes.

Given a network, it is possible to verify if it is a small world, by comparing it with a random graph of the same size. 
A random graph is a network with links randomly generated, based on a simple probabilistic model \cite{newmanHandbook}.
Different models can be employed to generate a random graph. According to one of the simplest methods, a random graph can be constructed by creating a set of $n$ isolated nodes; then, we consider every possible pair of nodes $x$, $y$, and we add a link $(x,y)$ with probability $p$, independently of other links. 
Random graphs exhibit a small average distance among nodes (varying typically as the logarithm of the number of nodes, $\sim \ln(n)$) along with a small clustering coefficient $\sim\frac{\text{mun links}}{n^2}$.

In practice, one can assess whether a network has a small average distance as for a random graph, but a significantly higher clustering coefficient. In this case, the network is a small world.

In the context of musical solos, a small world network (low distance, high clustering coefficient) corresponds to a solo where nodes are combined and played in various orders, with a significant amount of connections between notes that are in different clusters (or, in some sense, in different ``areas'' of the network). Conversely, a network that does not exhibit the small world phenomenon might can be obtained from a linear solo (low clustering coefficient, as we will see for the solo depicted in Figure \ref{fig:confNumb}), or some  particularly clustered network (but with high average path length), meaning that the player prefers playing certain groups of notes in his solo, moving from one cluster (of notes) to another through some few important notes (then, we should expect that these notes connecting different clusters have high centrality values, as discussed below).

\subsection{Centrality measures: betweenness}\label{sec:bet}

Betweenness centrality is a centrality measure that indicates to what extent a node lies on paths between other nodes.
A node with high betweenness has a large influence in the network. If for instance, we consider a communication network, a node with high betweenness has high control over information passing through the network.
In this context, a node $x$ with high betweenness represents a note that is played within many melodic ``phrases'' (in the music jargon, these are called ``licks''), since if we want to go from a note to another, we find that the path (which can be viewed as a typical phrase utilized by the performer) connecting these two notes passes through $x$.
In other words, the presence of nodes with high betweenness indicates that the player has a prevalence of passing through some main notes in his solos.

Betweenness of a node $x$ is defined as 
$$\text{bet}(x) = \sum_{y \neq x \neq z} \frac{\sigma_{xz}(x)}{\sigma_{xz}},$$ 
where $\sigma_{xz}$ is the total amount of shortest paths in the network going from $y$ to $z$, and $\sigma_{xz}(x)$ is the number of those paths passing through $x$.
This definition of betweenness centrality scales with the number of pairs of nodes, as implied by the summation indices.
When one wants to compare different networks, it is thus convenient to normalize this summation by dividing it by $n^2$ (with $n$ being the number of network nodes), which represents the total number of possible ordered node pairs \cite{Newman:2010}.

\subsection{Centrality measures: eigenvector centrality}

Eigenvector centrality is a centrality measure exploited to rank the influence of a node in a network, similarly to betweenness. 
It denotes the extent to which a node is an important node connected to other important nodes, i.e., it assesses how well connected a node is to the parts of the network with the greatest connectivity \cite{Newman:2010}.
In other words, this measure ranks nodes based on the concept that connections to nodes with higher ranks contribute more to the ranking of the node in question than equal connections to nodes with low ranks.
Thus, notes with high eigenvector represent those that are present in commonly played licks, since they are played together with other commonly exploited notes.

Since this centrality measure responds to the statement that ``a node is important if it is linked to by other important nodes'', we can say that the centrality $e_x$ of a node $x$ proportionally depends on the centrality of its neighbors, i.e.
$$e_x = \sum_y A_{xy} e_y.$$
To solve this problem, an iterative process can be exploited that computes the estimation of this centrality measure by starting from an arbitrary vector $e(0)$ of centrality measures for network nodes, and repeatedly computing the centralities
$$e(t) = A e(t-1) = A^t e(0),$$
where $e(t)$ is the vector of centrality measures of the nodes, computed at the $t$-th iteration of this process, and $A$ is the adjacency matrix \cite{Newman:2010}.
Now, if we write $e(0) = \sum_i c_i v_i$ as a linear combination of eigenvectors $v_i$, then the estimation $e(t)$ can be written in terms of this combination, and $A^t e(0)$ can be written using the eigenvalues $\lambda_i$ of $A$, associated to $v_i$, i.e.
$$e(t) = A^t \sum_i c_i v_i = \sum_i c_i \lambda_i^t v_i = \lambda_1^t \sum_i c_i \Big(\frac{\lambda_1}{\lambda_i}\Big)^t v_i,$$
with $\lambda_1$ being the largest eigenvalue. Thus, by repeating this process in the limit $t \to \infty$, $e(t) \to c_1 \lambda_1 v_1$.
To sum up, the centrality of nodes is proportional to the leading eigenvector of the adjacency matrix $A$, that is $Ae = \lambda_1 e$, or equivalently
$$e_x = \lambda_1^{-1} \sum_y A_{xy} e_y.$$

\subsection{Comparing musical solos}

Given all these metrics, a question is if it is possible to introduce a final measure of similarity between different solos. 
This is an aspect that would probably require further studies. 
An approach is to measure the similarity of solos by comparing the related networks. Network comparison is recognized as a challenging task and a vast literature exists \cite{AliakbaryMHM13,Lu:2014}. Networks can be compared by looking at their degree distributions. 
However, different networks with the same degree distribution may have distinct structural properties \cite{Grisi-FilhoOFA13}.
Thus, the focus on other metrics might provide important indications. In fact, we might compare two networks according to their density, clustering coefficient, average path lengths, centrality measures, or any other structural measure.

Not only, trying to compare the general structure of networks, approaches have been proposed that exploit graph isomorphism methods and heat equations on the graphs \cite{Lu:2014}. 
Then, graph kernels methods exploit machine learning to count similar subgraphs within two compared networks.  
As an example, in \cite{Gartner2003} a kernel method is exploited that measures the similarity of two graphs based on the length of all walks between each pair of nodes in the graphs.
Another proposed approach is based on counting small subgraphs (called motifs), that represent recurring, significant patterns of interconnections \cite{Milo824}.
% Another family of distance measures for network comparison exploits feature vectors to summarize graph topology . 
% For instance, in the attempt to identify a single integrated quantity, in \cite{AliakbaryMHM13} a network distance metric is proposed, which is named NetDistance. NetDistance is based on the use of a feature vector, and by exploiting machine learning a distance metric is computed that compares the structure of the networks and returns the dissimilarity of network topological features.
An alternative family of distance measures for network comparison exploits feature vectors to summarize the graph topology \cite{AliakbaryMHM13}. In particular, through feature vectors and machine learning techniques, a distance metric can computed to obtain the similarity (or dissimilarity) of network topological features.

While in this work we do not propose a novel similarity measure among solos and their related network, we claim that comparing the values of the different metrics might provide an overall view of the likeness of the generated networks, and the characteristics of related solos. With this in view, Table \ref{tab:metrics} summarizes the features provided by the metrics presented in this section.
It is of course a combined analysis (and comparison) of these metrics that allows understanding the specific features of a musical solo, as already mentioned in Section \ref{sec:sw}.

\begin{table*}[th]
\centering
\caption{Employed metrics and some considerations to analyze solos.}
\label{tab:metrics}
\scriptsize
\begin{tabular}{|| l || p{8cm} ||}
  \hline			
  \hline			
  \textit{Metrics} & \textit{Description} \\
\hline  
\hline			
  length of solo & how much a performer is inclined to elaborate the melodic line he is creating\\
\hline			
  number of nodes & it gives an idea of the diversity of exploited notes\\
\hline
  degree distribution & how much notes are connected; it also allows understanding if there are some notes which are more played than others\\
\hline
  distance & it gives an idea of how complex the solo is; larger networks might have higher distances; however, higher average distances mean that the player is used to move ``locally'' to notes he usually plays together\\
\hline
  clustering coefficient & how much notes are clustered, i.e., how much the performer plays notes in an interchangeable order\\
\hline
  betweenness & identifies those notes the player prefers passing through in his solo\\
\hline
  eigenvector centrality & identifies those notes the player prefers passing through in commonly played licks\\
  \hline  
  \hline			
\end{tabular}
\end{table*}

%%%%%%%%%%%%%%%%%%%%%%%
\section{An assessment on guitar solos}
\label{sec:assessment}
%%%%%%%%%%%%%%%%%%%%%%%%%

While the presented approach is applicable to all musical solos and tracks, regardless on the musical instrument, this study focuses on guitar solos. The reason is quite simple: guitar is probably the most popular musical instrument and there is a vast amount of information available on the Web.
Besides classic musical sheets based on the musical staff, alternative notation systems have been devised (e.g.~guitar tablature) and a wide set of software tools is available. 
This allowed creating a wide database of guitar solos, to be used for this study.
The idea of employing solos is due to the fact that during a solo a player creates a melody usually ``in real time''; during this process, it is reasonable to assert that he employs typical patterns (licks) he is used to utilize \cite{barrett,Boothroyd}.

A main clarification is that, even if these measures are quantitative, they do not aim at ranking the ability of performers or the ``beauty'' of a solo with respect to another. These are (subjective) opinions which are not under investigation here. 
The aim of these measurements is to extract some main characteristics of performers and their solos, that may serve to perform classifications, build applications for the automatic matching, identification or even automatic generation of media compositions respecting some musical genre or style.

\subsection{Database creation}

A database was created by downloading ($\sim150$) guitar scores available in several dedicated Web sites (e.g.~A-Z Guitar Tabs, 
\cite{guitartabs}, 
The Ultimate Guitar Tabs \cite{ultimate}).
It is important noting that these scores are cooperatively provided by users; thus, these sources might have some minor errors (that nevertheless, do not alter the general scope of this study). They represent an interesting source of music score sheets that, besides the general didactic contribution, promotes the development of novel application scenarios.
Since the focus here is on a limited number of performers, and solos were retrieved from user-generated databases, the amount of available solos was limited to few hundreds. 

Retrieved scores were available in Guitar Pro or Power Tab formats. 
Through the use of an existing python library, named PyGuitarPro \cite{pyguitar}, 
these scores have been manipulated so as to isolate the solo guitar part, extract the guitar solo and export it in a musicXML format \cite{musicXML}. 
Such a process was
mostly automatic (partially supervised in some circumstances). 
MusicXML is a standard open format for archiving, sharing and exchanging digital sheet music between applications.
As the name suggests, it describes digital score sheet music as XML documents that can be easily managed by means of a XML parser.
It allows describing complex score sheets composed of multiple instruments. Thus, through the implementation of a parser, it was possible extracting notes composing the solos of interest, and creating network nodes based on the notes' pitch, octave and duration.

Solos were isolated from the rest of the track, when usually the guitar plays a rhythmic (and thus repetitive) role.
The idea was to extrapolate the melodic line played by the guitar from the rest of the song. 
In certain cases, the track was completely instrumental. Thus, in this case the melodic line was entirely played by the performer. In other cases, the performer was continuously playing melodic lines during the song, in alternation with the main melodic line (i.e.~voice). 
In both these cases, a supervised process was performed in order to isolate the ``solo'' parts.

We consider aggregate measures involving the whole solos, that have different durations. 
The different durations of solos have an impact on some of the considered quantitative measures (e.g.~amount of played notes and number of nodes in the networks). 
This allows obtaining a general view of the solo, that considers its duration as a main feature. 
An alternative option might have been normalizing these values by dividing them over the duration of the solo. 
But in this case, two options would have been available for the duration, i.e.~a relative measure as the number of bars (segments of time used to provide regular reference points within a piece of music), or a fixed measure such as the number of seconds.

\subsection{Software for the analysis}

The software to perform the analysis was built by using Java, Octave and Python languages. 
An in-house software was developed in Java to handle the musicXML documents.
The open-source software JUNG (Java Universal Network/Graph) Framework has been utilized to manipulate networks and extract the metrics of interest \cite{jung}.
The Apache Commons Mathematics Library was exploited to perform the mathematics and statistics analysis \cite{apache}, together with libraries and functions provided by the Octave language.

\subsection{Selection of artists}
The selection of artists was based on three criteria, i.e.~the ''importance`` of the performer, the amount of songs available in the Web for that performer, the diversity (from a musical point of view) from other performers. 
The idea was to consider a wide range of different musical styles, to see if there are any differences in the corresponding obtained networks. 

We thus chose those musicians that, according to the general music criticism, have a unique playing style. 
Briefly, the artists are: 
\begin{itemize}
 \item Eric Clapton, a rock-blues guitar legend (the slogan ``Clapton is God'' testifies this); 
 \item David Gilmour, singer and guitarist of the famous Pink Floyd rock group, known for his melodic and intense guitar solos; 
 \item Jimi Hendrix, which is considered the most important electric guitar player of all times; 
 \item Allan Holdsworth, a fusion artist noted for his advanced style and his intricate solos;
 \item B.B.~King, a blues master quite often referred as ``The King of the Blues'' that inspired several generations of artists; 
 \item Pat Metheny, probably the most famous contemporary jazz guitar player;
 \item Steve Vai, a well known guitar virtuoso, famous for his ``unique'' approach to the instrument;
 \item Eddie Van Halen, which is considered one of the most influential hard-rock guitarists of the 20th century.
\end{itemize}

%%%%%%%%%%%%%%%%%%%%%%%%%
\section{Results}
\label{sec:res}
%%%%%%%%%%%%%%%%%%%%%%%%%

This section reports some aggregate measures for the considered performers. 
Each figure refers to a given metrics; in the figures, each column corresponds to a performer. For each performer, results for his considered tracks are reported as dots in the column, together with the mean value and standard deviation.
Moreover, tables are reported showing results of statistical hypothesis t-tests, so as to assess if the there are statistically significant differences among different performers.

\subsection{General Metrics: length of solos, number of nodes}

\begin{figure}[htbp]
   \centering
   \includegraphics[width=.8\linewidth]{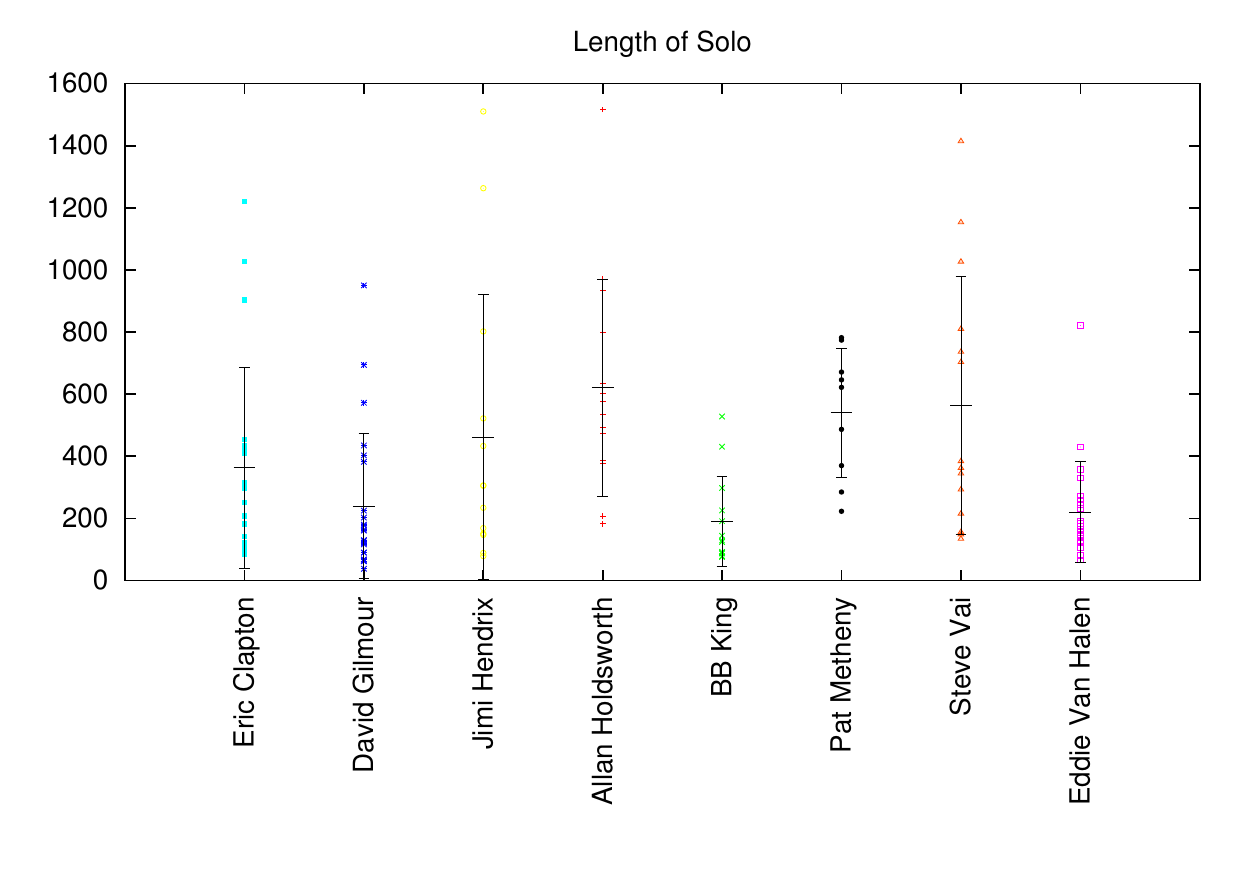}
   \caption{Length of solos}
   \label{fig:soloLength}
\end{figure}

Figure \ref{fig:soloLength} shows the length of solos for different tracks of the considered musicians.
As mentioned, this metrics is strongly related to the music genre, with jazz solos being usually longer than modern pop-rock ones.

\begin{table*}[th]
\centering
\caption{Length of solos -- the number is the smallest significance level at which one can reject the null hypothesis that the two means are equal in favor of the two-sided alternative that they are different. In particular, values in bold are those where there is a significant difference (p = 0.05). Acronyms are used instead of full names.}
\label{tab:soloLength}
\scriptsize
\begin{tabular}{|| l | l | l | l | l | l | l | l ||}
\hline
\hline
 & AH & BBK & DG & EVH & EC & JH & PM\\ 
 \hline
BBK & \textbf{0.0} &  &  &  &  &  & \\
\hline
DG & \textbf{0.0} & 0.45 &  &  &  &  & \\
\hline
EVH & \textbf{0.0} & 0.59 & 0.75 &  &  &  & \\
\hline
EC & \textbf{0.04} & 0.05 & 0.17 & 0.09 &  &  & \\
\hline
JH & 0.33 & 0.06 & 0.12 & 0.09 & 0.5 &  & \\
\hline
PM & 0.49 & \textbf{0.0} & \textbf{0.0} & \textbf{0.0} & 0.09 & 0.6 & \\
\hline
SV & 0.69 & \textbf{0.01} & \textbf{0.02} & \textbf{0.01} & 0.14 & 0.56 & 0.86\\
\hline
\hline
\end{tabular}
\end{table*}

Differences among these results can be appreciated by looking at Table \ref{tab:soloLength}, that shows the results of a statistical t-test made between the sets of length of solos, for each pair of considered musicians. A symmetric matrix is obtained, that is shown in the table (only the lower triangular part is reported). In particular, each value related to each pair of musicians is the smallest significance level at which one can reject the null hypothesis that the two means are equal, in favor of the two-sided alternative that they are different. Simply put, values in bold are those showing a significant difference among the two related musicians (p = 0.05).

It is confirmed that jazz and fusion musicians that usually play instrumental songs (i.e., Holdsworth, Metheny, Vai) do perform longer solos. Thus, there is a significant difference between these musicians and rock/blues ones. As concerns Jimi Hendrix, it shows common features with all other guitar players; hence it is not possible to reject the hypothesis that it is different to others (or better, that others share some characteristics of his style).

\begin{figure}[htbp]
   \centering
   \includegraphics[width=.8\linewidth]{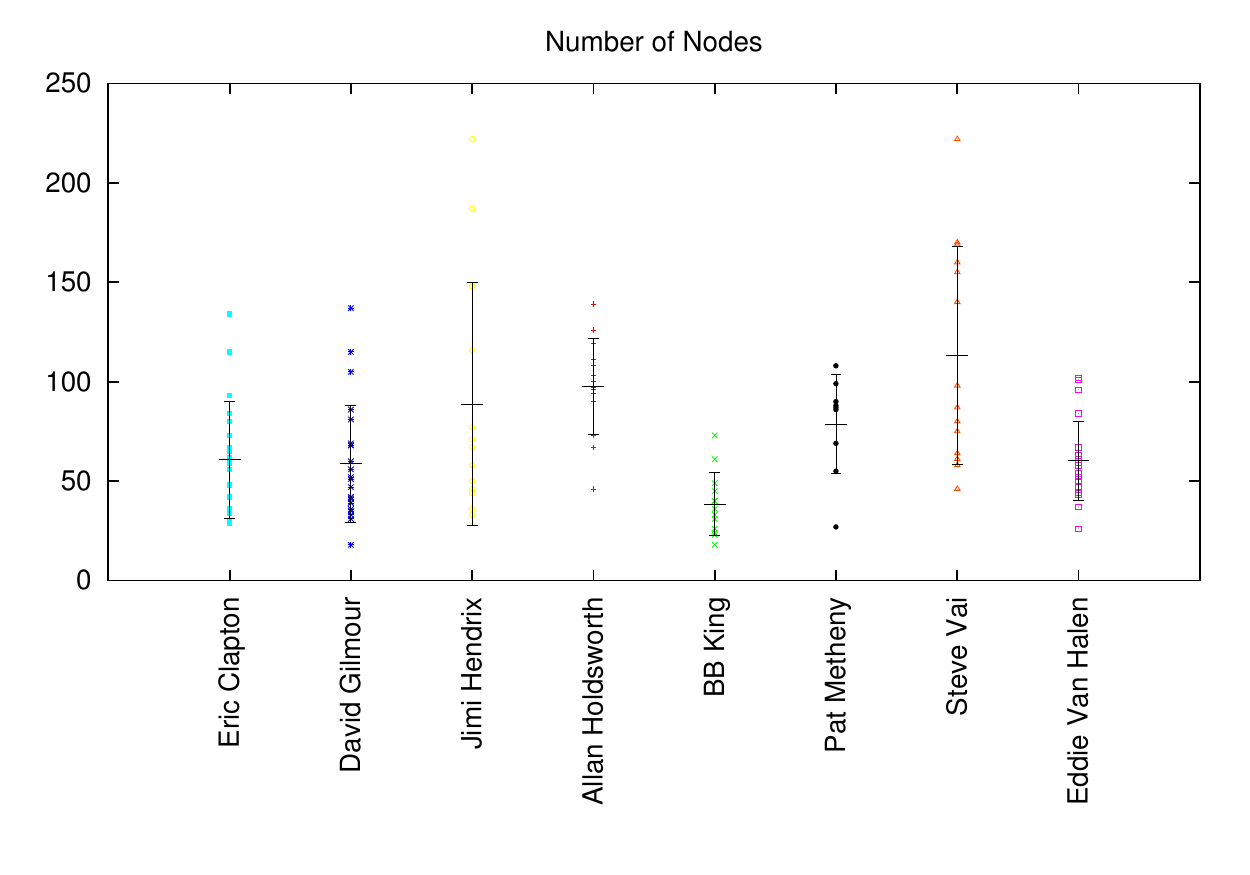}
   \caption{Number of Nodes}
   \label{fig:numNodes}
\end{figure}

\begin{table*}[th]
\centering
\caption{Number of nodes -- the number is the smallest significance level at which one can reject the null hypothesis that the two means are equal in favor of the two-sided alternative that they are different. In particular, values in bold are those where there is a significant difference (p = 0.05). Acronyms are used instead of full names.}
\label{tab:numNodes}
\scriptsize
\begin{tabular}{|| l | l | l | l | l | l | l | l ||}
\hline
\hline
 & AH & BBK & DG & EVH & EC & JH & PM\\ 
 \hline
BBK & \textbf{0.0} &  &  &  &  &  & \\
\hline
DG & \textbf{0.0} & \textbf{0.01} &  &  &  &  & \\
\hline
EVH & \textbf{0.0} & \textbf{0.0} & 0.84 &  &  &  & \\
\hline
EC & \textbf{0.0} & \textbf{0.01} & 0.83 & 0.96 &  &  & \\
\hline
JH & 0.63 & \textbf{0.01} & 0.12 & 0.12 & 0.14 &  & \\
\hline
PM & 0.09 & \textbf{0.0} & 0.07 & 0.07 & 0.1 & 0.6 & \\
\hline
SV & 0.35 & \textbf{0.0} & \textbf{0.0} & \textbf{0.0} & \textbf{0.0} & 0.29 & 0.05\\
\hline
\hline
\end{tabular}
\end{table*}

The metrics above can be analyzed together with the number of nodes, reported in Figure \ref{fig:numNodes}, that measures the number of different notes played during a solo. Table \ref{tab:numNodes} reports the results of the related statistical hypothesis t-test.
Those performers that have longer solos have, in general, higher amounts of nodes in their networks.
Based on this database, results from the figure suggest that Hendrix, Holdsworth and Vai have a richer vocabulary with respect to others. 
Moreover, they capture perfectly the style of B.B.~King, noted for his simpler (yet ``touching'') solos. In fact, his solos have a lower average amount of nodes (and lower lengths), with respect to others.

By looking at Table \ref{tab:numNodes}, B.B.~King is significantly different from others, Vai and Holdsworth differ from rock/blues musicians, while Hendrix and Metheny have a similarity with other musicians (excepting B.B.King).

\subsection{Degree distributions}

\begin{figure}[htbp]
   \centering
   \includegraphics[width=.7\textwidth]{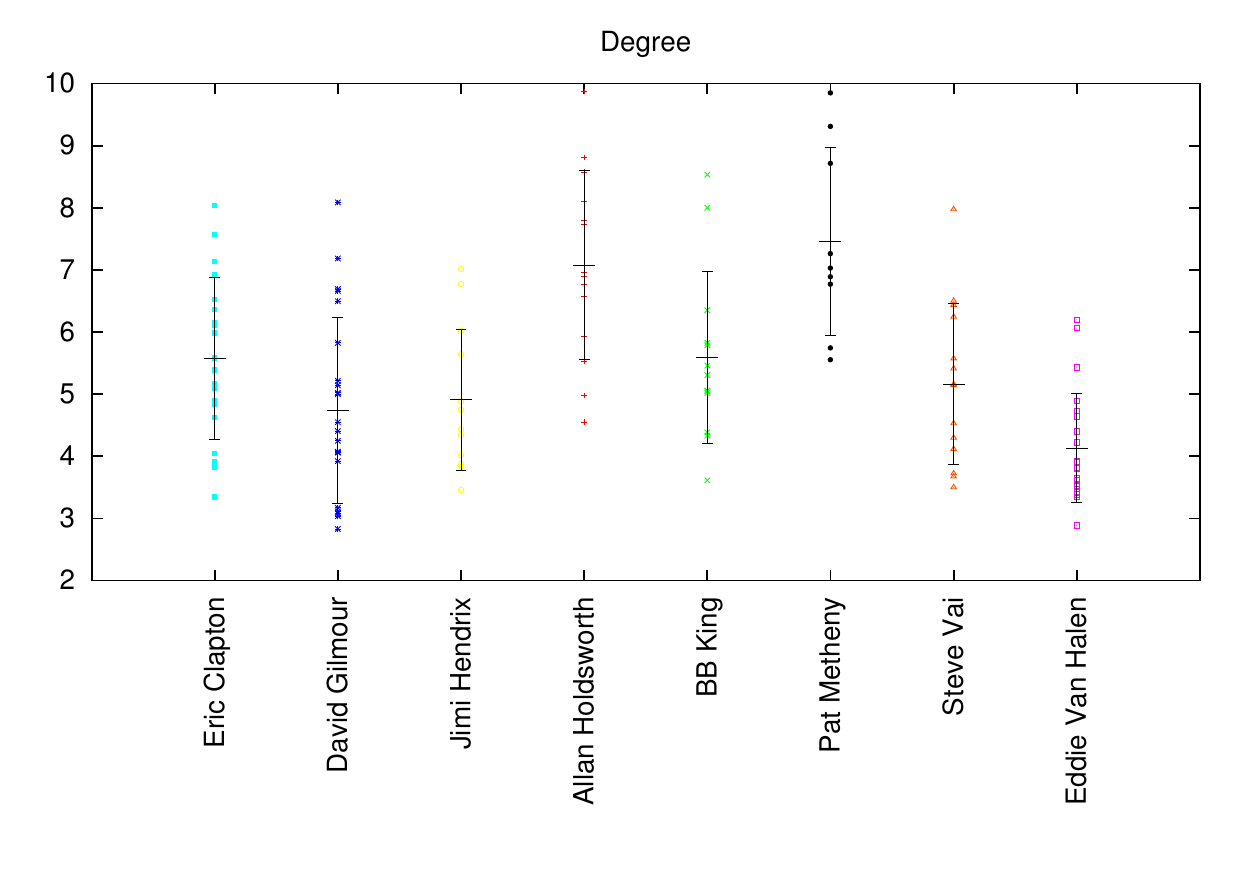}
   \includegraphics[width=.7\textwidth]{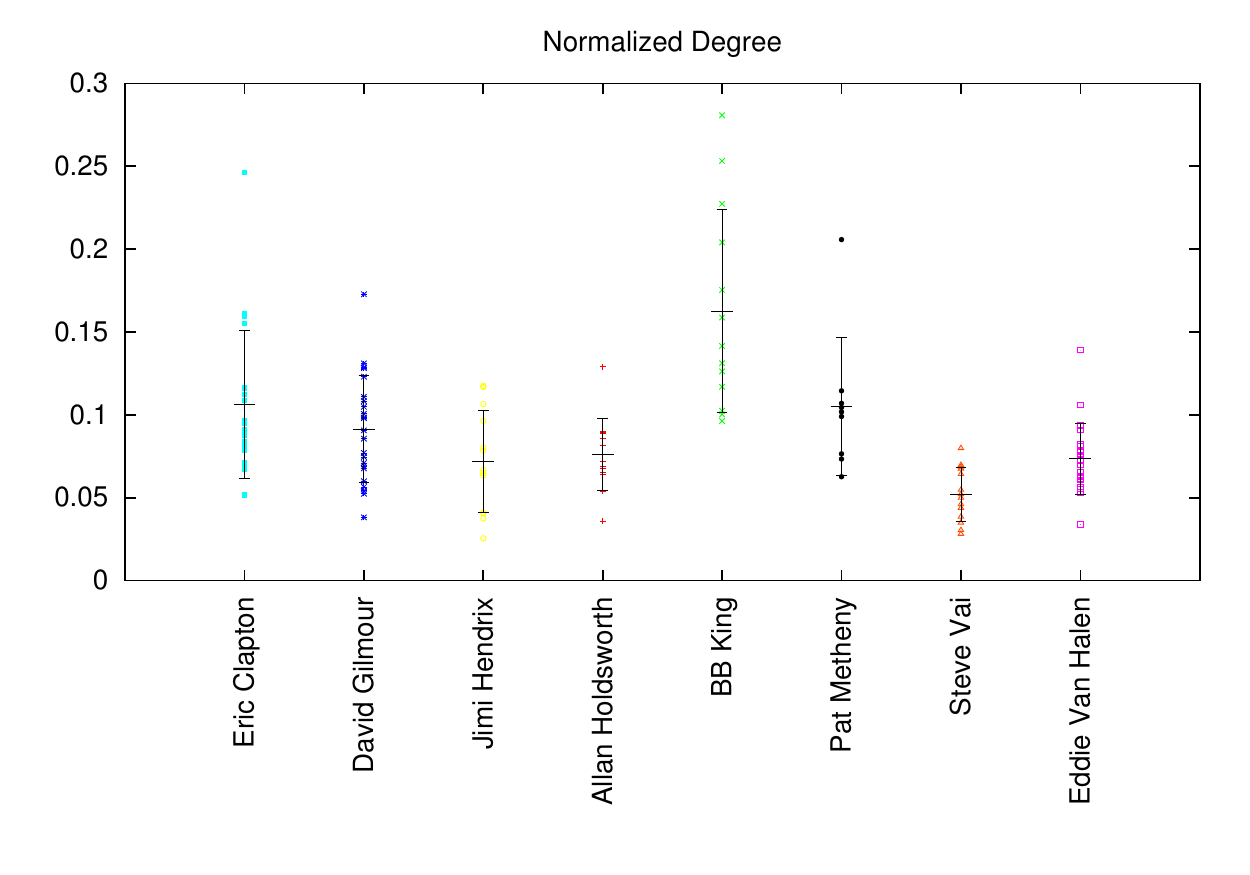}
   \includegraphics[width=.7\textwidth]{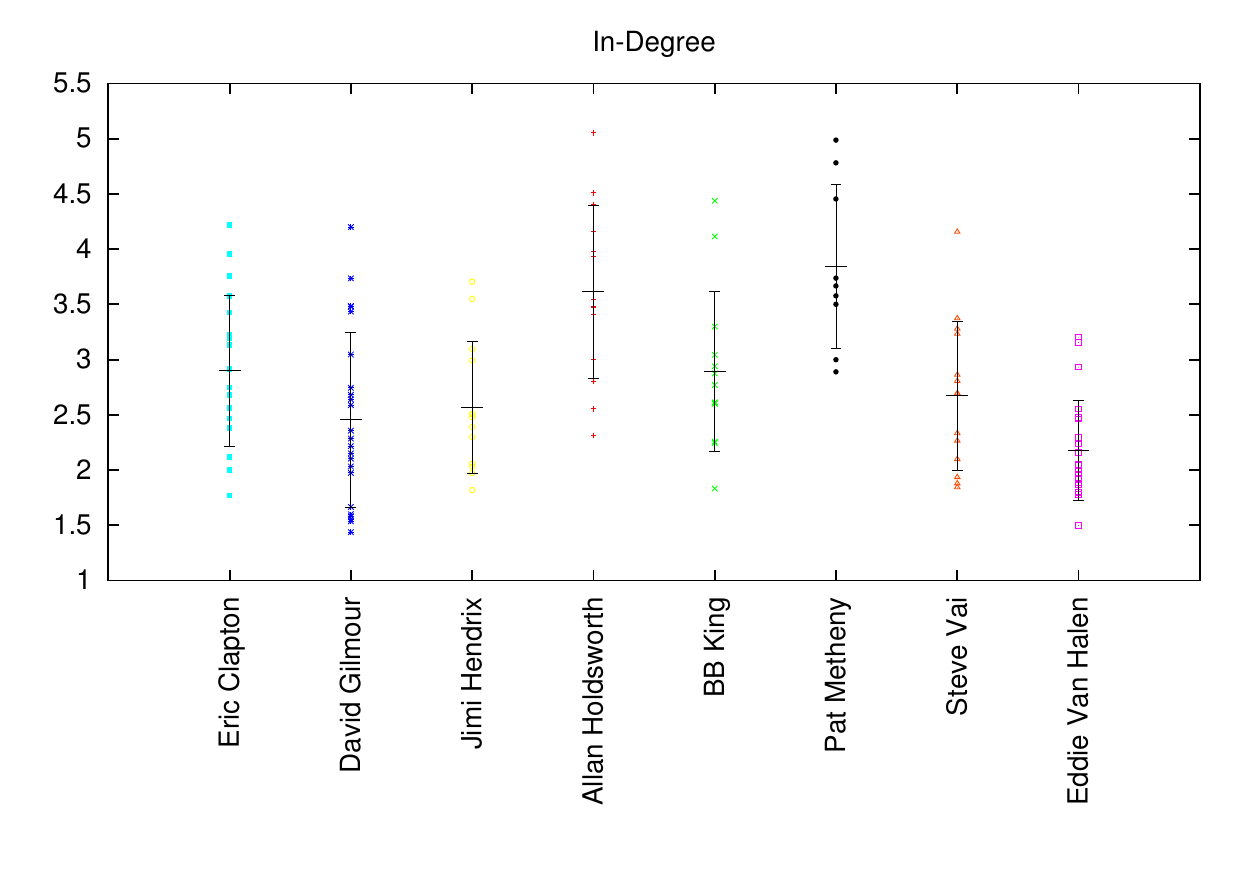}
   \caption{Degree}
   \label{fig:degree}
\end{figure}

Figure \ref{fig:degree} reports the average degrees of solos of the considered performers. 
Since we are observing averages of collected values, results for in-degrees are equal to those for out-degrees (each link leaves a vertex and enters another one; hence, counting the number of outgoing links and the number of incoming links produce the same result). Thus, we do not show out-degrees in the figure. 
Tables \ref{tab:degree}, \ref{tab:inDegree}, \ref{tab:normalizedDegree} show the results of the statistical hypothesis t-tests for degrees, in-degrees and normalized degrees, respectively. Tests for degrees and in-degrees provide quite similar results.

By looking at non-normalized degrees and in-degrees, Holdsworth and Metheny show higher average degrees; this is due probably to the fact that these two performers have high solo lengths. Their similarity (and statistical difference to others) is confirmed by the t-tests in Tables \ref{tab:degree}, \ref{tab:inDegree}.
Conversely, if we look at normalized degrees, shown in the related chart in Figure \ref{fig:degree} and Table  \ref{tab:normalizedDegree}, 
we can say that these two performers (as well as Hendrix and Van Halen) have lower average degrees, testifying their inclination to create various and complex solos. Results confirm that B.B.~King was used playing and repeating specific combinations of notes. In fact, his normalized degree is higher than others and the t-test shows a significant difference with other musicians.

\begin{table*}[th]
\centering
\caption{Degree -- the number is the smallest significance level at which one can reject the null hypothesis that the two means are equal in favor of the two-sided alternative that they are different. In particular, values in bold are those where there is a significant difference (p = 0.05). Acronyms are used instead of full names.}
\label{tab:degree}
\scriptsize
\begin{tabular}{|| l | l | l | l | l | l | l | l ||}
\hline
\hline
 & AH & BBK & DG & EVH & EC & JH & PM\\ 
 \hline
BBK & \textbf{0.01} &  &  &  &  &  & \\
\hline
DG & \textbf{0.0} & 0.1 &  &  &  &  & \\
\hline
EVH & \textbf{0.0} & \textbf{0.0} & 0.1 &  &  &  & \\
\hline
EC & \textbf{0.01} & 0.97 & 0.06 & \textbf{0.0} &  &  & \\
\hline
JH & \textbf{0.0} & 0.18 & 0.7 & 0.05 & 0.13 &  & \\
\hline
PM & 0.56 & \textbf{0.01} & \textbf{0.0} & \textbf{0.0} & \textbf{0.01} & \textbf{0.0} & \\
\hline
SV & \textbf{0.0} & 0.41 & 0.37 & \textbf{0.02} & 0.37 & 0.59 & \textbf{0.0}\\
\hline
\hline
\end{tabular}
\end{table*}

\begin{table*}[th]
\centering
\caption{In-Degree -- the number is the smallest significance level at which one can reject the null hypothesis that the two means are equal in favor of the two-sided alternative that they are different. In particular, values in bold are those where there is a significant difference (p = 0.05). Acronyms are used instead of full names.}
\label{tab:inDegree}
\scriptsize
\begin{tabular}{|| l | l | l | l | l | l | l | l ||}
\hline
\hline
 & AH & BBK & DG & EVH & EC & JH & PM\\ 
 \hline
BBK & \textbf{0.02} &  &  &  &  &  & \\
\hline
DG & \textbf{0.0} & 0.1 &  &  &  &  & \\
\hline
EVH & \textbf{0.0} & \textbf{0.0} & 0.15 &  &  &  & \\
\hline
EC & \textbf{0.01} & 0.98 & 0.06 & \textbf{0.0} &  &  & \\
\hline
JH & \textbf{0.0} & 0.22 & 0.64 & 0.05 & 0.15 &  & \\
\hline
PM & 0.49 & \textbf{0.01} & \textbf{0.0} & \textbf{0.0} & \textbf{0.01} & \textbf{0.0} & \\
\hline
SV & \textbf{0.0} & 0.42 & 0.38 & \textbf{0.02} & 0.35 & 0.67 & \textbf{0.0}\\
\hline
\hline
\end{tabular}
\end{table*}

\begin{table*}[th]
\centering
\caption{Normalized degree -- the number is the smallest significance level at which one can reject the null hypothesis that the two means are equal in favor of the two-sided alternative that they are different. In particular, values in bold are those where there is a significant difference (p = 0.05). Acronyms are used instead of full names.}
\label{tab:normalizedDegree}
\scriptsize
\begin{tabular}{|| l | l | l | l | l | l | l | l ||}
\hline
\hline
 & AH & BBK & DG & EVH & EC & JH & PM\\ 
 \hline
BBK & \textbf{0.0} &  &  &  &  &  & \\
\hline
DG & 0.09 & \textbf{0.0} &  &  &  &  & \\
\hline
EVH & 0.72 & \textbf{0.0} & \textbf{0.03} &  &  &  & \\
\hline
EC & \textbf{0.01} & \textbf{0.01} & 0.23 & \textbf{0.01} &  &  & \\
\hline
JH & 0.7 & \textbf{0.0} & 0.09 & 0.88 & \textbf{0.01} &  & \\
\hline
PM & 0.08 & \textbf{0.02} & 0.39 & 0.06 & 0.95 & 0.06 & \\
\hline
SV & \textbf{0.0} & \textbf{0.0} & \textbf{0.0} & \textbf{0.0} & \textbf{0.0} & 0.05 & \textbf{0.0}\\
\hline
\hline
\end{tabular}
\end{table*}

\begin{figure}[htbp]
   \centering
   \includegraphics[width=.7\textwidth]{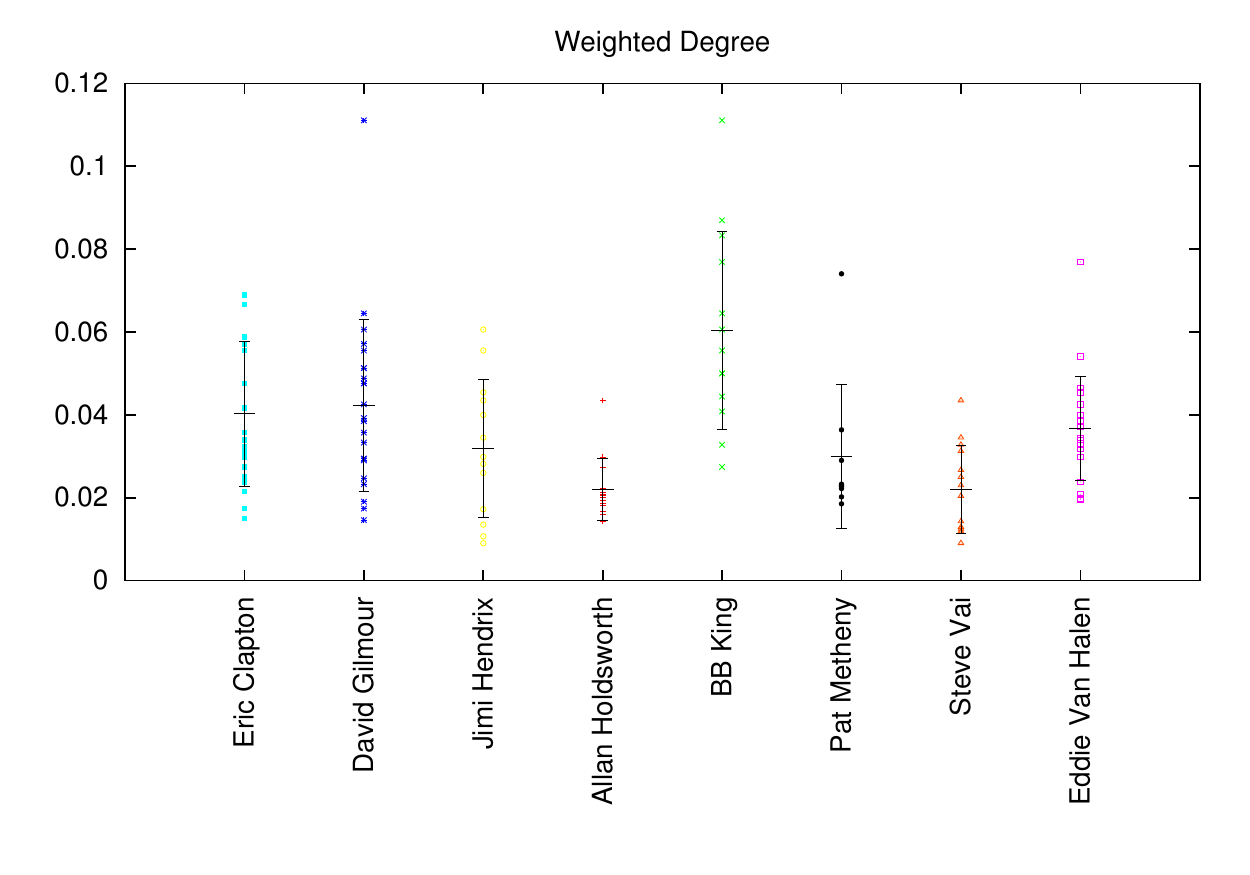}
   \includegraphics[width=.7\textwidth]{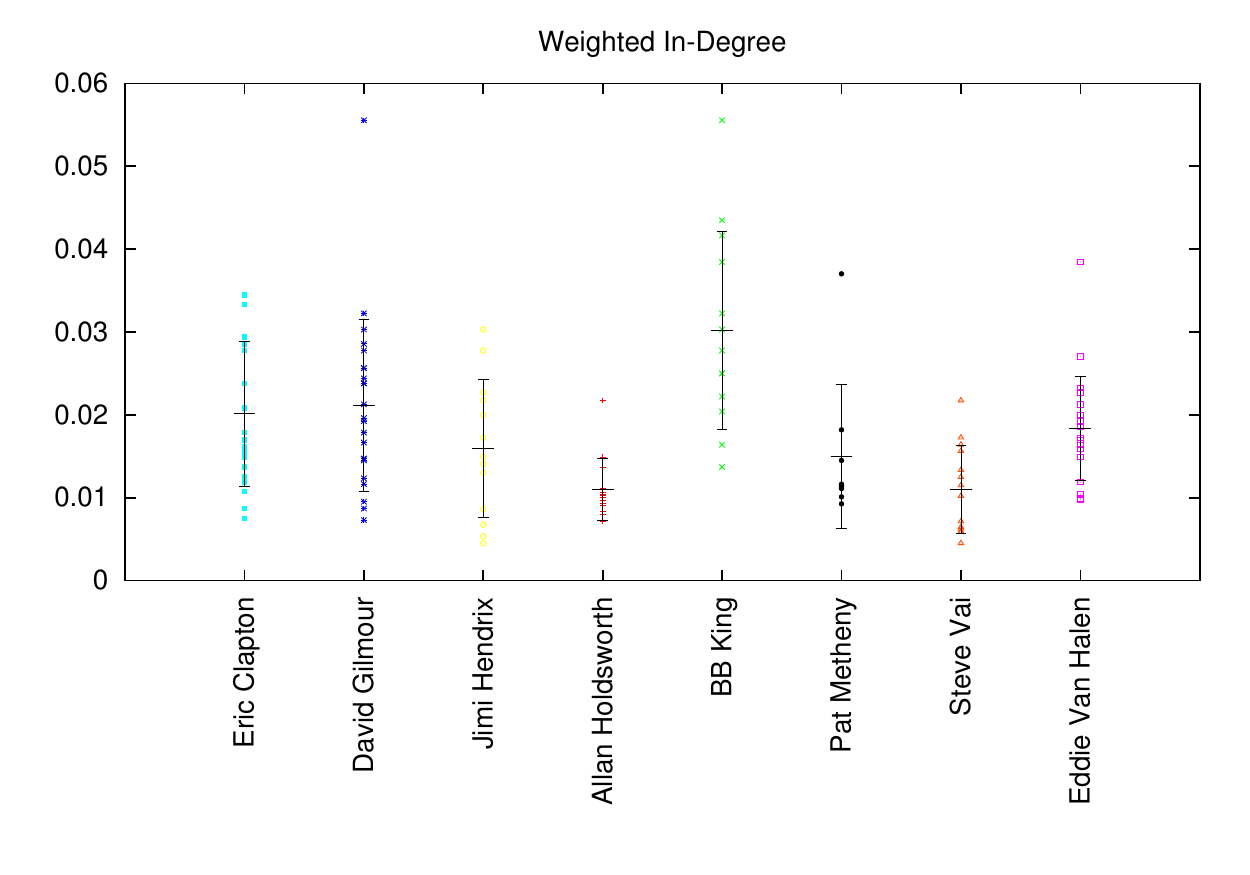}
   \caption{Weighted Degree}
   \label{fig:weightedDegree}
\end{figure}

As concerns weighted degrees, these are shown in Figure \ref{fig:weightedDegree}, Tables \ref{tab:wDegree} and \ref{tab:wInDegree}. 
These measures confirm the outcome for normalized degrees.

\begin{table*}[th]
\centering
\caption{Weighted degree -- the number is the smallest significance level at which one can reject the null hypothesis that the two means are equal in favor of the two-sided alternative that they are different. In particular, values in bold are those where there is a significant difference (p = 0.05). Acronyms are used instead of full names.}
\label{tab:wDegree}
\scriptsize
\begin{tabular}{|| l | l | l | l | l | l | l | l ||}
\hline
\hline
 & AH & BBK & DG & EVH & EC & JH & PM\\ 
 \hline
BBK & \textbf{0.0} &  &  &  &  &  & \\
\hline
DG & \textbf{0.0} & \textbf{0.03} &  &  &  &  & \\
\hline
EVH & \textbf{0.0} & \textbf{0.0} & 0.29 &  &  &  & \\
\hline
EC & \textbf{0.0} & \textbf{0.02} & 0.74 & 0.46 &  &  & \\
\hline
JH & 0.07 & \textbf{0.0} & 0.11 & 0.37 & 0.18 &  & \\
\hline
PM & 0.23 & \textbf{0.0} & 0.11 & 0.31 & 0.16 & 0.8 & \\
\hline
SV & 0.99 & \textbf{0.0} & \textbf{0.0} & \textbf{0.0} & \textbf{0.0} & 0.09 & 0.25\\
\hline
\hline
\end{tabular}
\end{table*}

\begin{table*}[th]
\centering
\caption{Weighted in-degree -- the number is the smallest significance level at which one can reject the null hypothesis that the two means are equal in favor of the two-sided alternative that they are different. In particular, values in bold are those where there is a significant difference (p = 0.05). Acronyms are used instead of full names.}
\label{tab:wInDegree}
\scriptsize
\begin{tabular}{|| l | l | l | l | l | l | l | l ||}
\hline
\hline
 & AH & BBK & DG & EVH & EC & JH & PM\\ 
 \hline
BBK & \textbf{0.0} &  &  &  &  &  & \\
\hline
DG & \textbf{0.0} & \textbf{0.03} &  &  &  &  & \\
\hline
EVH & \textbf{0.0} & \textbf{0.0} & 0.29 &  &  &  & \\
\hline
EC & \textbf{0.0} & \textbf{0.02} & 0.74 & 0.46 &  &  & \\
\hline
JH & 0.07 & \textbf{0.0} & 0.11 & 0.37 & 0.18 &  & \\
\hline
PM & 0.23 & \textbf{0.0} & 0.11 & 0.31 & 0.16 & 0.8 & \\
\hline
SV & 0.99 & \textbf{0.0} & \textbf{0.0} & \textbf{0.0} & \textbf{0.0} & 0.09 & 0.25\\
\hline
\hline
\end{tabular}
\end{table*}

\begin{figure*}[ht]
\centering
\includegraphics[width=.44\textwidth]{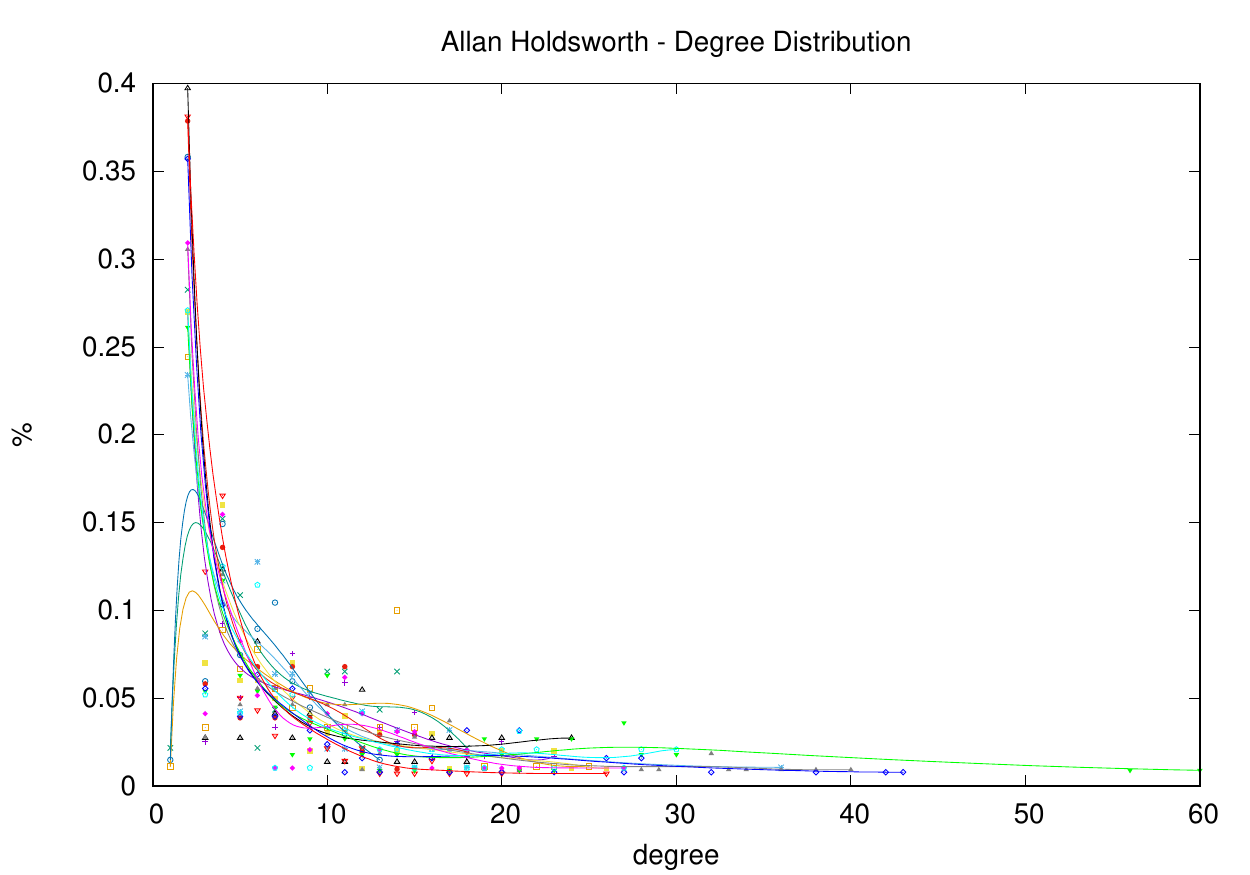}
\includegraphics[width=.44\textwidth]{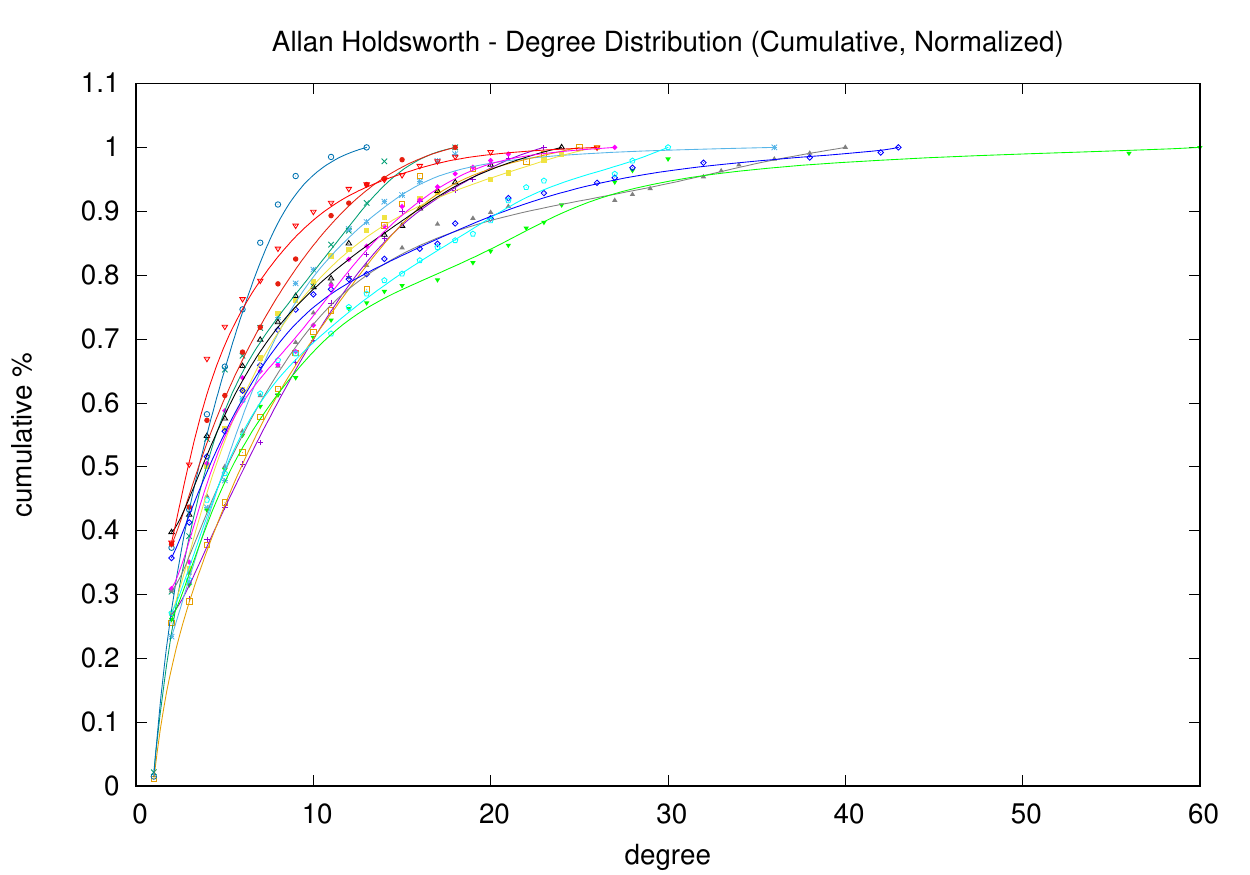}
\includegraphics[width=.44\textwidth]{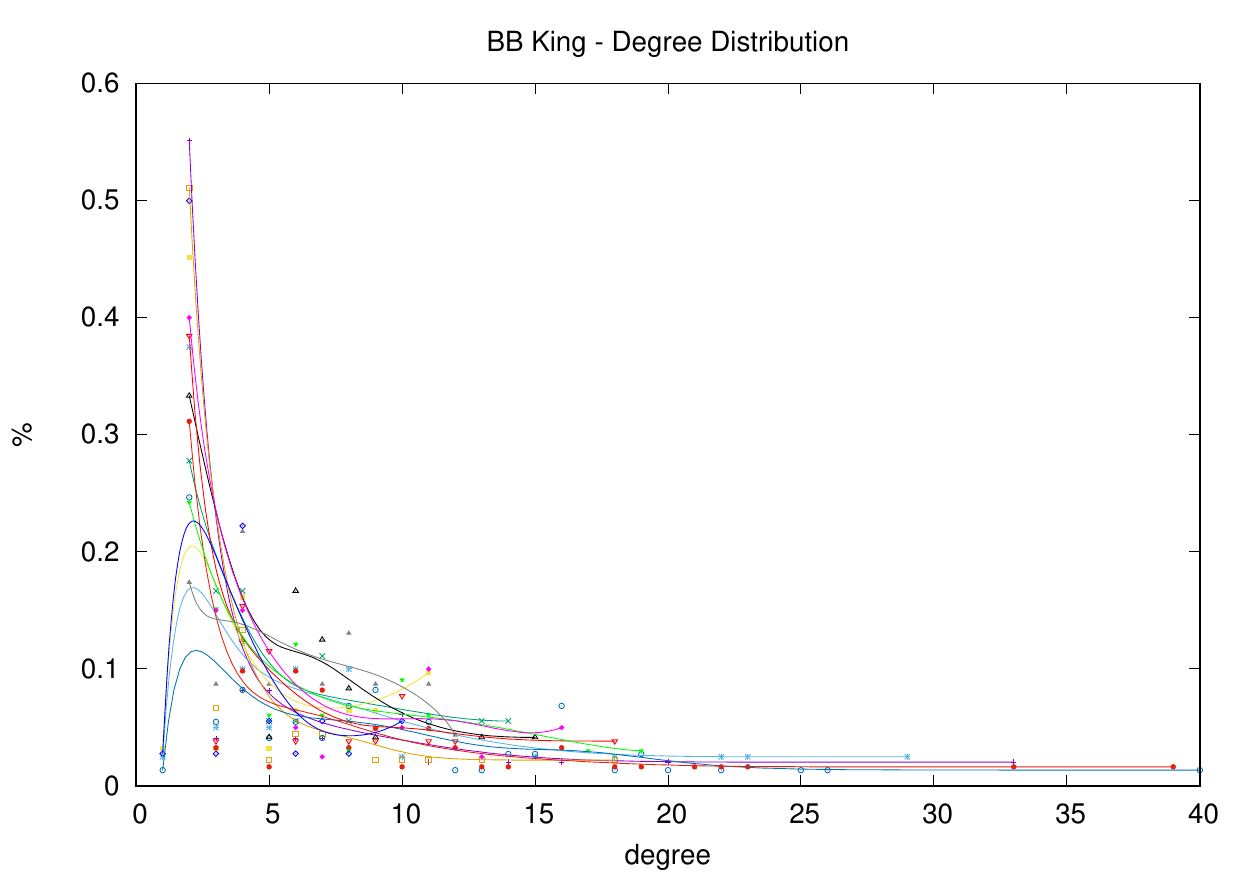}
\includegraphics[width=.44\textwidth]{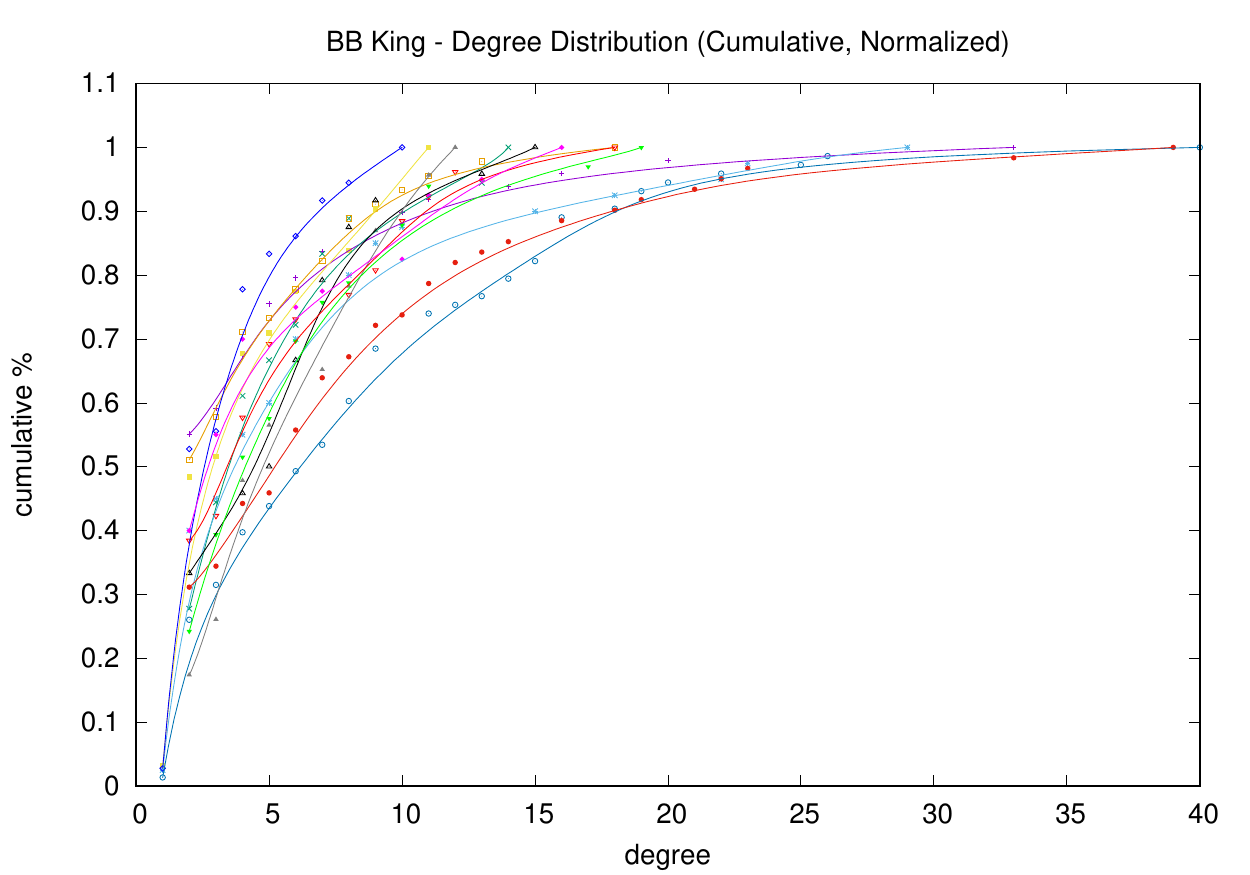}
\includegraphics[width=.44\textwidth]{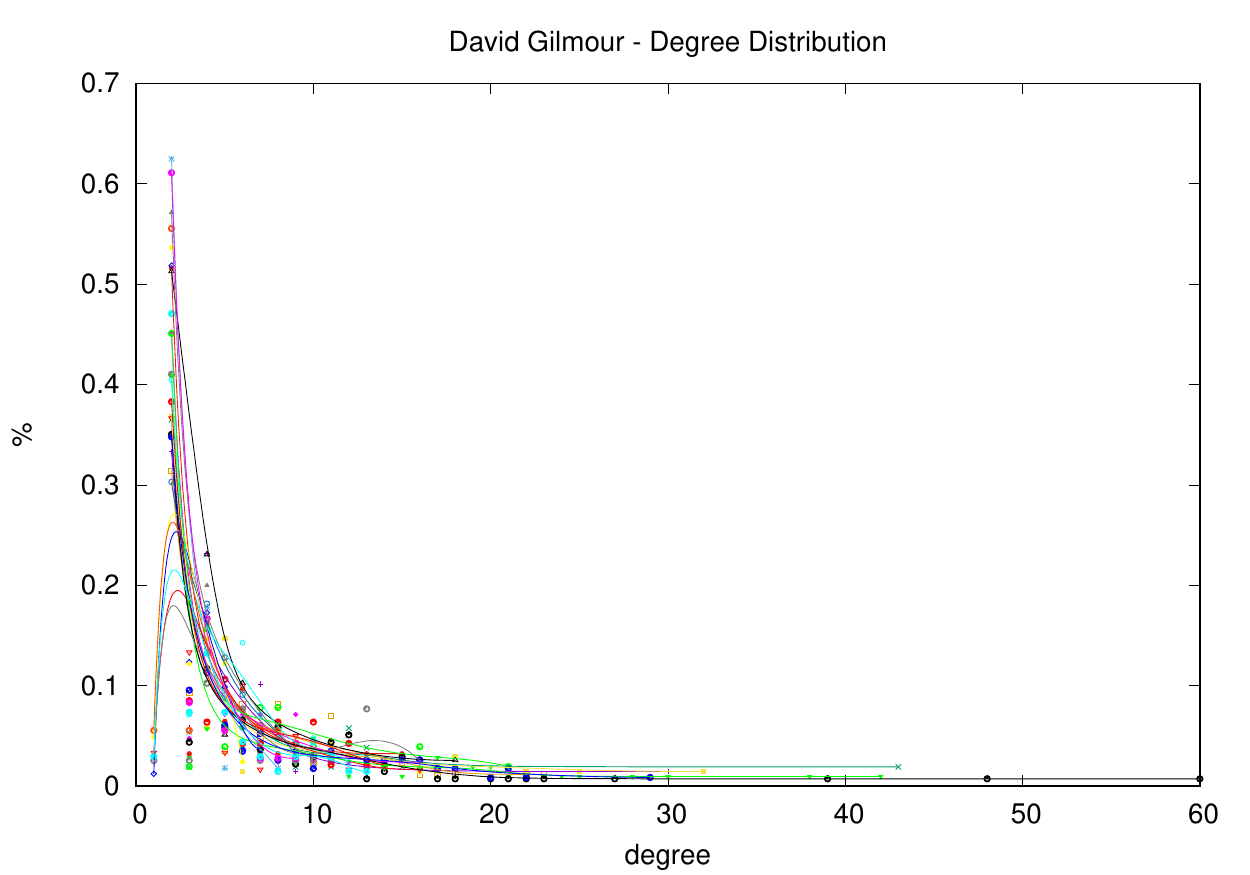}
\includegraphics[width=.44\textwidth]{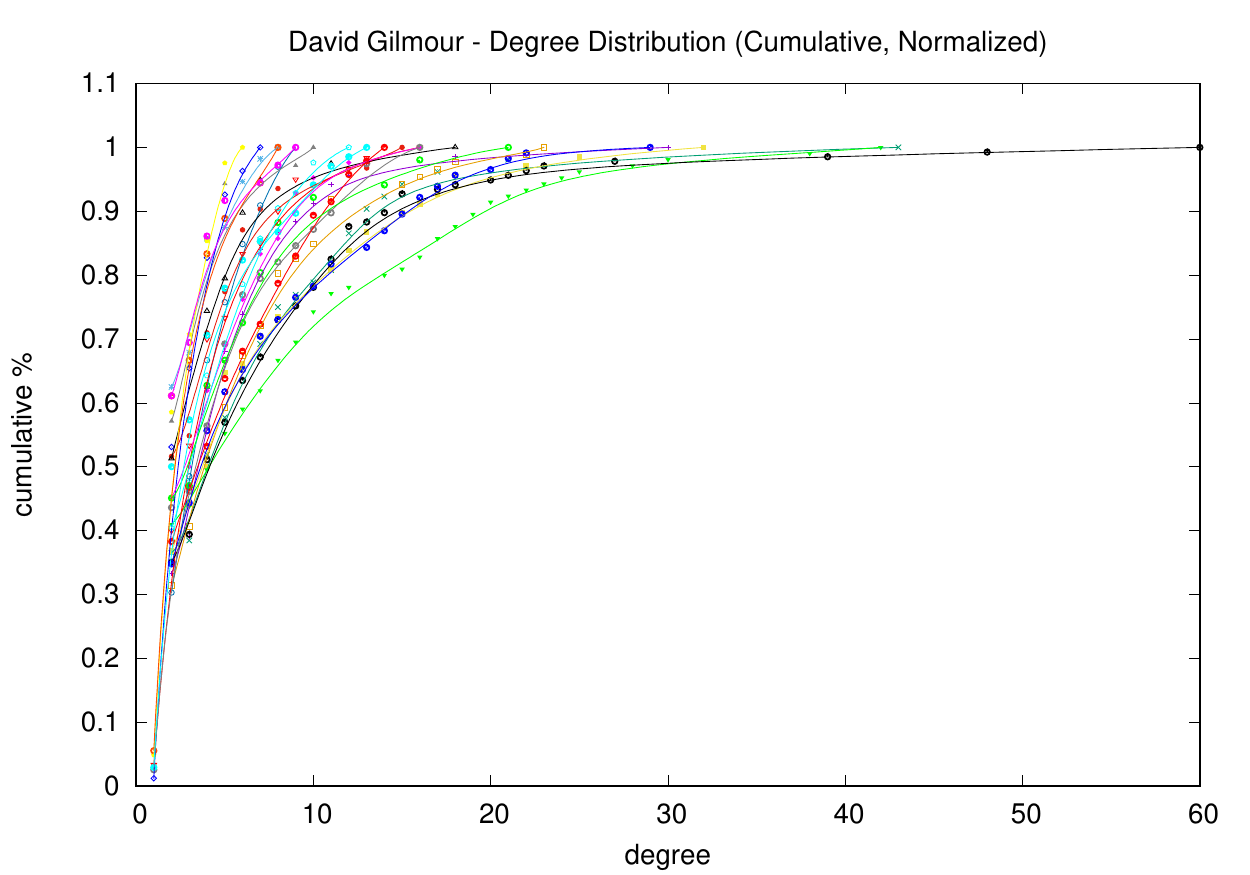}
\caption{Degree Distribution -- part 1. Each row corresponds to a given artist; the left side chart shows the degree distribution, while the right side chart shows the cumulative distribution. Points refer to values obtained for each of the tracks, while lines are the related Bézier interpolations.}
\label{fig:degreeDist_p1}
\end{figure*}
\begin{figure*}[ht]
\centering
\includegraphics[width=.44\textwidth]{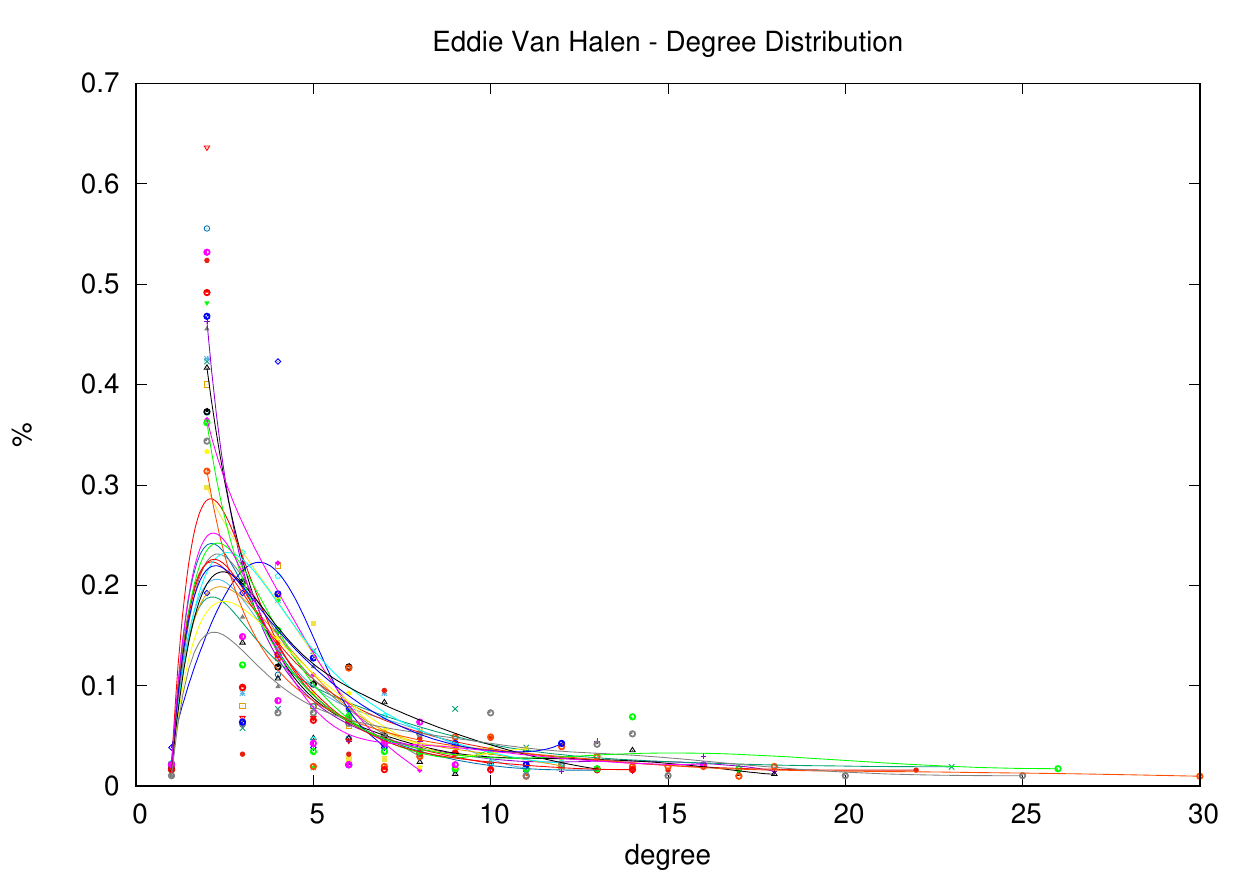}
\includegraphics[width=.44\textwidth]{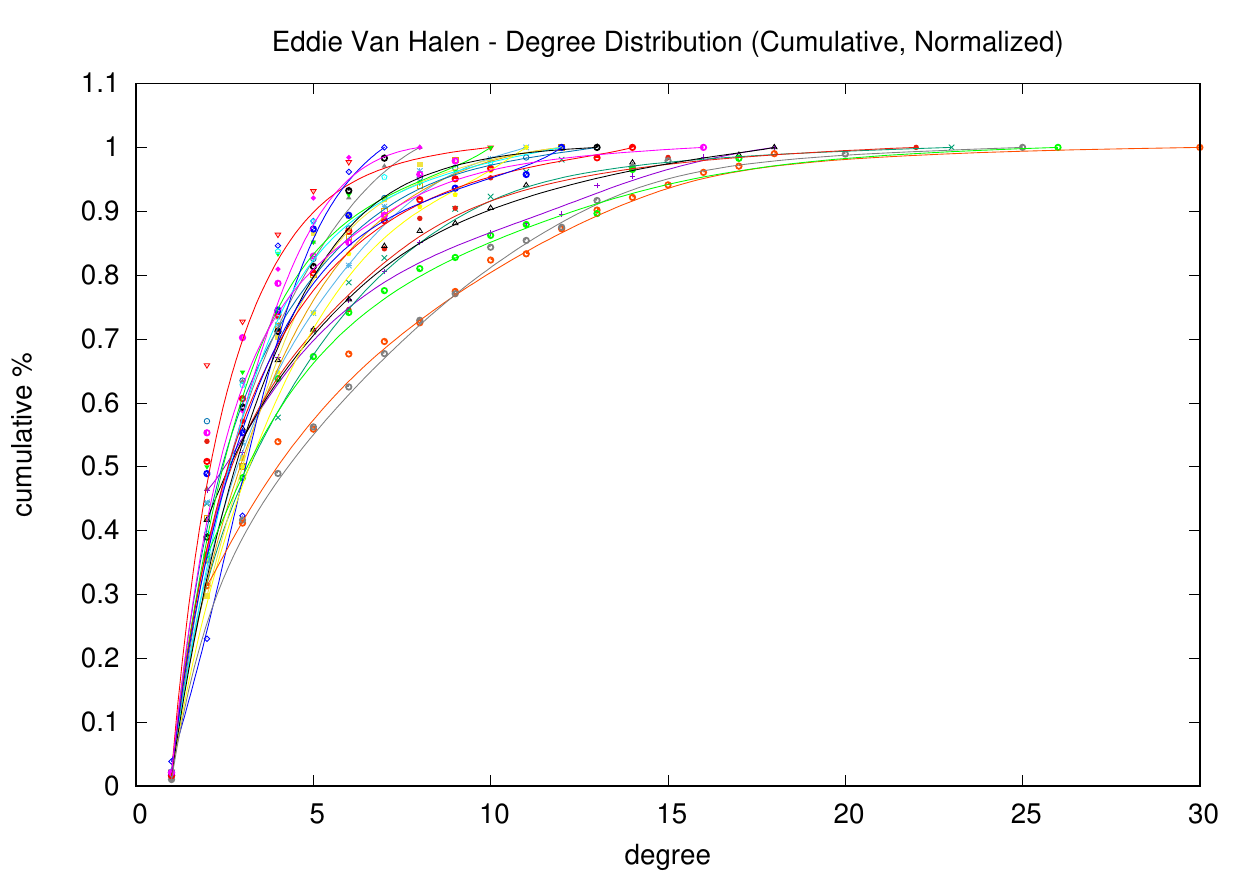}
\includegraphics[width=.44\textwidth]{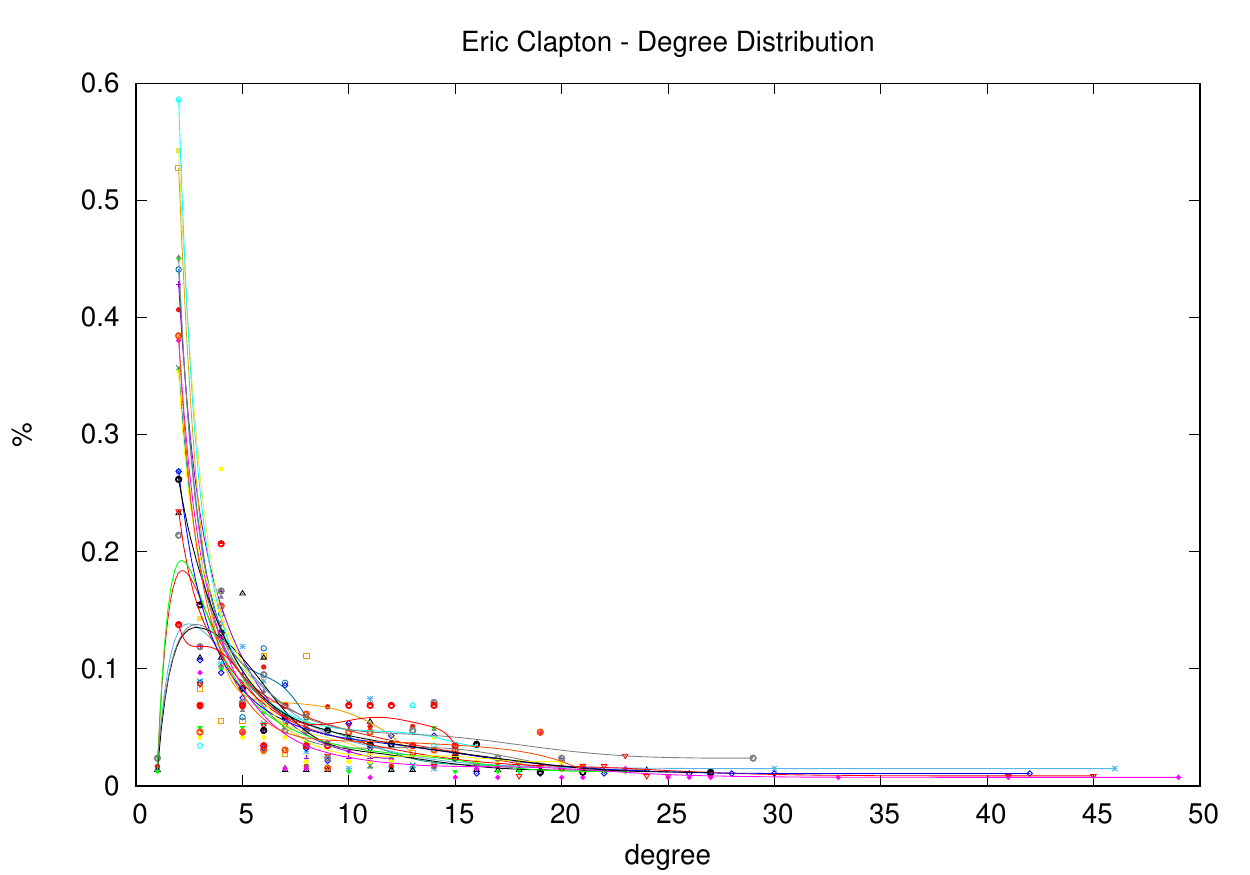}
\includegraphics[width=.44\textwidth]{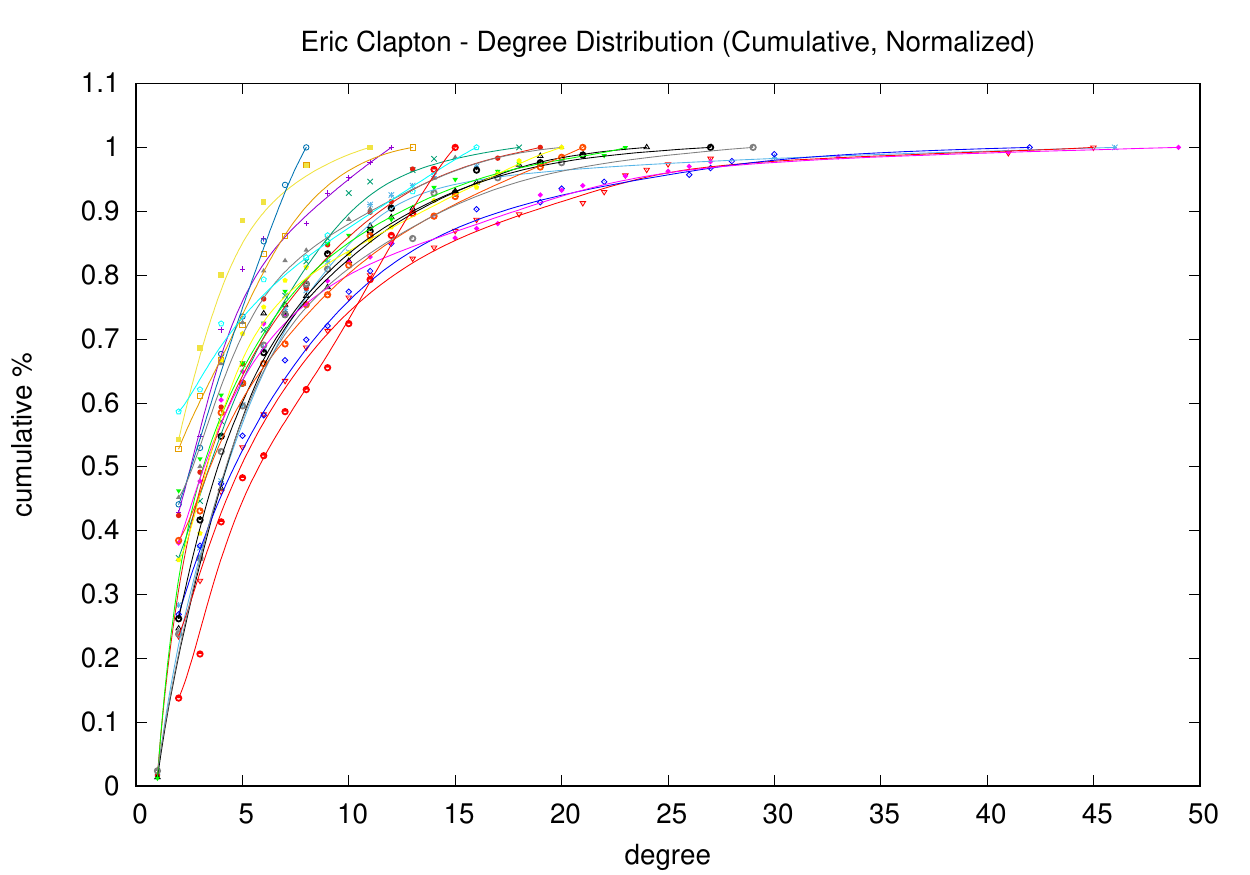}
\includegraphics[width=.44\textwidth]{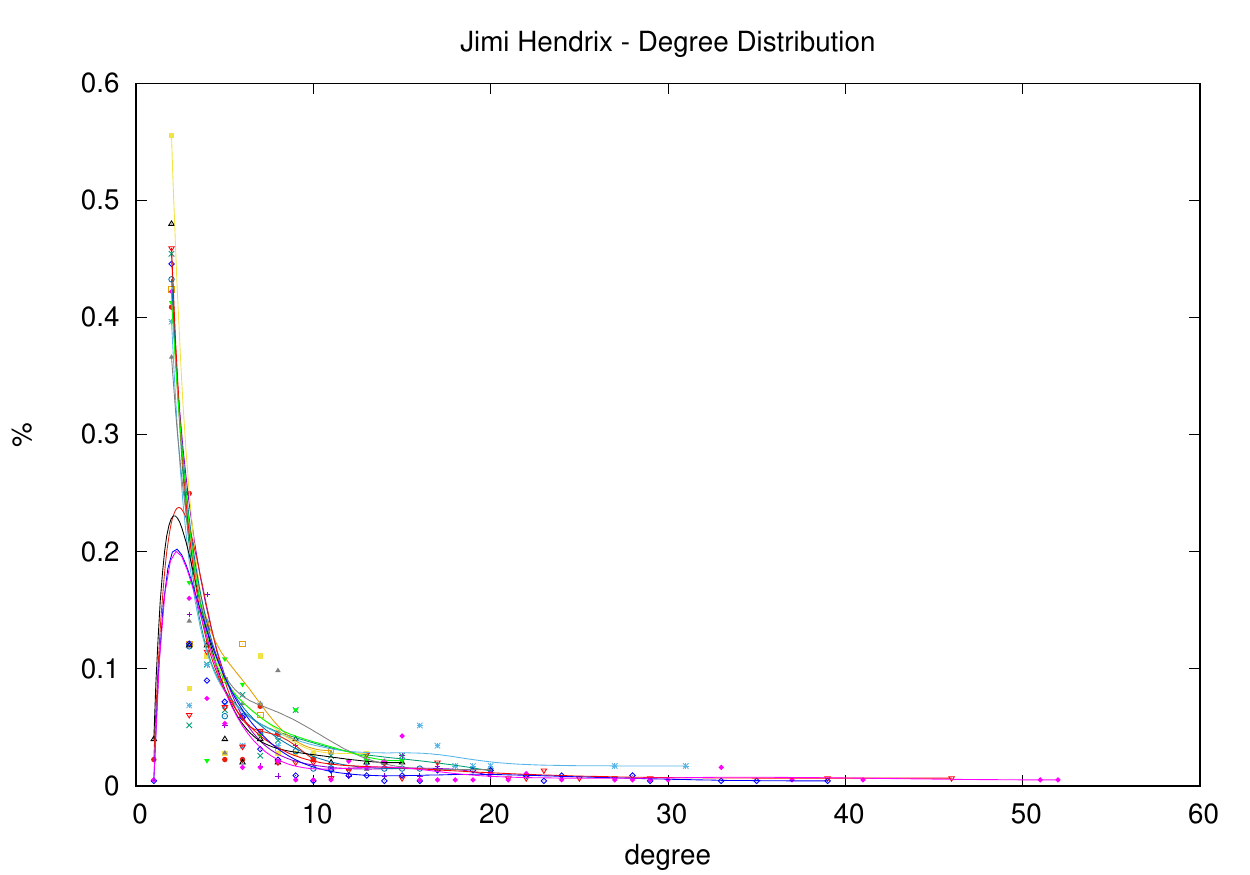}
\includegraphics[width=.44\textwidth]{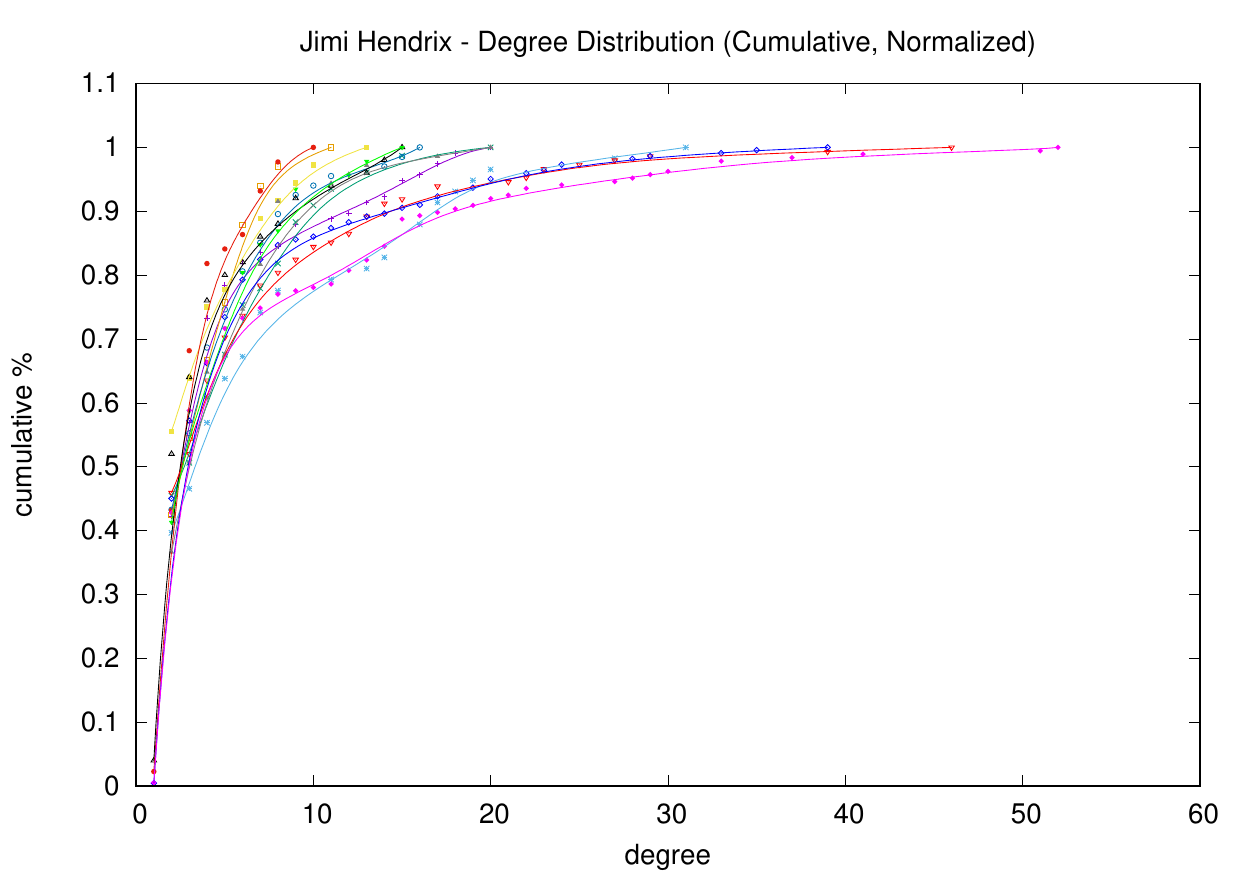}
\caption{Degree Distribution -- part 2. Each row corresponds to a given artist; the left side chart shows the degree distribution, while the right side chart shows the cumulative distribution. Points refer to values obtained for each of the tracks, while lines are the related Bézier interpolations.}
\label{fig:degreeDist_p2}
\end{figure*}
\begin{figure*}[ht]
\centering
\includegraphics[width=.44\textwidth]{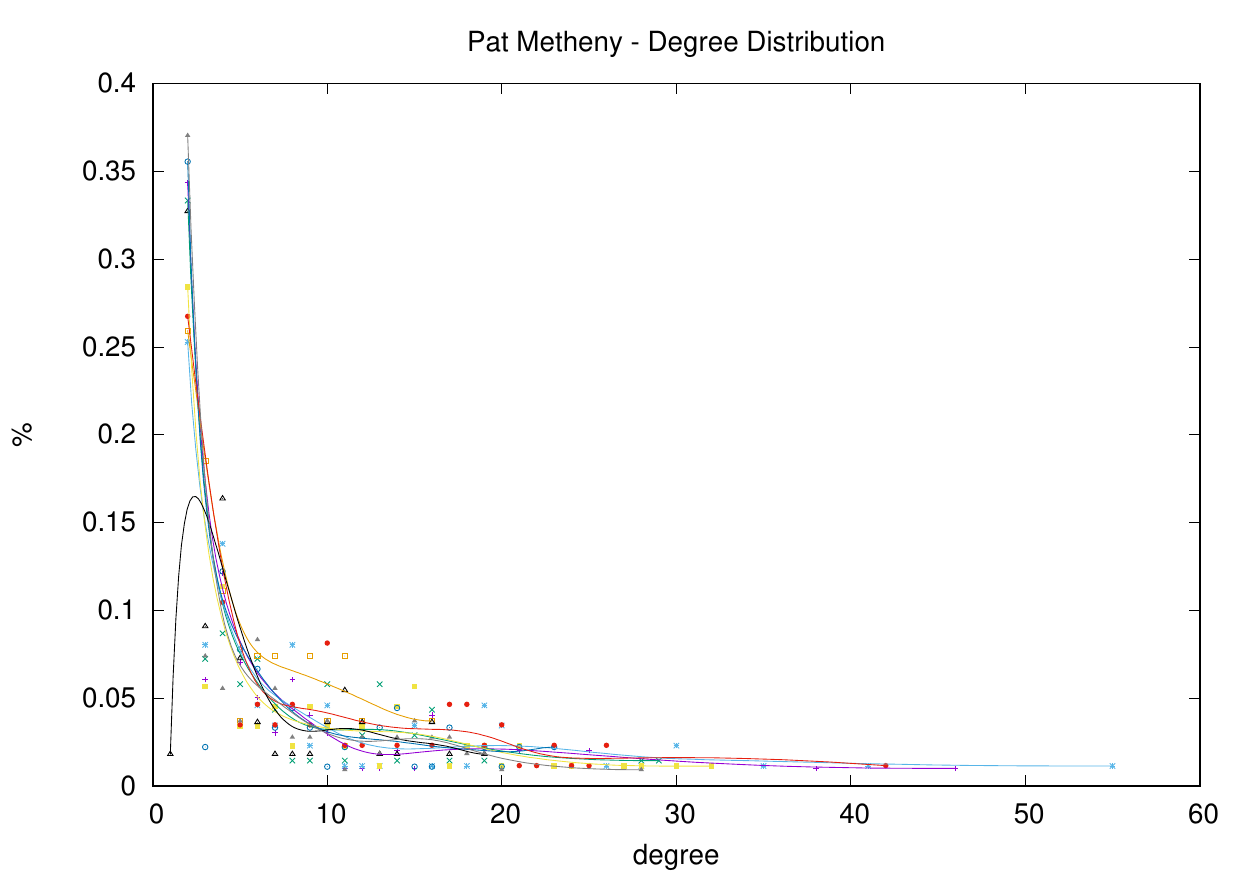}
\includegraphics[width=.44\textwidth]{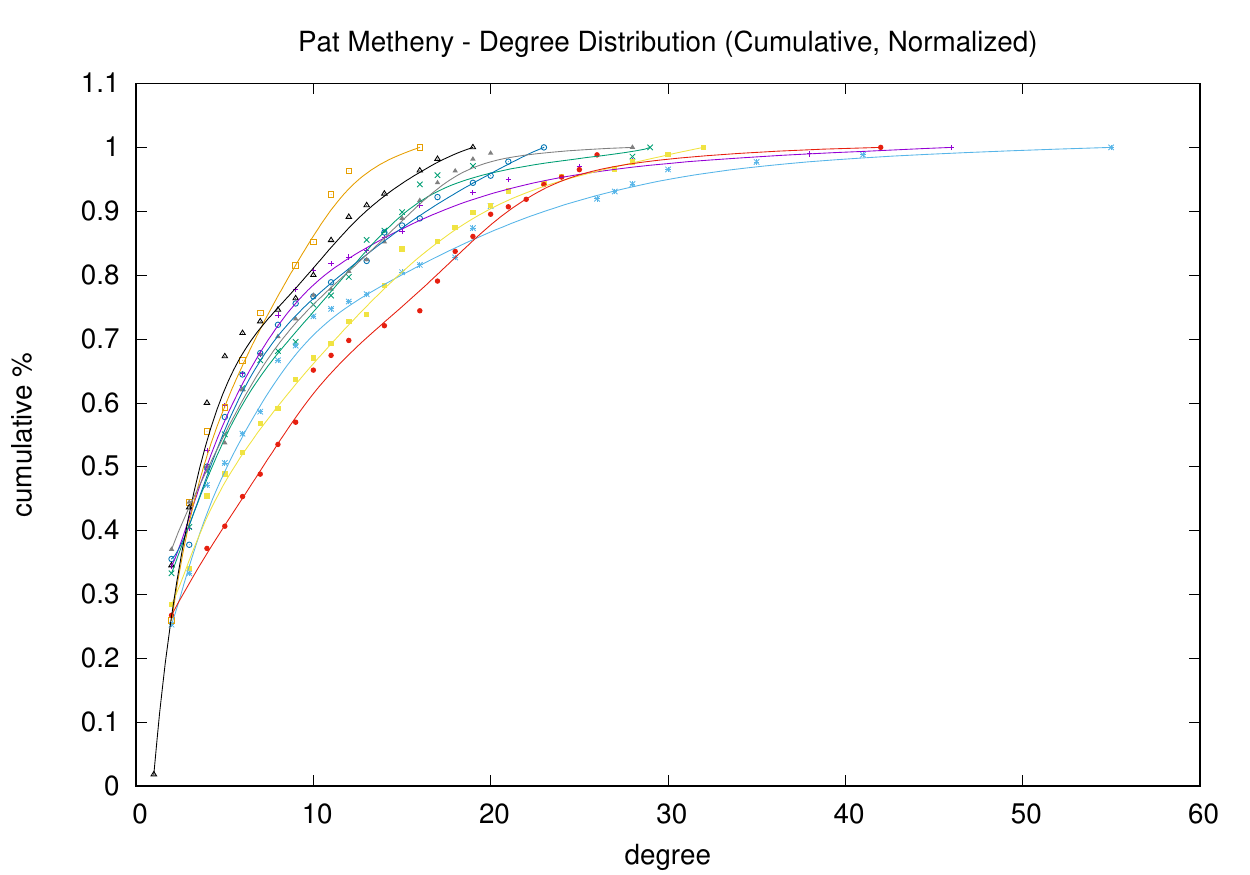}
\includegraphics[width=.44\textwidth]{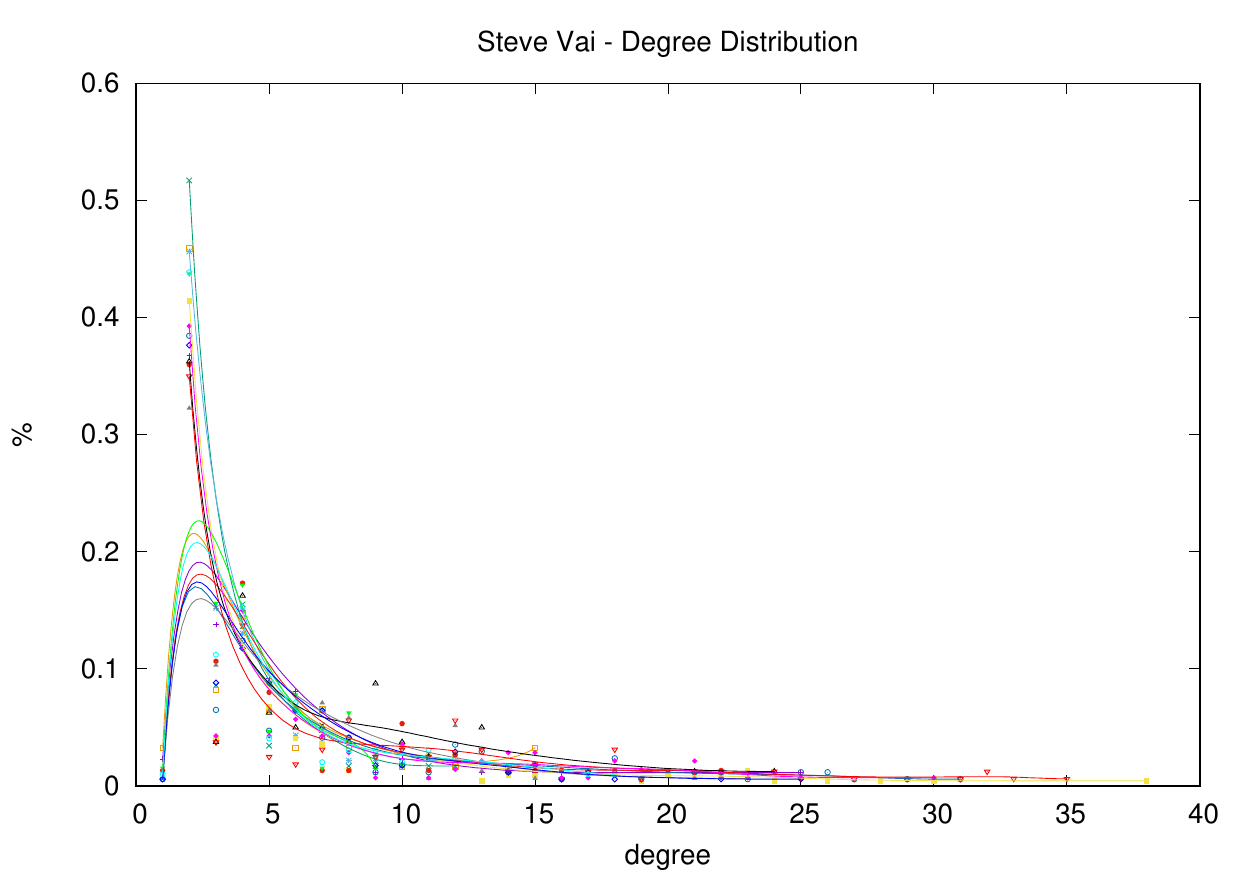}
\includegraphics[width=.44\textwidth]{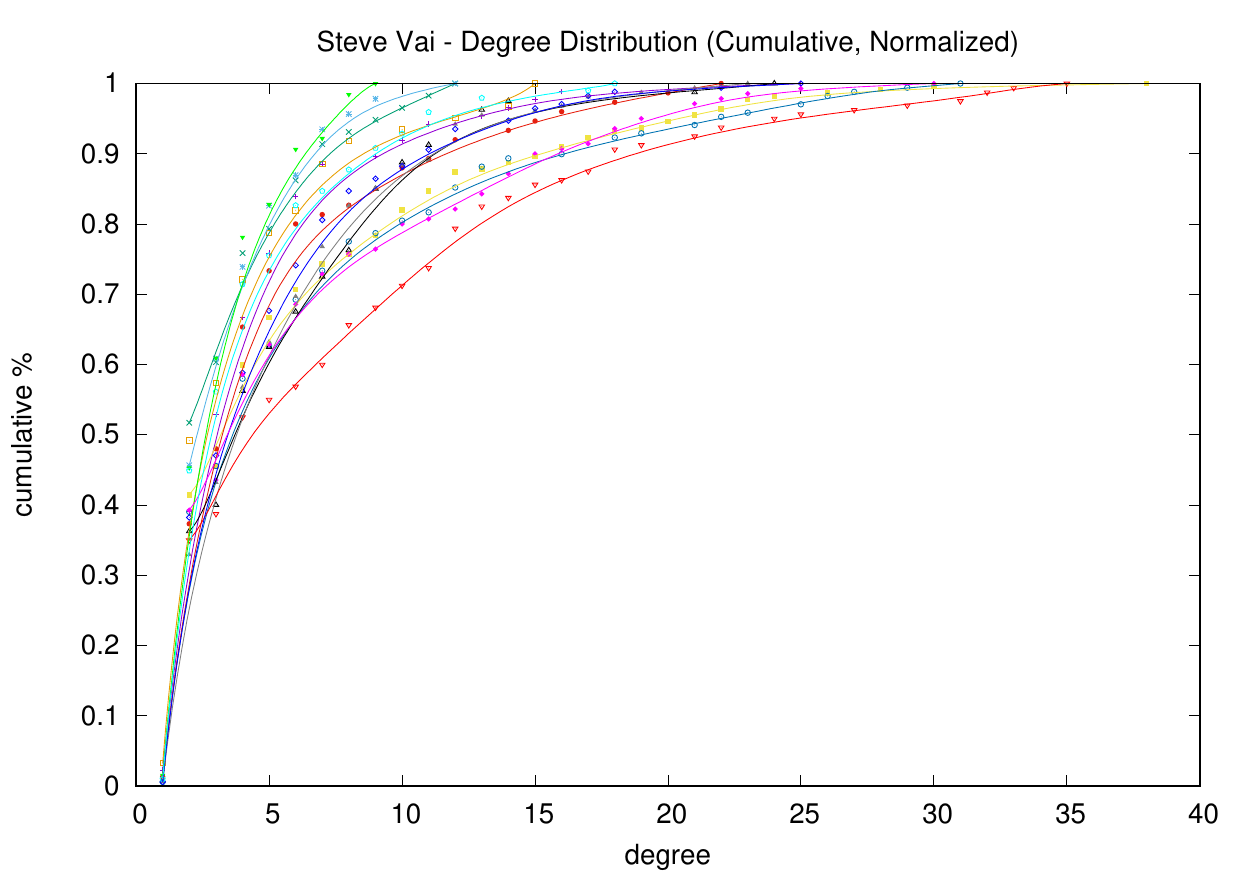}
\caption{Degree Distribution -- part 3. Each row corresponds to a given artist; the left side chart shows the degree distribution, while the right side chart shows the cumulative distribution. Points refer to values obtained for each of the tracks, while lines are the related Bézier interpolations.}
\label{fig:degreeDist_p3}
\end{figure*}

% 
% 
% \begin{figure*}[ht]
% \centering
% \includegraphics[width=.44\textwidth]{figs/degreeDist/degreeDistLog_Allan_Holdsworth_2nd-eps-converted-to.pdf}
% \includegraphics[width=.44\textwidth]{figs/degreeDist/degreeDistLog_BB_King_2nd-eps-converted-to.pdf}
% \includegraphics[width=.44\textwidth]{figs/degreeDist/degreeDistLog_David_Gilmour_2nd-eps-converted-to.pdf}
% \includegraphics[width=.44\textwidth]{figs/degreeDist/degreeDistLog_Eddie_Van_Halen_2nd-eps-converted-to.pdf}
% \includegraphics[width=.44\textwidth]{figs/degreeDist/degreeDistLog_Eric_Clapton_2nd-eps-converted-to.pdf}
% \includegraphics[width=.44\textwidth]{figs/degreeDist/degreeDistLog_Jimi_Hendrix_2nd-eps-converted-to.pdf}
% \caption{Degree Distribution (log-log scale) -- part 1}
% \label{fig:degreeDist_p1}
% \end{figure*}
% \begin{figure*}[ht]
% \centering
% \includegraphics[width=.44\textwidth]{figs/degreeDist/degreeDistLog_Pat_Metheny_2nd-eps-converted-to.pdf}
% \includegraphics[width=.44\textwidth]{figs/degreeDist/degreeDistLog_Steve_Vai_2nd-eps-converted-to.pdf}
% \caption{Degree Distribution (log-log scale) -- part 2}
% \label{fig:degreeDist_p2}
% \end{figure*}

Figures \ref{fig:degreeDist_p1}--\ref{fig:degreeDist_p3} show the degree distributions (and the corresponding cumulative distributions) of solos performed by each considered musician.
These charts suggest that there are no huge discrepancies among performers in the distribution shape, even if average values are in some case significantly different.

\subsection{Distances}

\begin{figure}[htbp]
   \centering
   \includegraphics[width=.8\linewidth]{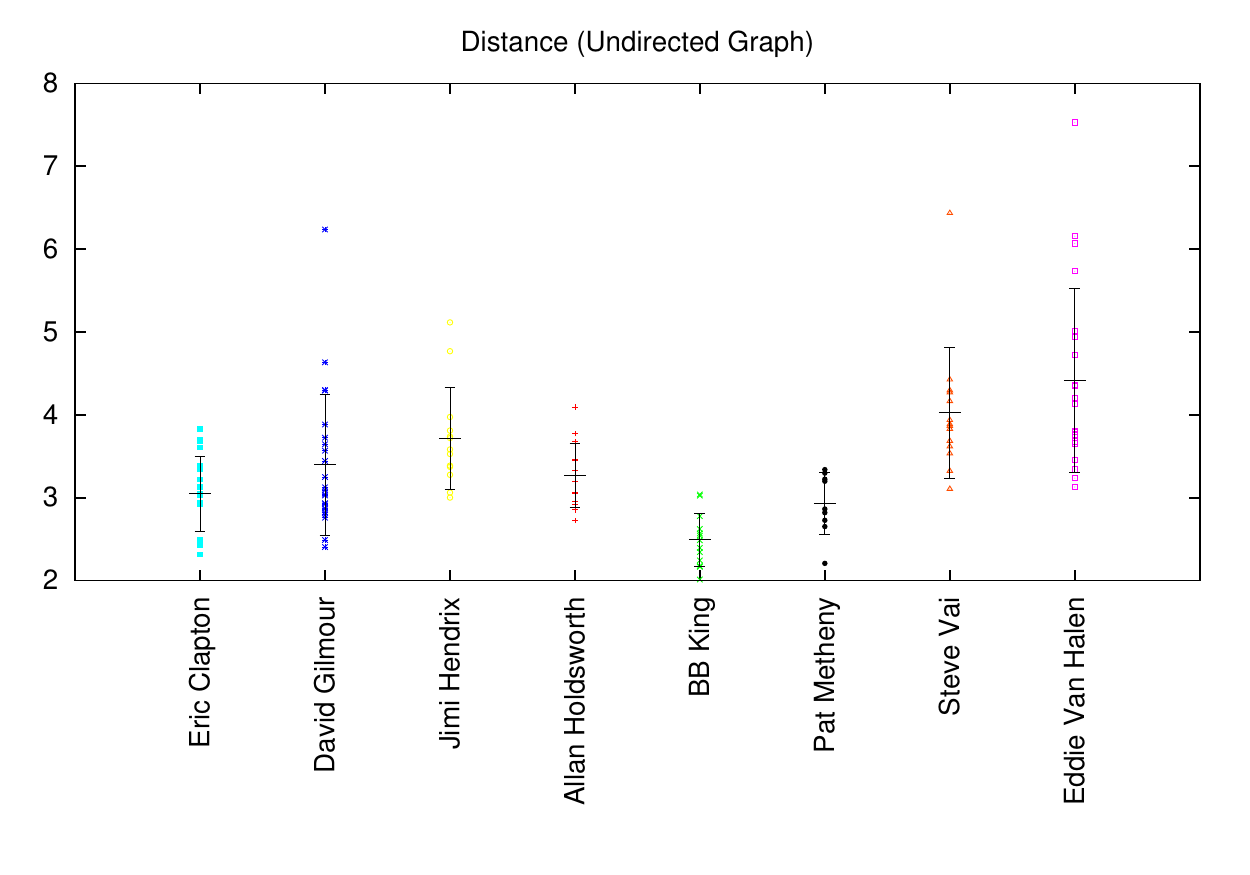}
   \caption{Distance (undirected network)}
   \label{fig:undirectedDistance}
\end{figure}

\begin{table*}[th]
\centering
\caption{Distance (undirected network) -- the number is the smallest significance level at which one can reject the null hypothesis that the two means are equal in favor of the two-sided alternative that they are different. In particular, values in bold are those where there is a significant difference (p = 0.05). Acronyms are used instead of full names.}
\label{tab:undirectedDistance}
\scriptsize
\begin{tabular}{|| l | l | l | l | l | l | l | l ||}
\hline
\hline
 & AH & BBK & DG & EVH & EC & JH & PM\\ 
 \hline
BBK & \textbf{0.0} &  &  &  &  &  & \\
\hline
DG & 0.54 & \textbf{0.0} &  &  &  &  & \\
\hline
EVH & \textbf{0.0} & \textbf{0.0} & \textbf{0.0} &  &  &  & \\
\hline
EC & 0.14 & \textbf{0.0} & 0.1 & \textbf{0.0} &  &  & \\
\hline
JH & \textbf{0.04} & \textbf{0.0} & 0.2 & \textbf{0.02} & \textbf{0.0} &  & \\
\hline
PM & 0.05 & \textbf{0.01} & \textbf{0.04} & \textbf{0.0} & 0.45 & \textbf{0.0} & \\
\hline
SV & \textbf{0.0} & \textbf{0.0} & \textbf{0.03} & 0.23 & \textbf{0.0} & 0.27 & \textbf{0.0}\\
\hline
\hline
\end{tabular}
\end{table*}

Figure \ref{fig:undirectedDistance} shows the average distances obtained for the considered performers. Outcomes confirm the claims reported for the degrees.
In general, classic blues performers (B.B.~King) have simpler and smaller networks, thus resulting in lower distances with respect to rock virtuoso performers (Eddie Van Halen, Steve Vai), meaning that the latter ones create more complex network structures. Results of t-tests, shown in Table \ref{tab:undirectedDistance}, confirm this claim.
Indeed, this outcome is in complete accordance with the common opinions of music experts \cite{Pressing,covach1997understanding,smith,zenni}.

\subsection{Clustering coefficient}

\begin{figure}[htbp]
   \centering
   \includegraphics[width=.8\linewidth]{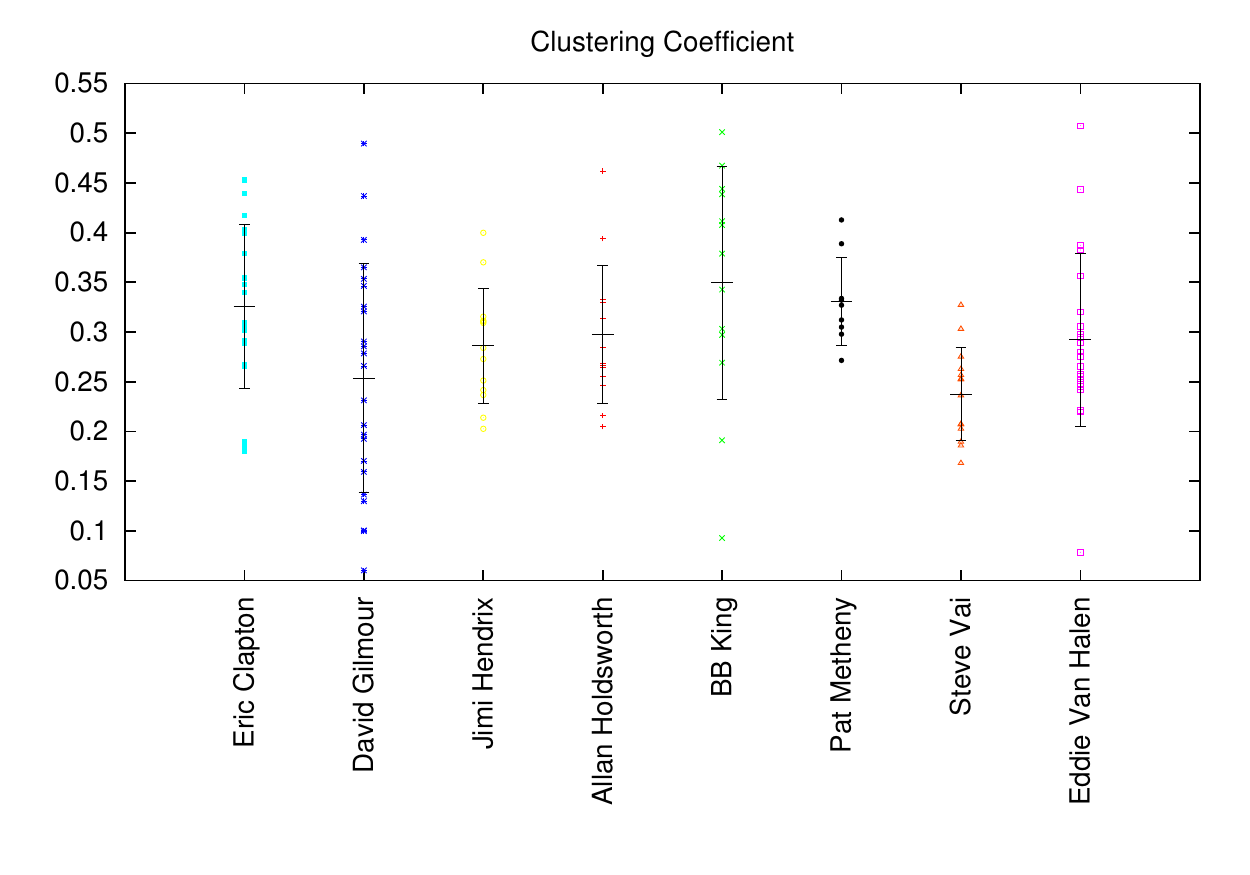}
   \caption{Clustering coefficient}
   \label{fig:clusteringCoefficient}
\end{figure}

\begin{table*}[th]
\centering
\caption{Clustering coefficient -- the number is the smallest significance level at which one can reject the null hypothesis that the two means are equal in favor of the two-sided alternative that they are different. In particular, values in bold are those where there is a significant difference (p = 0.05). Acronyms are used instead of full names.}
\label{tab:clusteringCoefficient}
\scriptsize
\begin{tabular}{|| l | l | l | l | l | l | l | l ||}
\hline
\hline
 & AH & BBK & DG & EVH & EC & JH & PM\\ 
 \hline
BBK & 0.18 &  &  &  &  &  & \\
\hline
DG & 0.15 & \textbf{0.03} &  &  &  &  & \\
\hline
EVH & 0.82 & 0.14 & 0.21 &  &  &  & \\
\hline
EC & 0.29 & 0.53 & \textbf{0.02} & 0.2 &  &  & \\
\hline
JH & 0.63 & 0.1 & 0.27 & 0.81 & 0.11 &  & \\
\hline
PM & 0.18 & 0.61 & \textbf{0.01} & 0.11 & 0.82 & 0.05 & \\
\hline
SV & \textbf{0.01} & \textbf{0.01} & 0.56 & \textbf{0.02} & \textbf{0.0} & \textbf{0.03} & \textbf{0.0}\\
\hline
\hline
\end{tabular}
\end{table*}

Figure \ref{fig:clusteringCoefficient} shows the measured clustering coefficient for the considered performers. In this case, there are no big differences among groups of performers (in terms of musical genres); David Gilmour has a lower average clustering coefficient, but with a high standard deviation. Steve Vai has a lower clustering coefficient, with smaller deviation. Indeed, the difference between Steve Vai and others is confirmed in Table \ref{tab:clusteringCoefficient} showing the results of t-tests for this metrics.
This might indicate (from the considered solos) a higher inclination to create melodic (even if quite elaborate) solos.
It is also worth citing that the obtained results for the clustering coefficient are in line with results reported in \cite{Liu2010126}, where an average clustering coefficient around $0.3$ was measured for the aggregate networks embodying different music pieces by the same composer.

\subsection{Is there any small world property?}

\begin{table*}[th]
\centering
\caption{Small world property: comparison between the clustering coefficient (column ``cc'') and the average distance (column ``avg dist'') of the considered network, and the clustering coefficient (column ``cc (RG)'') and the average distance (column ``avg dist (RG)'') of the corresponding random graph.}
% \caption{Small world property: comparison between solo networks and corresponding random graphs.}
\label{tab:sw}
% \small
\scriptsize
\begin{tabular}{|| l || c | c | c | c ||}
  \hline			
  \hline			
%   song & clus coeff & clus coeff (RG) & avg distance & avg distance (RG)  \\
  song & cc & cc (RG) & avg dist & avg dist (RG)  \\
  \hline  
  \hline			
  B.B.~King -- Rock me baby & 0.41 & 0.11	& 2.17 & 3.26 \\
  \hline			
  D.~Gilmour (Pink Floyd) -- Comfortably numb (1st solo) & 0.06 &	0.03 &	4.30 &	4.03 \\
  \hline			
  E.~Clapton (Cream) -- Crossroads (2nd solo) & 0.40 &	0.04 &	3.68 &	4.29 \\
  \hline			
  J.~Hendrix -- Red House & 0.24 &	0.02 &	3.37 &	5.00 \\
  \hline  
  \hline			
\end{tabular}
\end{table*}

Table \ref{tab:sw} assesses if some particular networks (solos) exhibit a small-world phenomenon, by comparing the clustering coefficient and average distance of these networks with those of a random graph of the same size. The considered networks (solos) are those shown in Figure \ref{fig:someNets}. For each solo, column ``cc'' of the table reports the clustering coefficient measured for the corresponding network; column ``cc (RG)'' shows the clustering coefficient for a random graph of the same size; column ``avg dist'' provides the average distance among nodes in the networks (i.e.~the average shortest path); finally, column ``avg dist (RG)'' reports the average distance for a random graph of the same size.

It is possible to observe that two solos can be classified as small worlds, namely ``Crossroads'' by E.~Clapton and ``Red House'' by J.~Hendrix. In fact, these two solos have a clustering coefficient significantly higher than their corresponding random graphs, and the average distances are lower but comparable to those obtained for random graphs.
As to ``Rock me baby'', by B.B.~King, its clustering coefficient is lower than that of its corresponding random graph (and its average distance is lower as well), but this difference is not evident as for the other two solos mentioned above.
Finally, a small world property is not evident for ``Comfortably numb (1st solo)'' by D.~Gilmour (Pink Floyd). Indeed, this solo is a (touching) linear melody, without an intricate structure, and this is evident by the pictorial representation of the network reported in Figure \ref{fig:confNumb}.

Further tests have been accomplished for other solos. Results demonstrate that there are different outcomes (i.e., some solos are small worlds, other do not), depending on the solo, as with the four solos considered above.

\subsection{Centrality measures: betweenness}

\begin{figure*}[ht]
\centering
\includegraphics[width=.44\textwidth]{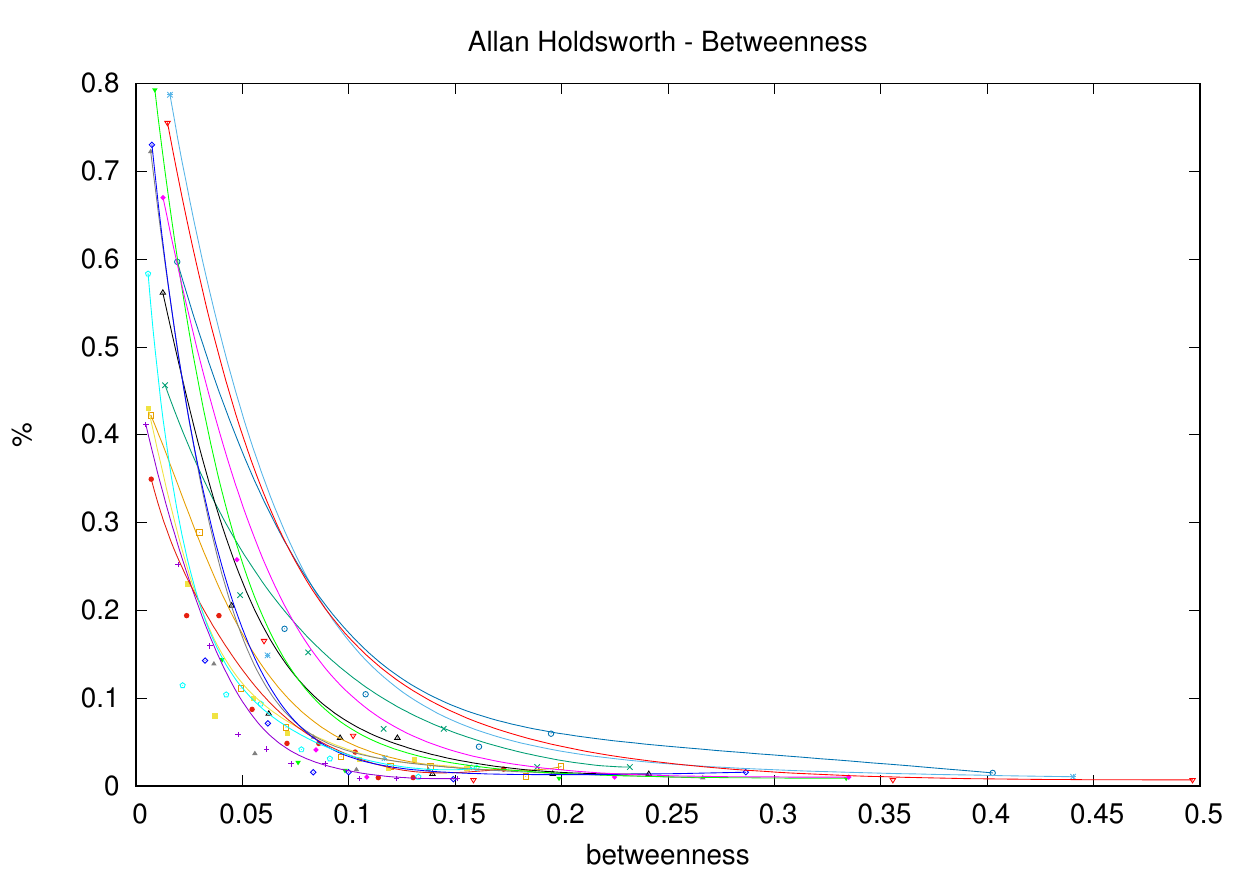}
\includegraphics[width=.44\textwidth]{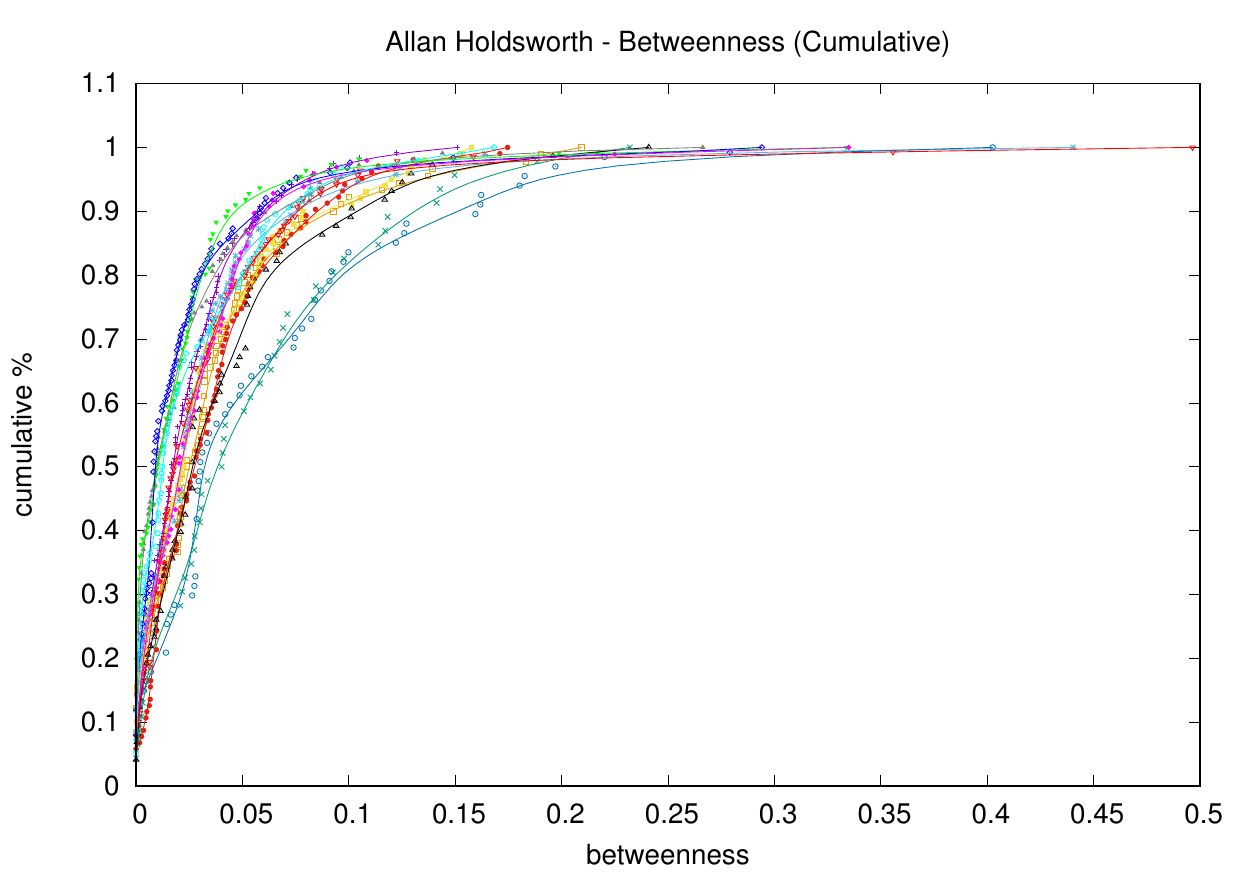}
\includegraphics[width=.44\textwidth]{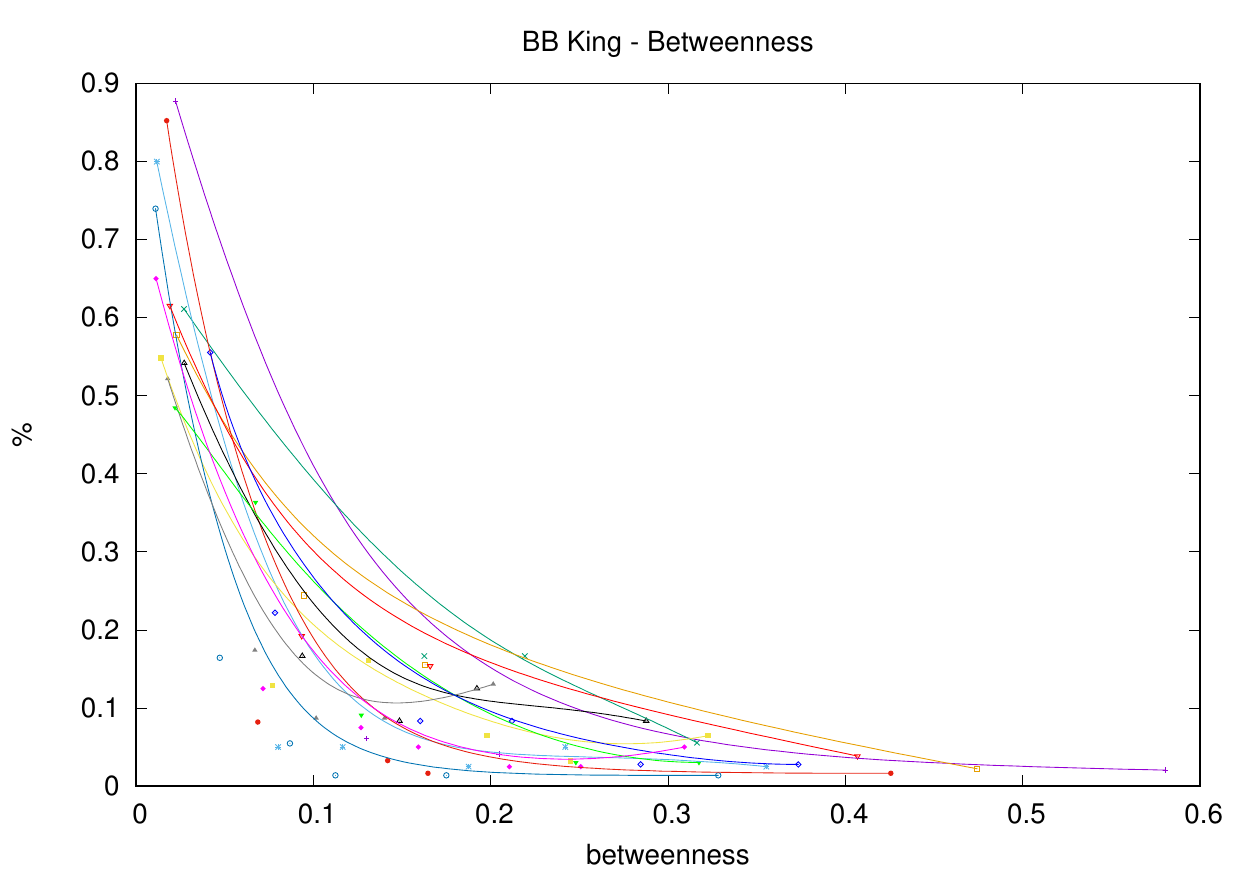}
\includegraphics[width=.44\textwidth]{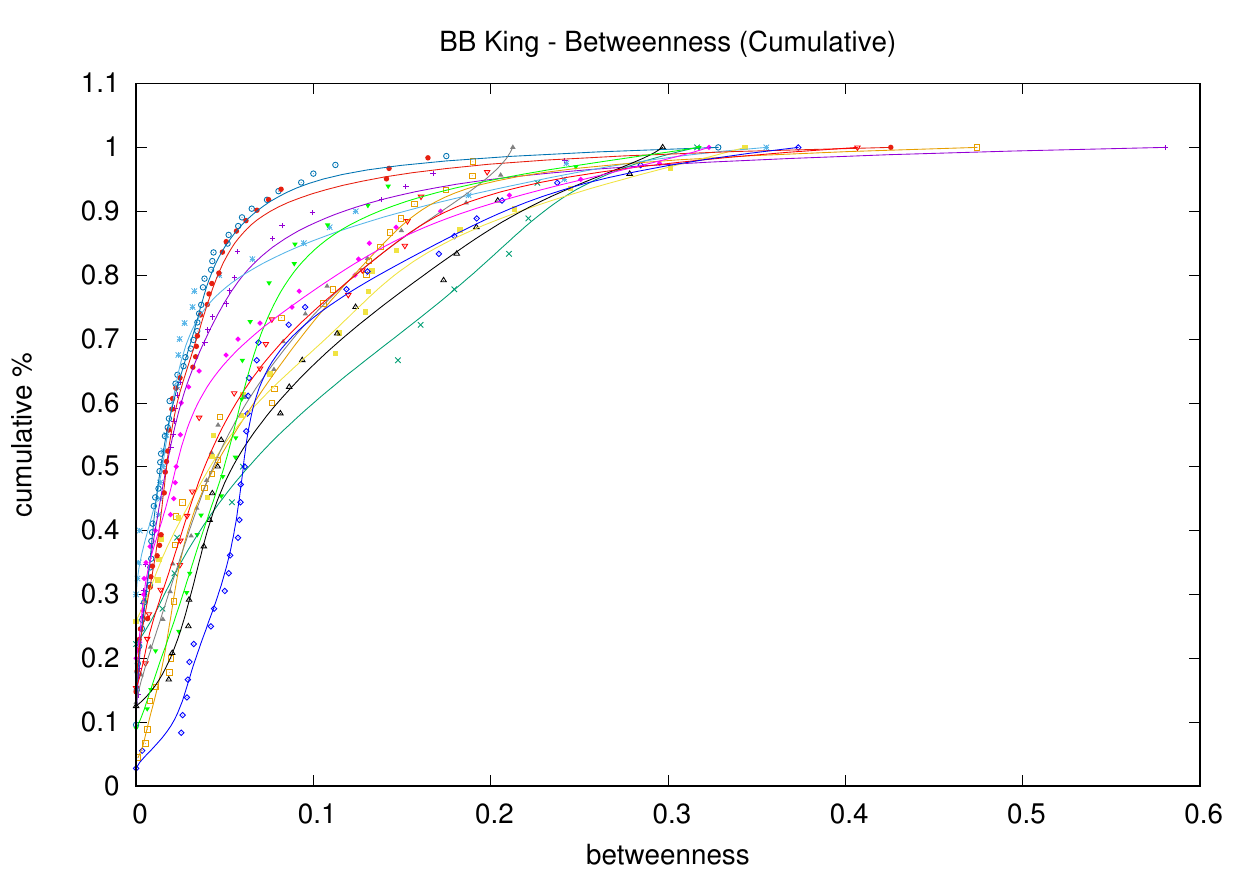}
\includegraphics[width=.44\textwidth]{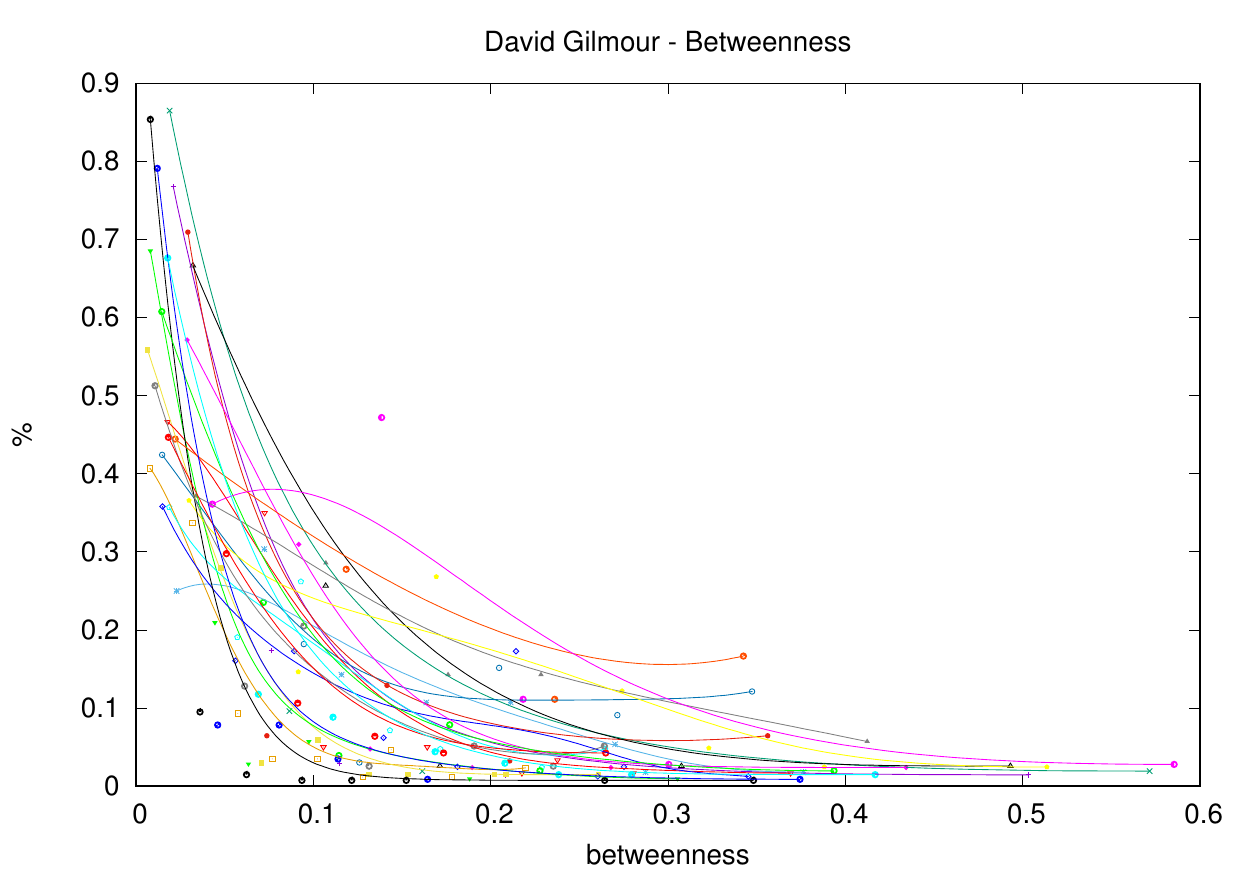}
\includegraphics[width=.44\textwidth]{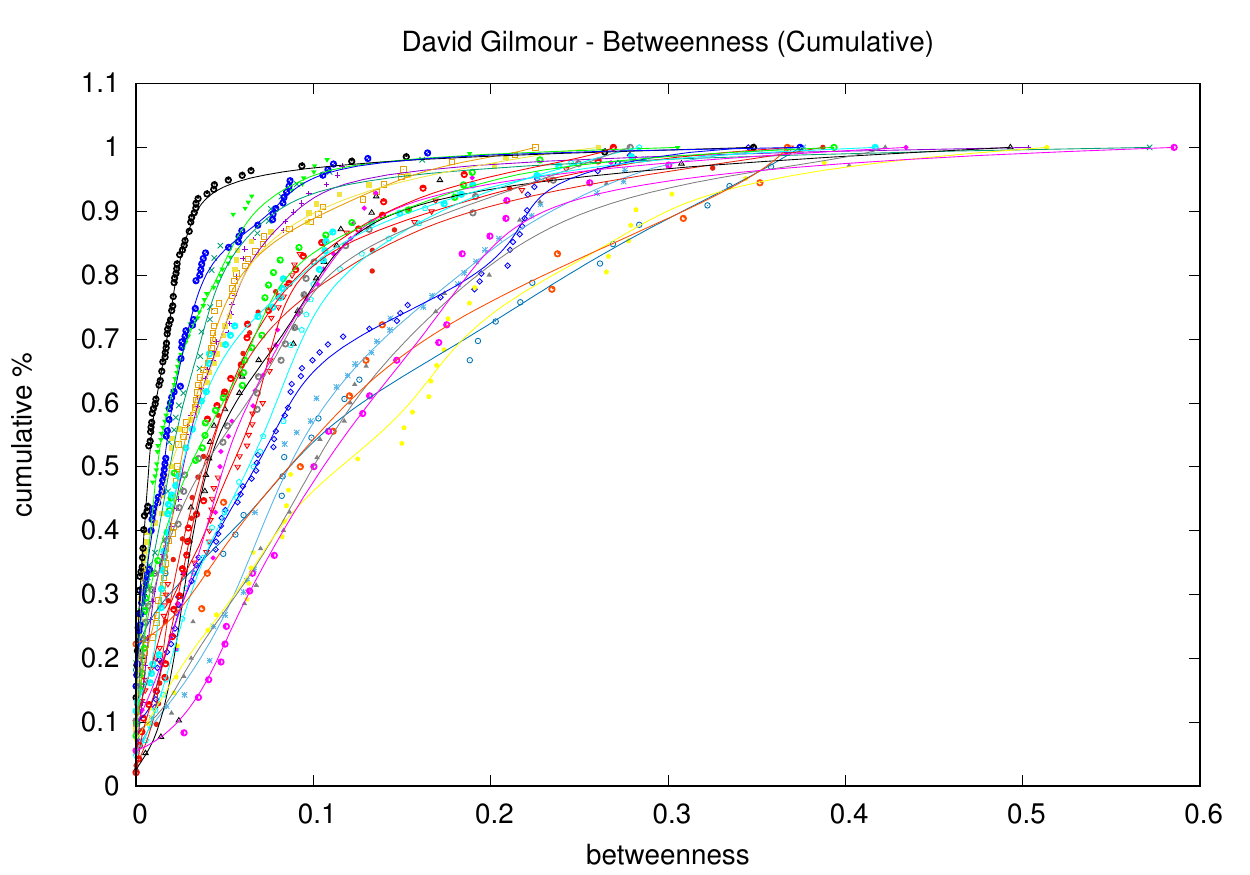}
\caption{Betweenness -- part 1. Each row corresponds to a given artist; the left side chart shows the degree distribution, while the right side chart shows the cumulative distribution. Points refer to values obtained for each of the tracks, while lines are the related Bézier interpolations.}
\label{fig:betweenneess_p1}
\end{figure*}

\begin{figure*}[ht]
\centering
\includegraphics[width=.44\textwidth]{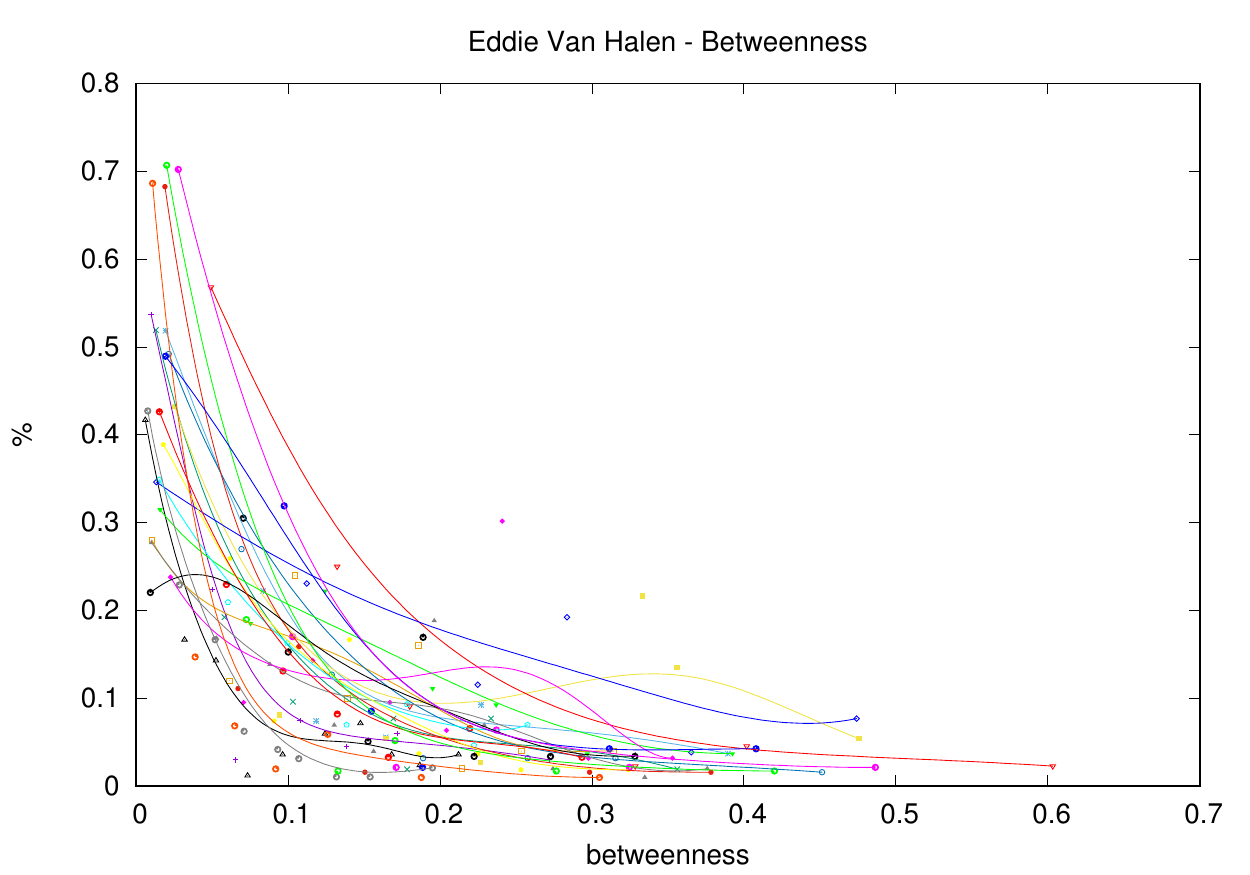}
\includegraphics[width=.44\textwidth]{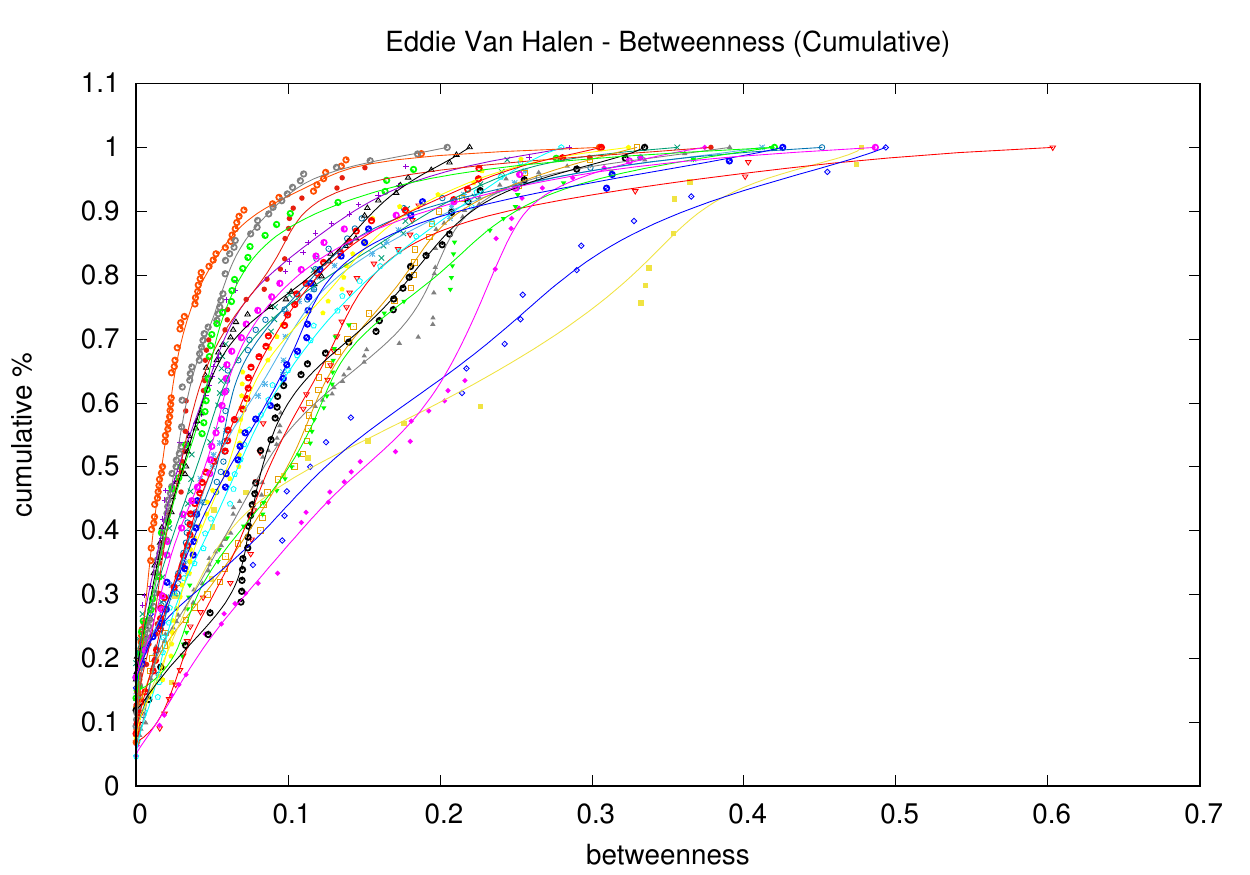}
\includegraphics[width=.44\textwidth]{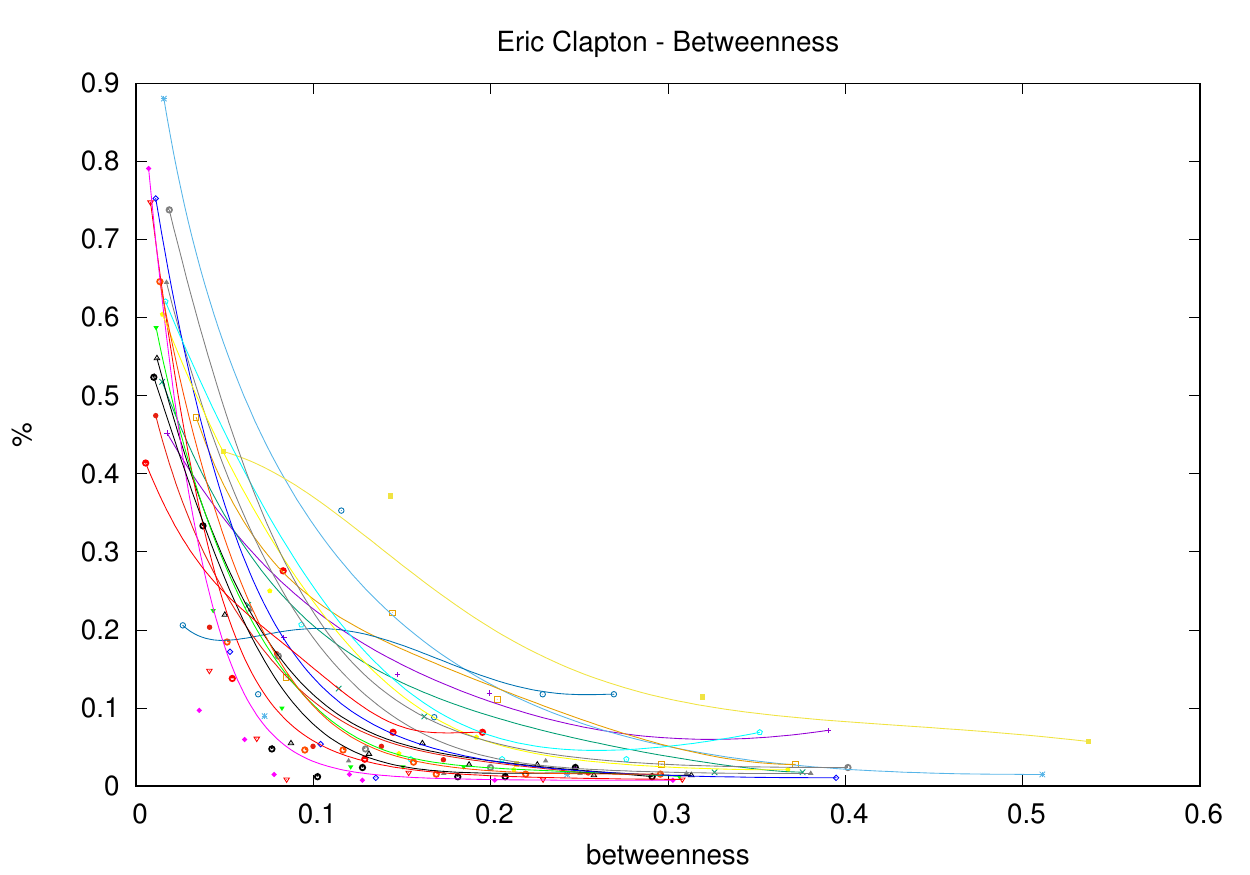}
\includegraphics[width=.44\textwidth]{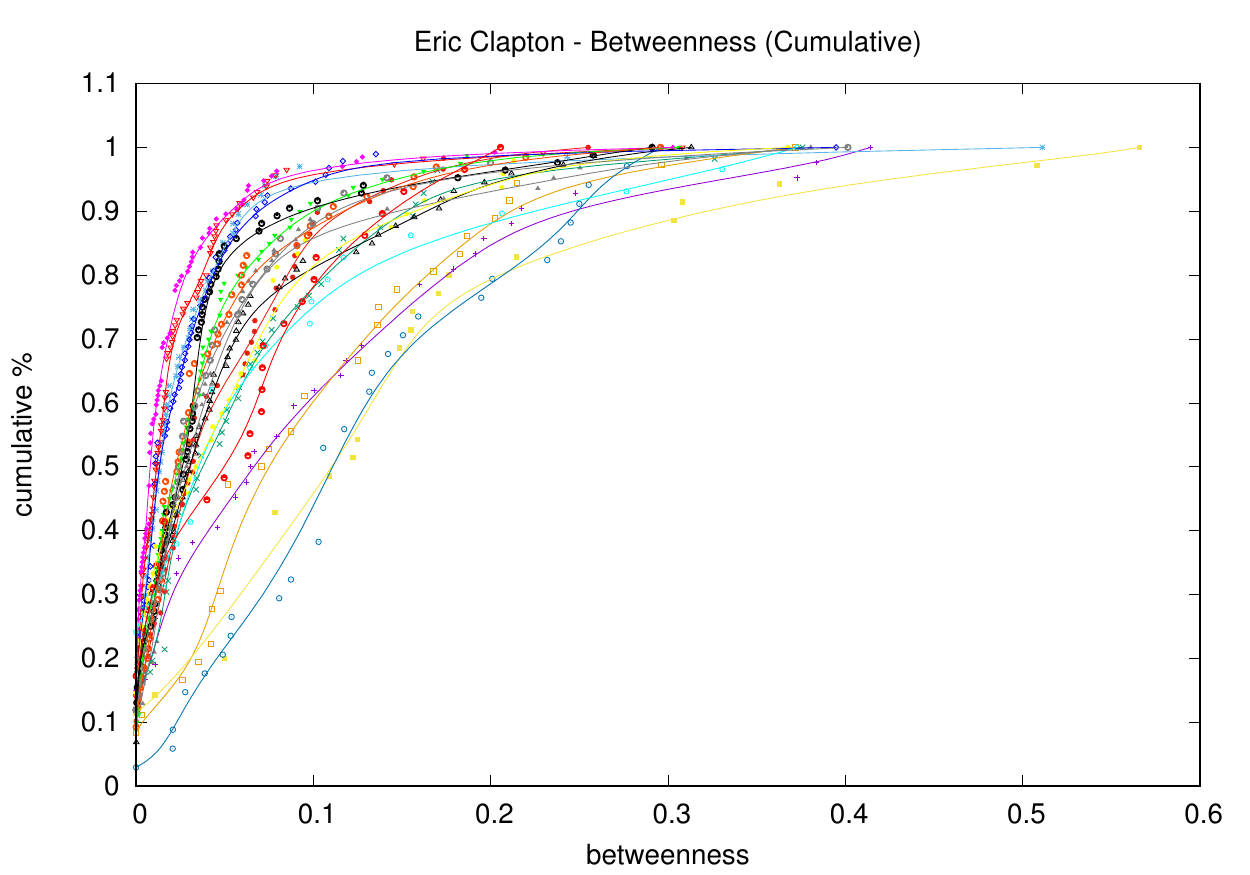}
\includegraphics[width=.44\textwidth]{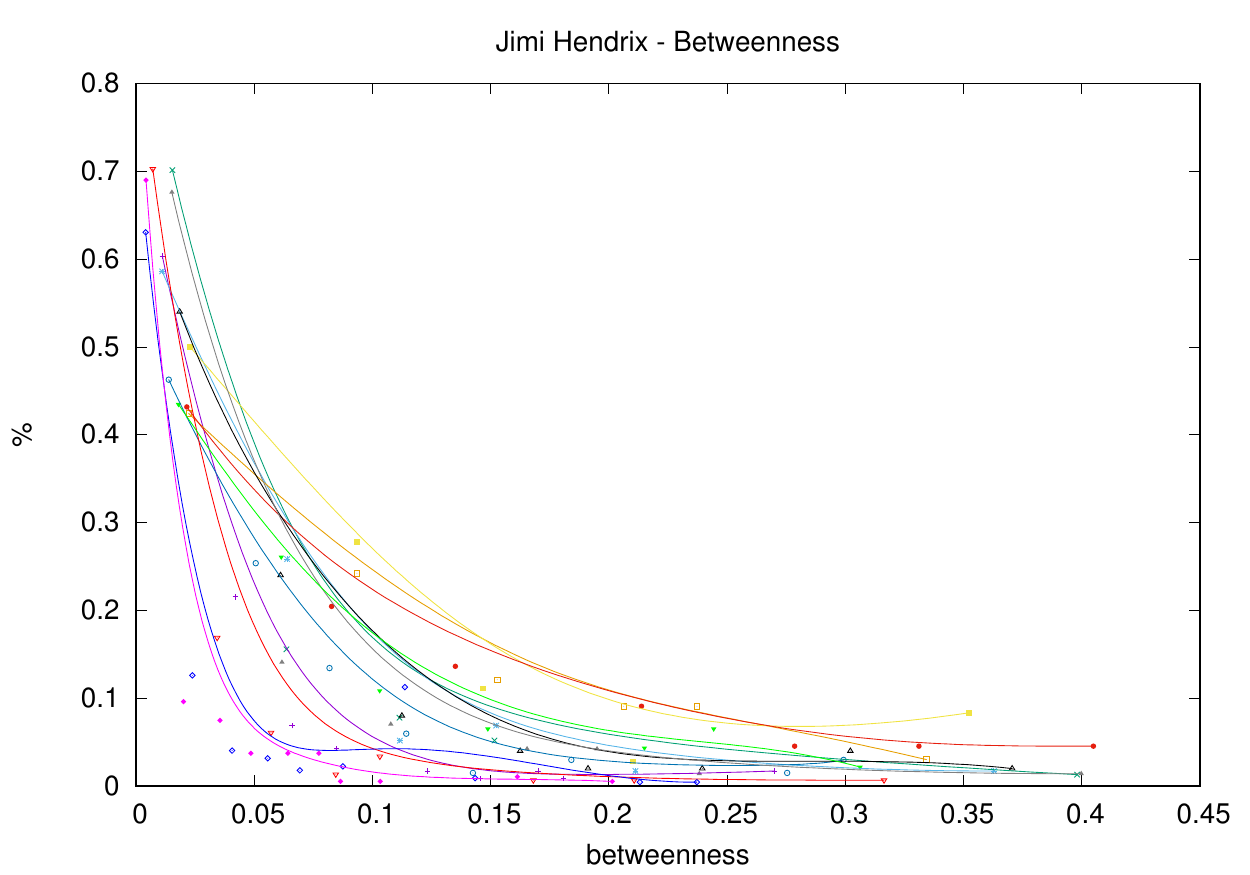}
\includegraphics[width=.44\textwidth]{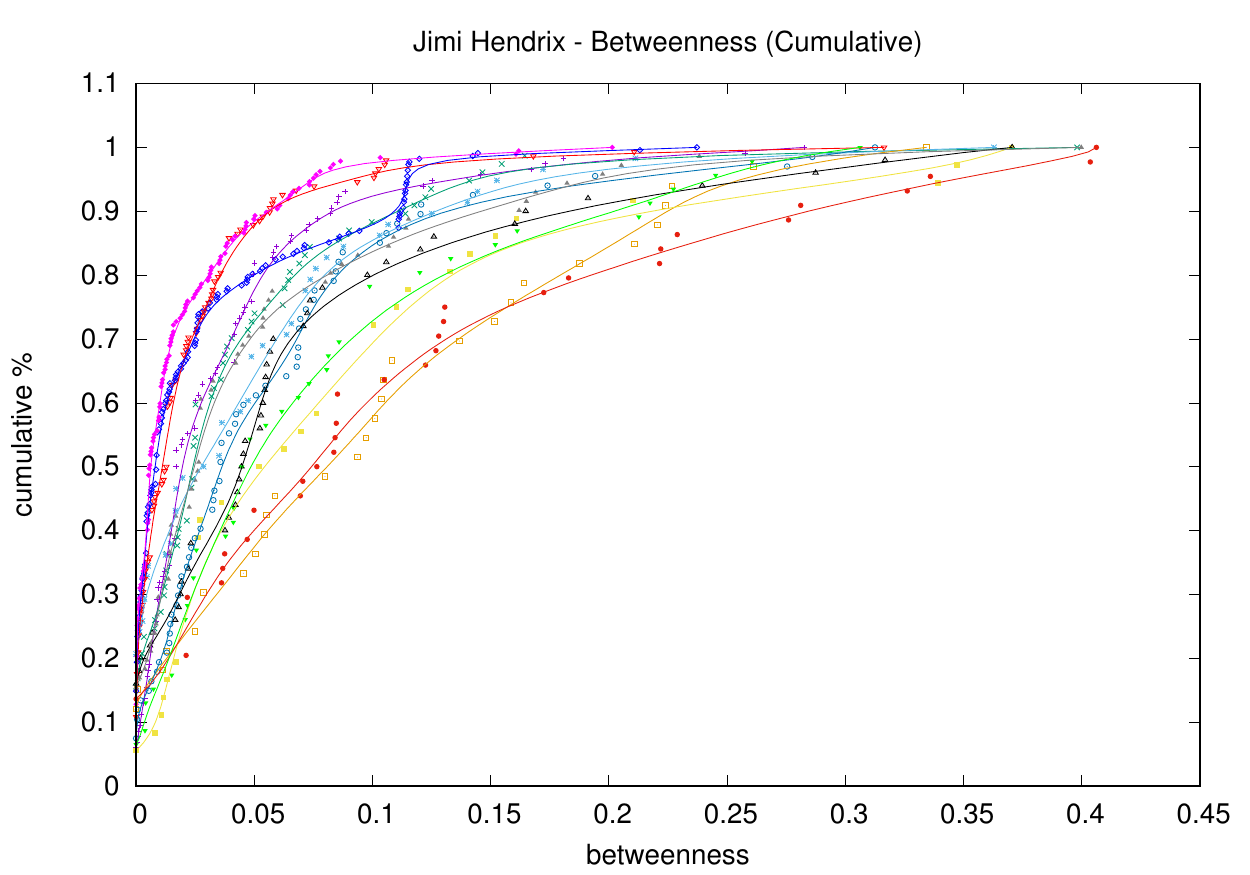}
\caption{Betweenness -- part 2. Each row corresponds to a given artist; the left side chart shows the degree distribution, while the right side chart shows the cumulative distribution. Points refer to values obtained for each of the tracks, while lines are the related Bézier interpolations.}
\label{fig:betweenneess_p2}
\end{figure*}

\begin{figure*}[ht]
\centering
\includegraphics[width=.44\textwidth]{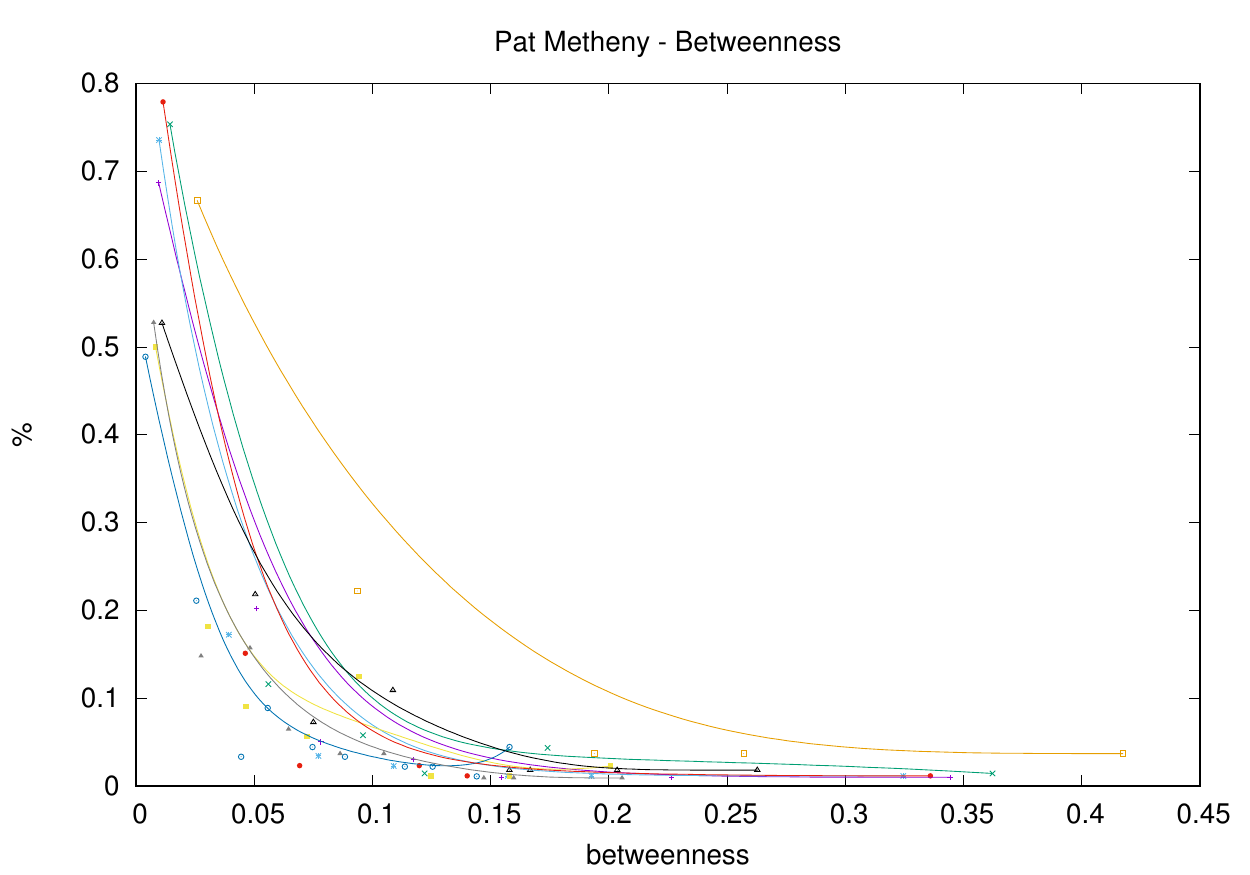}
\includegraphics[width=.44\textwidth]{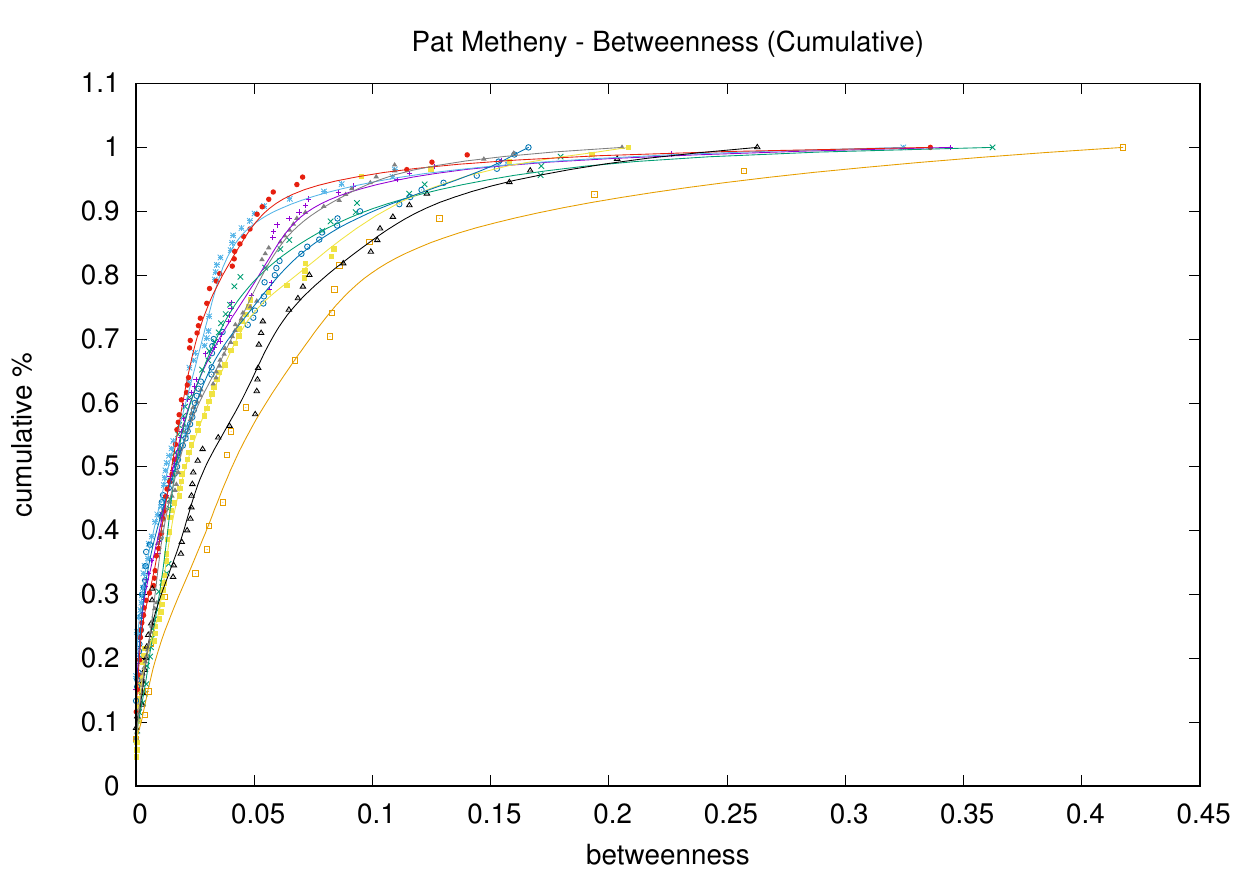}
\includegraphics[width=.44\textwidth]{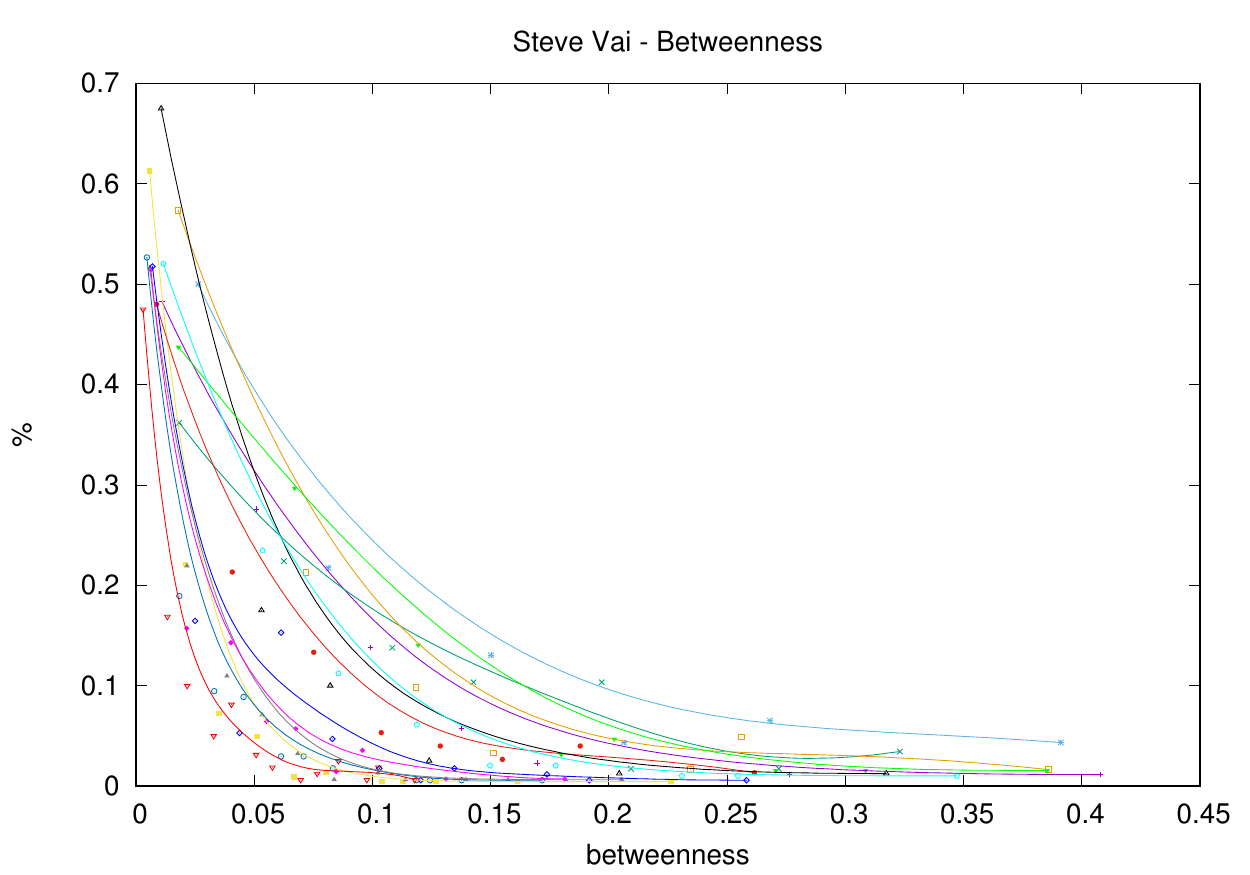}
\includegraphics[width=.44\textwidth]{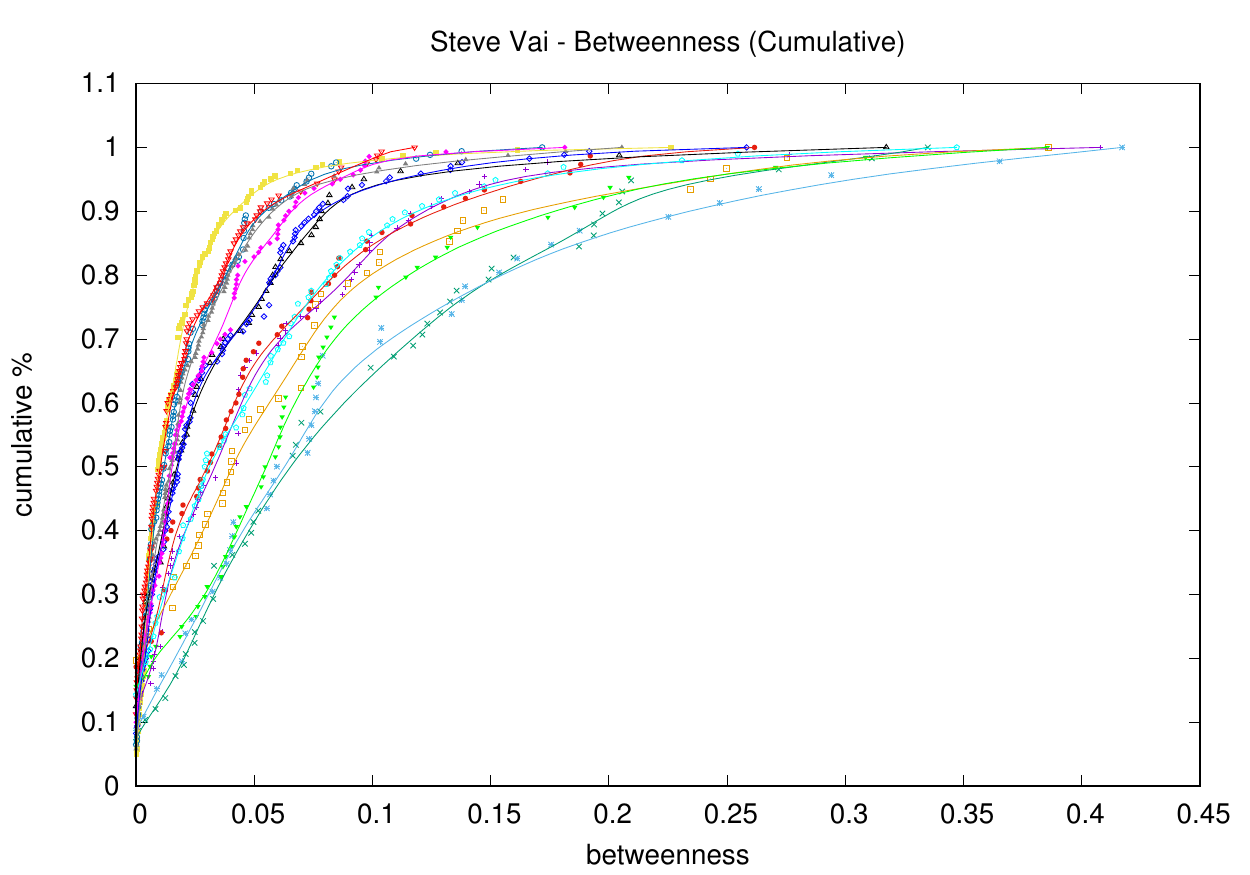}
\caption{Betweenness -- part 3. Each row corresponds to a given artist; the left side chart shows the degree distribution, while the right side chart shows the cumulative distribution. Points refer to values obtained for each of the tracks, while lines are the related Bézier interpolations.}
\label{fig:betweenneess_p3}
\end{figure*}

Figures \ref{fig:betweenneess_p1}--\ref{fig:betweenneess_p3} show, for each considered performer, the distribution of the betweenness centrality measure for the considered solos (left-side charts) and the cumulative distribution (right-side charts). In the charts, values on the $x$-axis are the betweenness values, and the values on the $y$-axis is the distribution (or cumulative distribution).
The distribution allows understanding if there is a non-negligible presence of nodes with a betweenness higher than others, i.e., if the performer is used to build his licks by exploiting some main, preferred notes.

Charts show important differences among performers. In fact, it appears that, mainly, jazz/fusion performers (Holdsworth, Metheny) have nodes with low betweenness; curves in the distribution decrease quite rapidly, and the curves in the cumulative distribution reach the $1$ value quite rapidly.
Other rock-blues performers have higher portions of nodes with higher betweenness. 
In particular, a higher variability is evident for the betweenness distributions of different solos of the same (rock-blues) artist, while such a variability is limited for the mentioned jazz/fusion performers.


\begin{figure*}[ht]
\centering
\includegraphics[width=.44\textwidth]{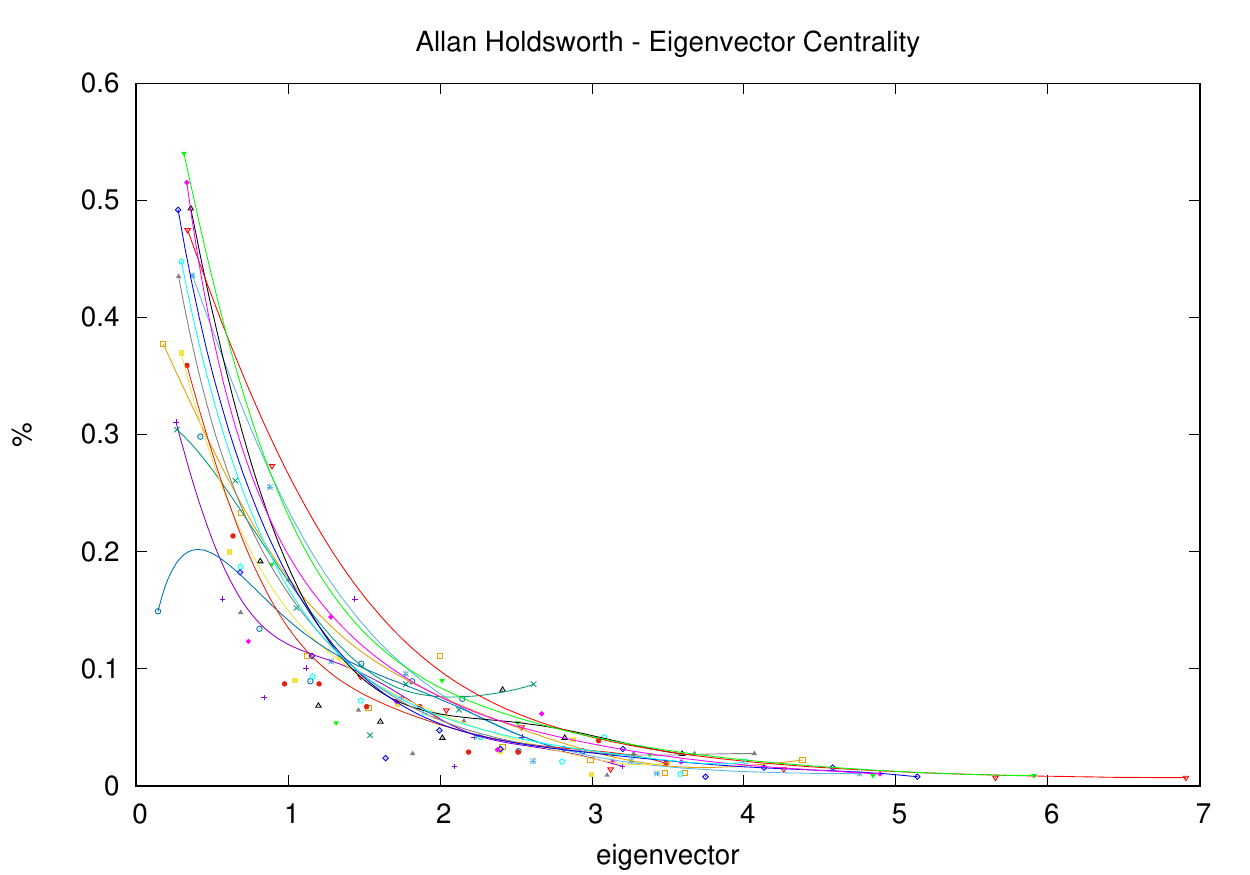}
\includegraphics[width=.44\textwidth]{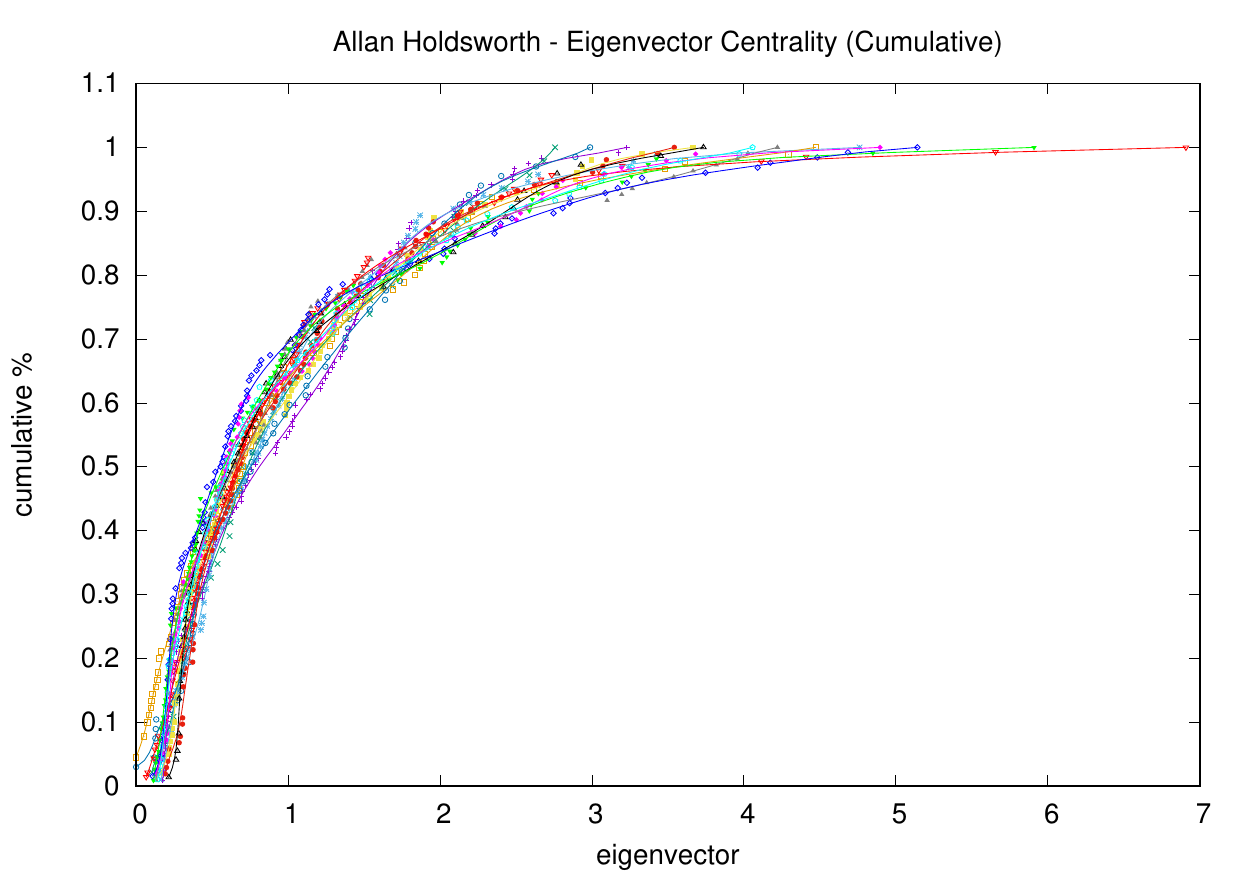}
\includegraphics[width=.44\textwidth]{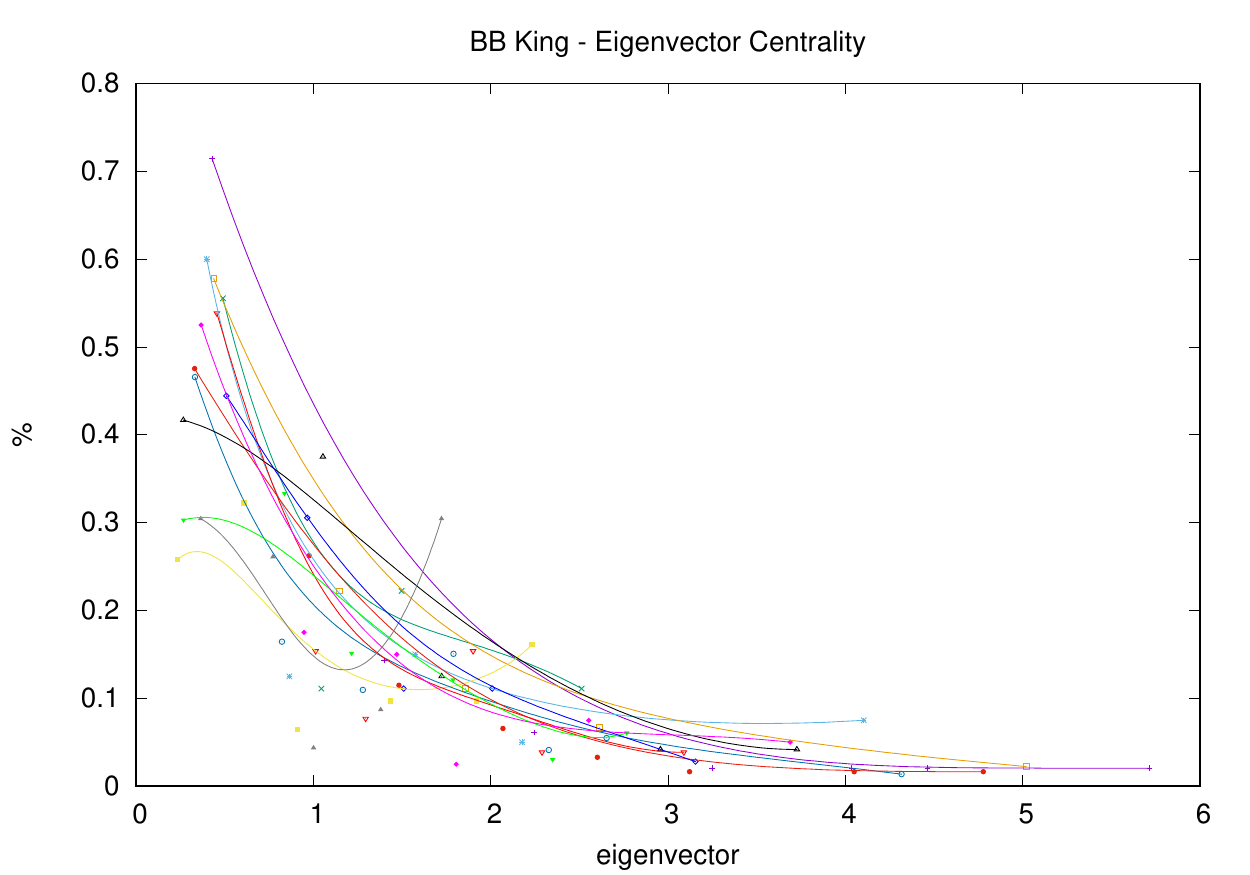}
\includegraphics[width=.44\textwidth]{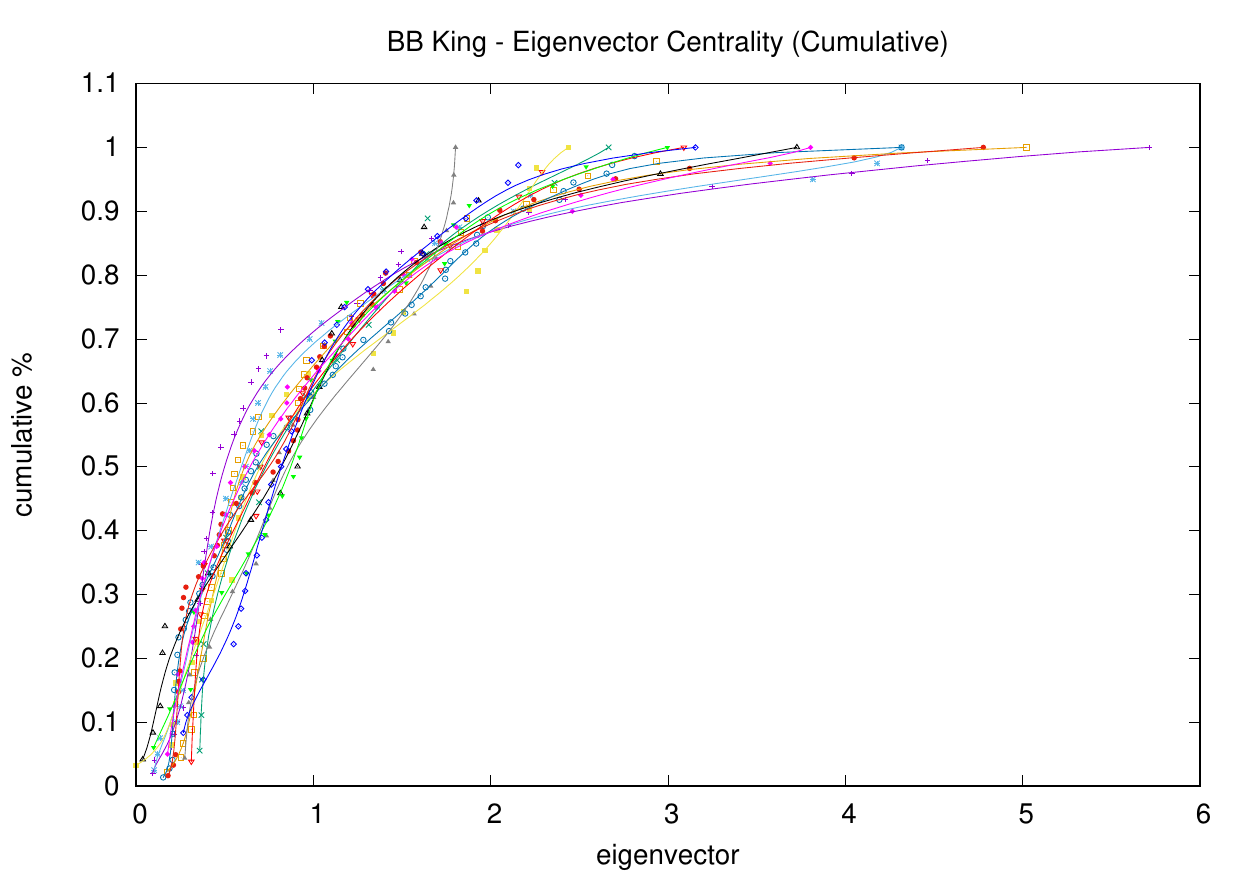}
\includegraphics[width=.44\textwidth]{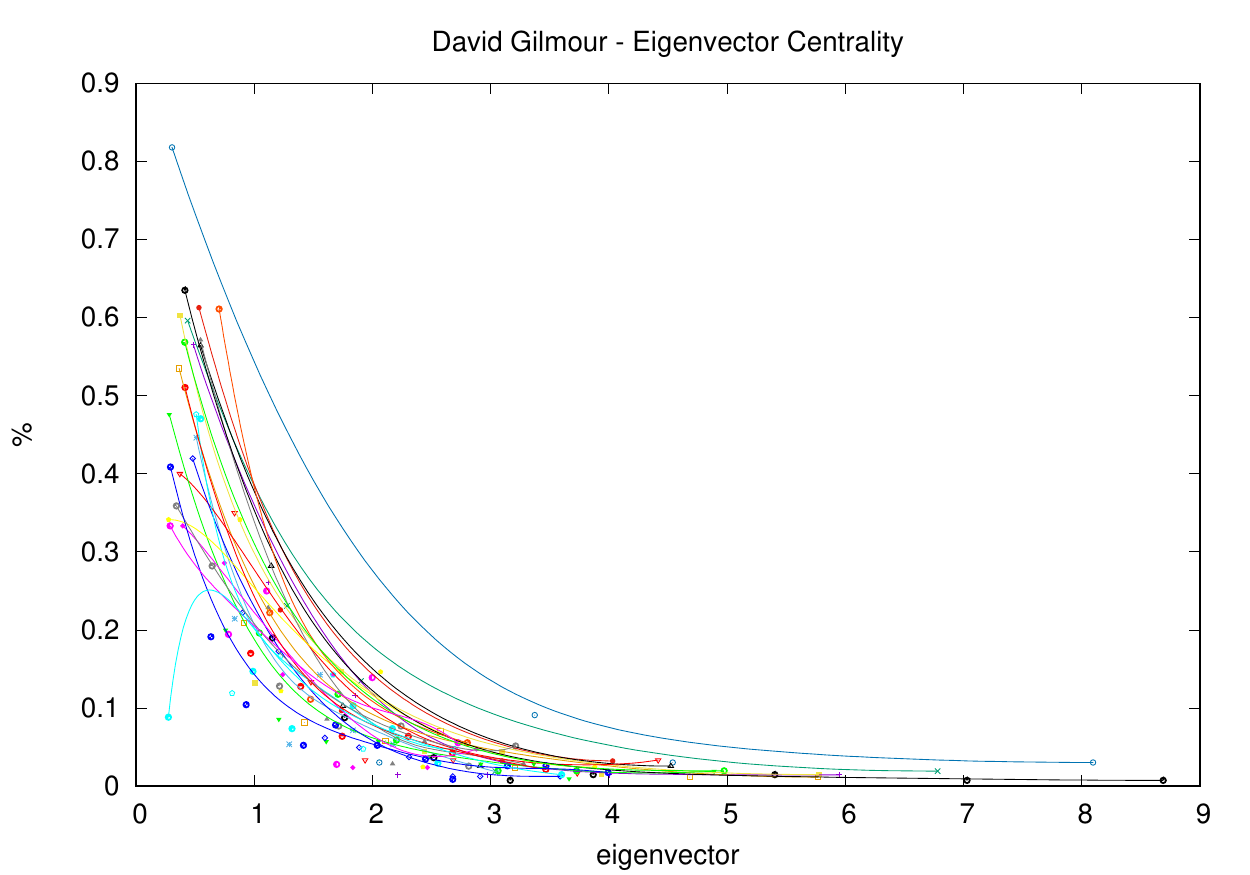}
\includegraphics[width=.44\textwidth]{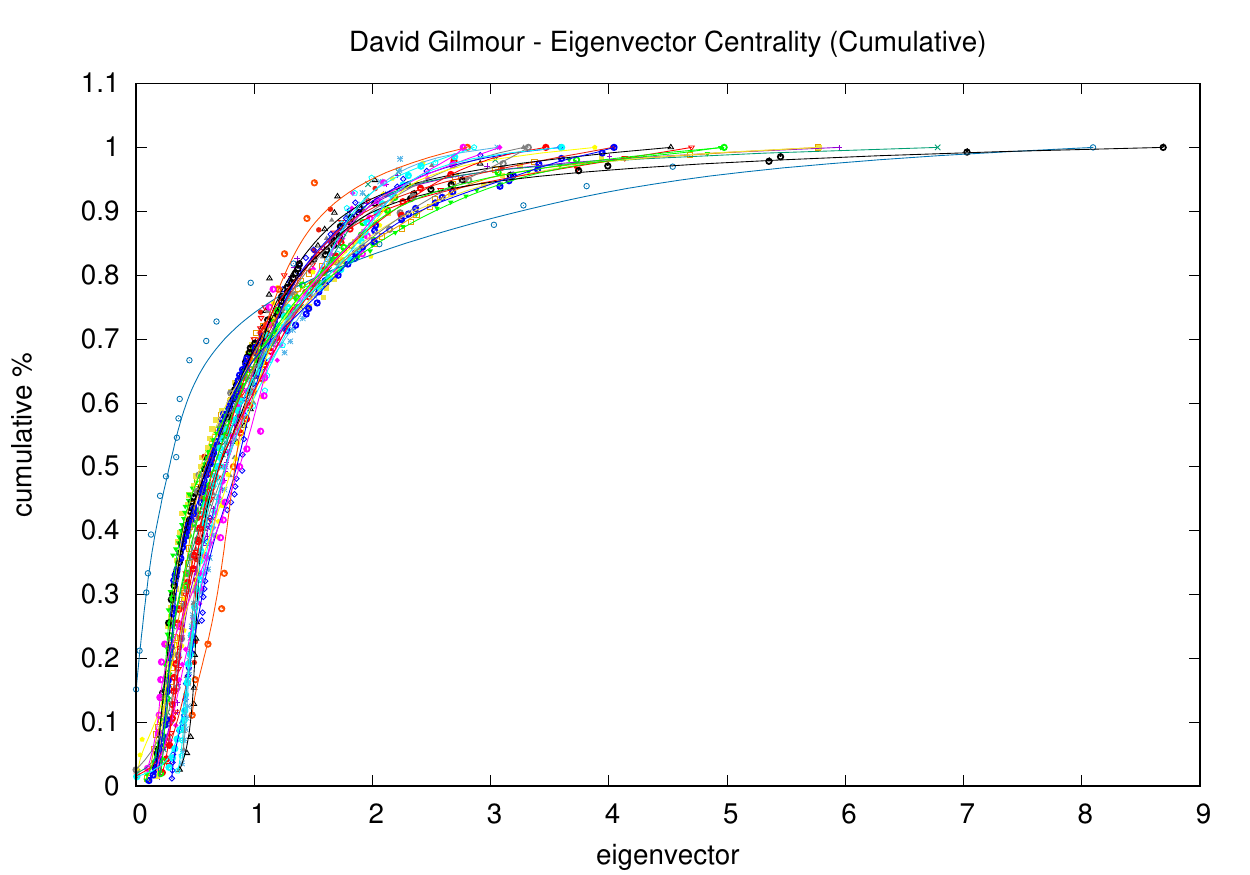}
\caption{Eigenvector Centrality. Each row corresponds to a given artist; the left side chart shows the degree distribution, while the right side chart shows the cumulative distribution. Points refer to values obtained for each of the tracks, while lines are the related Bézier interpolations.}
\label{fig:eigenvector_p1}
\end{figure*}

We also measured the weighted betweenness centrality, a measure which is similar to the betweenness centrality, but in the formula reported in Section \ref{sec:bet} the summation considers weighted links to measure the shortest paths. In this case, the differences among performers is (slightly) more evident, but in essence results are comparable to those for the betweenness. For this reason and for the sake of brevity, we do not report these charts.

\subsection{Centrality measures: eigenvector centrality}

% Figures \ref{fig:eigenvector_p1}--\ref{fig:eigenvector_p3} show 
Figure \ref{fig:eigenvector_p1} shows 
the distributions (left-side charts) and the cumulative distributions (right-side charts) of the eigenvector centrality for solos of the first three performers (in alphabetical order). In this case, we show only the first three ones in order to limit the total amount of figures, since no important differences are evident from the distributions. Rock-blues musicians appear to have slightly higher portions of nodes with low eigenvector centralities. However, the trend is almost similar for all musicians.
These results suggest that there are notes more present than others in commonly played licks, and this is a common practice for all the considered performers.

\subsection{Analysis of concatenated solos}

This subsection presents results related to networks obtained by concatenating different solos of the same artist, following the approach employed in \cite{Liu2010126}.
The idea is that musical solos can result in networks smaller than typical ones employed in complex network theory.
Through concatenation, we obtain one track (and one network) per artist composed of a number of notes higher than 20K. 
In general, some of the features and metrics related to the concatenated network are different to those of single solos. 

The rationale here is to obtain results that can be compared with those presented in \cite{Liu2010126}. 
In fact, they found out that these concatenated networks exhibit a scale-free structure, small-world phenomenon, mean shortest distances around $3$ and clustering coefficients around $0.3$.

\begin{table*}[th]
\centering
\caption{Small world property of concatenated solos: comparison between the clustering coefficient (column ``cc'') and the average distance (column ``avg dist'') of the considered network, and the clustering coefficient (column ``cc (RG)'') and the average distance (column ``avg dist (RG)'') of the corresponding random graph.}
% \caption{Small world property: comparison between solo networks and corresponding random graphs.}
\label{tab:sw_concat}
% \small
\scriptsize
\begin{tabular}{|| l || c | c | c | c ||}
  \hline			
  \hline			
  artist & cc & cc (RG) & avg dist & avg dist (RG)  \\
  \hline  
  \hline			
Allan Holdsworth	&	0.41	&	0.03	&	2.97	&	5.77\\
BB King	&	0.44	&	0.03	&	2.54	&	5.24\\
David Gilmour	&	0.38	&	0.02	&	3.01	&	6.03\\
Eddie Van Halen	&	0.33	&	0.01	&	3.44	&	6.05\\
Eric Clapton	&	0.44	&	0.01	&	3.05	&	6.06\\
Jimi Hendrix	&	0.32	&	0.01	&	4.04	&	6.39\\
Pat Metheny	&	0.38	&	0.03	&	2.91	&	5.51\\
Steve Vai	&	0.30	&	0.01	&	3.35	&	6.30\\
  \hline  
  \hline			
\end{tabular}
\end{table*}

Table \ref{tab:sw_concat} assesses if the concatenated network obtained for each artist is a small-world, by comparing the clustering coefficient and average distance of these networks with those of random graphs of the same size. 
We previously observed that when considering single solos, a small world phenomenon was not always present. As concerns concatenated networks, it is evident that all networks are small-worlds. This is in accordance with results in \cite{Liu2010126}.

\begin{table*}[th]
\centering
\caption{Clustering coefficient of concatenated solos vs average clustering coefficient of separate solos. 
Artists sorted by increasing values of the clustering coefficient of the concatenated network.}
\label{tab:cc_concat}
\scriptsize
\begin{tabular}{|| l || c | c ||}
  \hline			
  \hline	
artist	&	cc concatenated	net &	 average cc\\
  \hline  
  \hline	
Steve Vai	&	0.30	&	0.24\\
Jimi Hendrix	&	0.32	&	0.29\\
Eddie Van Halen	&	0.33	&	0.29\\
David Gilmour	&	0.38	&	0.25\\
Pat Metheny	&	0.38	&	0.33\\
Allan Holdsworth	&	0.41	&	0.30\\
BB King		&	0.44	&	0.35\\
Eric Clapton	&	0.44	&	0.33\\
  \hline  
  \hline			
\end{tabular}
\end{table*}

In any case, if we compare the clustering coefficient of the concatenated network with the average clustering coefficient of separate solos of the same artist, we notice that the former value is always higher that the latter. Results are shown in Table \ref{tab:cc_concat}.
It is thus unclear whether this measure for concatenated solos can represent a general value characterizing the artist.

\begin{table*}[th]
\centering
\caption{Average degree of concatenated solos vs average degree of separate solos.
Artists sorted by increasing values of the average degree of the concatenated network.}
\label{tab:deg_concat}
\scriptsize
\begin{tabular}{|| l || c | c ||}
  \hline			
  \hline	
artist	&	average degree concatenated net	&	 average degree\\
  \hline  
  \hline	
Jimi Hendrix	&	8.98	&	4.91\\
Eddie Van Halen	&	10.85	&	4.13\\
BB King	&	11.75	&	5.59\\
Eric Clapton	&	12.34	&	5.57\\
Steve Vai	&	12.47	&	5.16\\
David Gilmour	&	12.88	&	4.74\\
Pat Metheny	&	15.87	&	7.46\\
Allan Holdsworth	&	18.34	&	7.08\\
  \hline  
  \hline			
\end{tabular}
\end{table*}

The difference between the concatenated network and the separate networks is more evident if we consider the degree. Table \ref{tab:deg_concat} compares the average degree of the concatenated network with the average of all the average degrees measured for separate networks. It is possible to notice a high gap between these measures, in all cases.

\begin{table*}[th]
\centering
\caption{Average distance of concatenated solos vs average distance of separate solos.
Artists sorted by increasing values of the average distance of the concatenated network.}
% \caption{Small world property: comparison between solo networks and corresponding random graphs.}
\label{tab:apl_concat}
% \small
\scriptsize
\begin{tabular}{|| l || c | c ||}
  \hline			
  \hline	
artist	&	avg dist concatenated net	&	 avg dist\\
  \hline  
  \hline	
BB King	&	 2.54	&	 2.49\\
Pat Metheny	&	 2.91	&	 2.93\\
Allan Holdsworth	&	 2.97	&	 3.27\\
David Gilmour	&	 3.01	&	 3.40\\
Eric Clapton	&	 3.05	&	 3.05\\
Steve Vai	&	 3.35	&	 4.03\\
Eddie Van Halen	&	 3.44	&	 4.42\\
Jimi Hendrix	&	 4.04	&	 3.72\\
  \hline  
  \hline			
\end{tabular}
\end{table*}

Table \ref{tab:apl_concat} compares the average distance of the concatenated network and the average of the average distances of separate networks of all artists. In this case there are slight differences. It is interesting to observe that the average distance of the concatenated networks is in general around $3$, as in \cite{Liu2010126}.

Based on these results, it is possible to conclude that the concatenation of different solos of an artist is a simple strategy to collect all the licks played in his works. 
It might be exploited as a means to capture his whole ``musical vocabulary'', to be employed during a music generation process \cite{Liu2010126}.
Thus, it might be manipulated in order to automatically extract common patterns and licks played by the artist in his works; indeed, the (non-automatic) ``manual'' version of such a technique is broadly employed in modern music didactics.

On the other hand, the focus on single tracks and solos is probably more accurate, in general. 
Indeed, melodic lines have their own ``meaning'' and intention. 
Melodies are influenced by the underlying harmonic structure, the tempo and the general mood of the specific track. 
Thus, what the performer plays is strictly dependent on the track; this results in a melody that has properties which can be different to those for other tracks. 
Results obtained on the assessment on single tracks confirm this claim. 
Conversely, the concatenation of musical tracks changes the value of certain (average) metrics, since it aggregates several melodies into a single one.

Finally, while the presented assessment focused on the peculiar characteristics of artists, one might decide to study other aspects of a set of solos, such as their similarity. 
Another example is the study of differences among songs or music genres; e.g., melodic songs ``inspire'' artists to create melodies with certain features, while blues melodies have other features. In this case, it is clear that the networks obtained from the tracks are the elements to inspect.

To sum up, depending on the analysis, one should identify the most proper musical element (e.g., entire track, isolated solo, concatenation of solos, etc.) to be treated as a network, so as to optimize the analysis.

%%%%%%%%%%%%%%%%%%%%%%%%%
\section{Conclusions}
\label{sec:conc}
%%%%%%%%%%%%%%%%%%%%%%%%%

We presented an approach to model melodies (and music pieces in general) as networks. 
To this aim, several methods can be exploited; in this work, a network is created based on a melody/solo, by representing notes as network nodes, and links are added between two nodes if the related notes are played in succession.

A database of different solos played by eight important guitar players has been analyzed through this approach. Some main metrics, typical of complex network theory, have been computed and analyzed, i.e.~length of the solo, number of nodes, degree distribution, distance, clustering coefficient, betweenness and eigenvector centrality measures. Outcomes are in complete accordance with the common opinions of music experts.
Indeed, there are works where aspects on improvisation and musical styles are considered \cite{Pressing,covach1997understanding,smith,zenni}. 
While in agreement to the discussion in this work, there are no quantitative measures that can be compared with our results. 
A contribution of this work is that of considering some mathematical metrics that can be employed during the classification of a musician.
Summing up, jazz guitar players and rock ``virtuoso'' players do create solos corresponding to more complex structures.
Conversely, those solos that are usually considered as ``melodic'' have a simpler structure. Blues performers such as B.B.~King are different from other musicians (as demonstrated via statistical tests).
This testifies the correctness of the presented approach. 

We argue that the use of a mathematical modeling of a solo (or a music track in general) provides a general and compact way to analyze music.
We noticed interesting differences among performers in the degree distributions, distance measures, clustering coefficient and betweenness.
Moreover, we assessed that some networks/solos are small worlds, while others, usually melodic ones, do not.
This suggests that further studies might be done to classify solos, regardless of the performers that played them.

The provided insights stimulate new questions on the possibility to fully characterize and capture the artistic traits and skills of a musician, through mathematical concepts.
The application scenarios of this study are related to the possible development of innovative applications related to music classification and categorization. 
Moreover, the proposed framework can be exploited as a tool during the automatic generation of music. If we are able to harness the main characteristics of a musician, it would be possible to combine this approach to some artificial intelligence machinery and generate, for example, a solo ``à la'' Miles Davis. 
As an example, these results suggest that 
a ``bluesy'' solo should have a high clustering coefficient and low distance;
a ``modern rock'' solo should have high distance and an average degree lower than other genres,
while a ``melodic'' solo should have a simple network structure.
This might have interesting applications in music 
didactics, multimedia entertainment, and digital music generation.

Another interesting further work is related to the clustering of musical solos, so that to group them, provide support for speculative analysis and capture the typical characteristics of a given ``type of solo'', regardless of the particular musician that might have played it. 

%%%%%%%%%%%%%%%%%%%%
%% BIBLIOGRAFIA
%%%%%%%%%%%%%%%%%%%%

% \bibliographystyle{elsarticle-num}
\bibliographystyle{elsart-num-sort}

\begin{thebibliography}{10}
\expandafter\ifx\csname url\endcsname\relax
  \def\url#1{\texttt{#1}}\fi
\expandafter\ifx\csname urlprefix\endcsname\relax\def\urlprefix{URL }\fi

\bibitem{guitartabs}
Guitare tab! web site.
\newline\urlprefix\url{http://www.guitaretab.com/}

\bibitem{apache}
{Math - Commons-Math: The Apache Commons Mathematics Library}.
\newline\urlprefix\url{http://commons.apache.org/math/}

\bibitem{musicXML}
musicxml web site.
\newline\urlprefix\url{http://www.musicxml.com/}

\bibitem{pyguitar}
Pyguitarpro web site.
\newline\urlprefix\url{http://pyguitarpro.readthedocs.org/en/latest/}

\bibitem{ultimate}
Ultimate guitar tab web site.
\newline\urlprefix\url{http://www.ultimate-guitar.com}

\bibitem{AliakbaryMHM13}
S.~Aliakbary, S.~Motallebi, J.~Habibi, A.~Movaghar, Learning an integrated
  distance metric for comparing structure of complex networks, CoRR
  abs/1307.3626.
\newline\urlprefix\url{http://arxiv.org/abs/1307.3626}

\bibitem{Anglade09genreclassification}
A.~Anglade, Q.~Mary, R.~Ramirez, S.~Dixon, Q.~Mary, Genre classification using
  harmony rules induced from automatic chord transcriptions, in: In 10th
  International Society for Music Information Retrieval Conference, 2009.

\bibitem{barrett}
S.~Barrett, Kind of blue and the economy of modal jazz, Popular Music 25 (2006)
  185--200.
\newline\urlprefix\url{http://ir.lib.uwo.ca/notabene/vol3/iss1/5}

\bibitem{Berenzweig:2004}
A.~Berenzweig, B.~Logan, D.~P.~W. Ellis, B.~P.~W. Whitman, A large-scale
  evaluation of acoustic and subjective music-similarity measures, Comput.
  Music J. 28~(2) (2004) 63--76.
\newline\urlprefix\url{http://dx.doi.org/10.1162/014892604323112257}

\bibitem{BiemannRW12}
C.~Biemann, S.~Roos, K.~Weihe, Quantifying semantics using complex network
  analysis, in: {COLING} 2012, 24th International Conference on Computational
  Linguistics, Proceedings of the Conference: Technical Papers, 8-15 December
  2012, Mumbai, India, 2012, pp. 263--278.
\newline\urlprefix\url{http://aclweb.org/anthology/C/C12/C12-1017.pdf}

\bibitem{Boccaletti2006175}
S.~Boccaletti, V.~Latora, Y.~Moreno, M.~Chavez, D.-U. Hwang, Complex networks:
  Structure and dynamics, Physics Reports 424~(4–5) (2006) 175 -- 308.
\newline\urlprefix\url{http://www.sciencedirect.com/science/article/pii/S037015730500462X}

\bibitem{Boothroyd}
M.~Boothroyd, Modal jazz and miles davis: George russell's influence and the
  melodic inspiration behind modal jazz, Canadian Undergraduate Journal of
  Musicology 3.
\newline\urlprefix\url{http://ir.lib.uwo.ca/notabene/vol3/iss1/5}

\bibitem{Cancho2261}
R.~F.~i. Cancho, R.~V. Sol{\'e}, The small world of human language, Proceedings
  of the Royal Society of London B: Biological Sciences 268~(1482) (2001)
  2261--2265.

\bibitem{Cilibrasi:2004}
R.~Cilibrasi, P.~Vit\'{a}nyi, R.~De~Wolf, Algorithmic clustering of music based
  on string compression, Comput. Music J. 28~(4) (2004) 49--67.
\newline\urlprefix\url{http://dx.doi.org/10.1162/0148926042728449}

\bibitem{Cong2014598}
J.~Cong, H.~Liu, Approaching human language with complex networks, Physics of
  Life Reviews 11~(4) (2014) 598 -- 618.
\newline\urlprefix\url{http://www.sciencedirect.com/science/article/pii/S1571064514000578}

\bibitem{covach1997understanding}
J.~Covach, G.~M. Boone, Understanding Rock: Essays in Musical Analysis, Oxford
  University Press, 1997.

\bibitem{icme16}
S.~Ferretti, Guitar solos as networks, in: 2016 IEEE International Conference
  on Multimedia and Expo (ICME), 2016, pp. 1--6.

\bibitem{Gartner2003}
T.~G{\"a}rtner, P.~Flach, S.~Wrobel, On Graph Kernels: Hardness Results and
  Efficient Alternatives, Springer Berlin Heidelberg, Berlin, Heidelberg, 2003,
  pp. 129--143.

\bibitem{doi:10.1080/09298210802479268}
R.~O. Gjerdingen, D.~Perrott, Scanning the dial: The rapid recognition of music
  genres, Journal of New Music Research 37~(2) (2008) 93--100.
\newline\urlprefix\url{http://dx.doi.org/10.1080/09298210802479268}

\bibitem{Grabska}
I.~Grabska-Gradzińska, A.~Kulig, J.~Kwapien, S.~Drozdz, Complex network
  analysis of literary and scientific texts, International Journal of Modern
  Physics C 23~(07).
\newline\urlprefix\url{http://www.worldscientific.com/doi/abs/10.1142/S0129183112500519}

\bibitem{Grisi-FilhoOFA13}
J.~H.~H. Grisi-Filho, R.~Ossada, F.~Ferreira, M.~Amaku, Scale-free networks
  with the same degree distribution: Different structural properties., CoRR
  abs/1306.0233.
\newline\urlprefix\url{http://dblp.uni-trier.de/db/journals/corr/corr1306.html\#Grisi-FilhoOFA13}

\bibitem{Knopke2011}
I.~Knopke, F.~J{\"u}rgensen, chap. Symbolic Data Mining in Musicology, Chapman
  \& Hall/CRC Data Mining and Knowledge Discovery Series, CRC Press, 2011, pp.
  327--345, 0.

\bibitem{Li2011}
T.~Li, L.~Li, chap. Music Data Mining, Chapman \& Hall/CRC Data Mining and
  Knowledge Discovery Series, CRC Press, 2011, pp. 3--42, 0.

\bibitem{Liu2010126}
X.~F. Liu, C.~K. Tse, M.~Small, Complex network structure of musical
  compositions: Algorithmic generation of appealing music, Physica A:
  Statistical Mechanics and its Applications 389~(1) (2010) 126 -- 132.
\newline\urlprefix\url{http://www.sciencedirect.com/science/article/pii/S0378437109006827}

\bibitem{Lu:2014}
S.~Lu, J.~Kang, W.~Gong, D.~Towsley, Complex network comparison using random
  walks, in: Proceedings of the 23rd International Conference on World Wide
  Web, WWW '14 Companion, ACM, New York, NY, USA, 2014, pp. 727--730.
\newline\urlprefix\url{http://doi.acm.org/10.1145/2567948.2579363}

\bibitem{Manaris:2005}
B.~Manaris, J.~Romero, P.~Machado, D.~Krehbiel, T.~Hirzel, W.~Pharr, R.~B.
  Davis, Zipf's law, music classification, and aesthetics, Comput. Music J.
  29~(1) (2005) 55--69.
\newline\urlprefix\url{http://dx.doi.org/10.1162/comj.2005.29.1.55}

\bibitem{McKayF06}
C.~McKay, I.~Fujinaga, Musical genre classification: Is it worth pursuing and
  how can it be improved?, in: ISMIR, 2006, pp. 101--106.

\bibitem{Milo824}
R.~Milo, S.~Shen-Orr, S.~Itzkovitz, N.~Kashtan, D.~Chklovskii, U.~Alon, Network
  motifs: Simple building blocks of complex networks, Science 298~(5594) (2002)
  824--827.
\newline\urlprefix\url{http://science.sciencemag.org/content/298/5594/824}

\bibitem{Newman:2010}
M.~Newman, Networks: An Introduction, Oxford University Press, Inc., New York,
  NY, USA, 2010.

\bibitem{newmanHandbook}
M.~E.~J. Newman, Random graphs as models of networks, in: Handbook of Graphs
  and Networks, Wiley-VCH Verlag, 2005, pp. 35--68.

\bibitem{jung}
J.~O'Madadhain, D.~Fisher, S.~White, Y.~Boey, {The JUNG (Java Universal
  Network/Graph) Framework}, Tech. rep., UCI-ICS (Oct. 2003).
\newline\urlprefix\url{http://www.datalab.uci.edu/papers/JUNG\_tech\_report.html}

\bibitem{pardo}
T.~Pardo, L.~Antiqueira, M.~das Gracas~Nunes, O.~N. Oliveira,
  L.~da~Fontoura~Costa, Using complex networks for language processing: The
  case of summary evaluation, in: Communications, Circuits and Systems
  Proceedings, 2006 International Conference on, vol.~4, 2006, pp. 2678--2682.

\bibitem{Patra:2013}
B.~G. Patra, D.~Das, S.~Bandyopadhyay, Unsupervised approach to hindi music
  mood classification, in: Proceedings of the First International Conference on
  Mining Intelligence and Knowledge Exploration - Volume 8284, MIKE 2013,
  Springer-Verlag New York, Inc., New York, NY, USA, 2013, pp. 62--69.

\bibitem{Prather:1996}
R.~E. Prather, Harmonic analysis from the computer representation of a musical
  score, Commun. ACM 39~(12es).
\newline\urlprefix\url{http://doi.acm.org/10.1145/272682.272716}

\bibitem{Pressing}
J.~Pressing, Improvisation: methods and models, in: Generative Processes in
  Music: the Psychology of Performance, Improvisation and Composition, Oxford
  University Press, Oxford, UK, 2001.

\bibitem{Ryynanen:2008}
M.~P. Ryyn\"{a}nen, A.~P. Klapuri, Automatic transcription of melody, bass
  line, and chords in polyphonic music, Comput. Music J. 32~(3) (2008) 72--86.
\newline\urlprefix\url{http://dx.doi.org/10.1162/comj.2008.32.3.72}

\bibitem{SchnitzerEtAl_2009}
D.~Schnitzer, A.~Flexer, G.~Widmer, A filter-and-refine indexing method for
  fast similarity search in millions of music tracks, in: Proceedings of the
  10th International Society for Music Information Retrieval Conference, Kobe,
  Japan, 2009, pp. 537--542.

\bibitem{smith}
S.~Smith, Jazz Theory Book, 2008.
\newline\urlprefix\url{http://www.cs.uml.edu/~stu/JazzTheory.pdf}

\bibitem{Su:2012}
D.~Su, P.~Fung, Personalized music emotion classification via active learning,
  in: Proceedings of the Second International ACM Workshop on Music Information
  Retrieval with User-centered and Multimodal Strategies, MIRUM '12, ACM, New
  York, NY, USA, 2012, pp. 57--62.

\bibitem{watts1998cds}
D.~J. Watts, S.~H. Strogatz, {Collective dynamics of'small-world'networks.},
  Nature 393~(6684) (1998) 409--10.

\bibitem{ASI:ASI20876}
M.~A. Winget, Annotations on musical scores by performing musicians:
  Collaborative models, interactive methods, and music digital library tool
  development, Journal of the American Society for Information Science and
  Technology 59~(12) (2008) 1878--1897.
\newline\urlprefix\url{http://dx.doi.org/10.1002/asi.20876}

\bibitem{Yang:2006}
Y.-H. Yang, C.-C. Liu, H.~H. Chen, Music emotion classification: A fuzzy
  approach, in: Proceedings of the 14th ACM International Conference on
  Multimedia, MM '06, ACM, New York, NY, USA, 2006, pp. 81--84.
\newline\urlprefix\url{http://doi.acm.org/10.1145/1180639.1180665}

\bibitem{zenni}
S.~Zenni, The aesthetics of duke ellington's suites: The case of "togo brava",
  Black Music Research Journal 21~(1) (2001) 1--28.
\newline\urlprefix\url{http://www.jstor.org/stable/3181592}

\end{thebibliography}

% \balancecolumns

\end{document}